\RequirePackage{ifpdf}
\documentclass{JHEP3}

\usepackage{graphicx}
\usepackage{cite}
\usepackage{booktabs}

\newcommand{\as}{\alpha_s}

\newcommand{\be}{\begin{equation}}

\newcommand{\ee}{\end{equation}}
\newcommand{\bea}{\begin{eqnarray}}
\newcommand{\eea}{\end{eqnarray}}

\newcommand{\smallz}{{\scriptscriptstyle Z}}
\newcommand{\mz}{m_\smallz}
\newcommand{\smallw}{{\scriptscriptstyle W}}
\newcommand{\smallv}{{\scriptscriptstyle V}}
\newcommand{\mv}{m_\smallv}
\newcommand{\mw}{m_\smallw}
\newcommand{\mwc}{m_{\smallw 0}}

\newcommand{\mwj}{m_{\smallw, j}}

\newcommand{\mb}{m_b}

\newcommand{\ptw}{p_\perp^{\ell\nu}}
\newcommand{\ptz}{p_\perp^{\ell^+\ell^-}}
\newcommand{\ptv}{p_\perp^{\smallv}}

\newcommand{\qsh}{Q_{sh}}
\newcommand{\mush}{\mu_{sh}}
\newcommand{\gev}{\textrm{GeV}}
\newcommand{\mev}{\textrm{MeV}}
\newcommand{\kfactor}{$K$-\textrm{factor}\,}

\newcommand{\mcnlo}{{\sc MC@NLO}\,\,}
\newcommand{\mcnlonospace}{{\sc MC@NLO}}
\newcommand{\amcnlo}{{\sc MadGraph5\_aMC@NLO}\,\,}
\newcommand{\amcnlonospace}{{\sc MadGraph5\_aMC@NLO}}
\newcommand{\powheg}{{\sc POWHEG}\,\,}
\newcommand{\powhegnospace}{{\sc POWHEG}}
\newcommand{\powhegbox}{{\sc POWHEG-BOX}\,\,}
\newcommand{\powhegboxnospace}{{\sc POWHEG-BOX}}
\newcommand{\pythia}{{\sc Pythia8}\,\,}
\newcommand{\pythianospace}{{\sc Pythia8}}
\newcommand{\herwig}{{\sc Herwig++}\,\,}
\newcommand{\herwignospace}{{\sc Herwig++}}
\newcommand{\sherpa} {{\sc Sherpa}\,\,}

\newcommand{\fivebveto}{{\rm 5FS-Bveto\,\,}}
\newcommand{\fivebvetonospace}{{\rm 5FS-Bveto}}
\newcommand{\dyqt}  {{\sc DYqT}}
\newcommand{\dyres}  {{\sc DYRes}}
\newcommand{\resbos}  {{\sc ResBos}}

\title{Lepton-pair production in association with a $b\bar b$ pair
  and the determination of the $W$ boson mass}

\author{ Emanuele Bagnaschi,\\
Deutsches Elektronen-Synchrotron (DESY), Notkestra{\ss}e 85, 22607 Hamburg, Germany  \\
Email:~\email{emanuele.bagnaschi@desy.de}}

\author{ Fabio Maltoni,\\
  Centre for Cosmology, Particle Physics and Phenomenology (CP3),
Universit\'e catholique de Louvain, Chemin du Cyclotron 2, 1348 Louvain-La-Neuve, Belgium
  \\
Email:~\email{fabio.maltoni@uclouvain.be}}

\author{ Alessandro Vicini,\\
Tif lab, Dipartimento di Fisica, Universit\`a degli Studi di Milano \emph{and}\\
INFN, Sezione di Milano,
Via Celoria 16, I-20133 Milano, Italy\\
Email:~\email{alessandro.vicini@mi.infn.it}}

\author{ Marco Zaro\\
          Nikhef, Science Park 105, NL-1098 XG Amsterdam, The Netherlands, \emph{and}\\
          Sorbonne Universit\'es, UPMC Univ. Paris 06, UMR 7589, LPTHE, F-75005, Paris, France, \emph{and}
          CNRS, UMR 7589, LPTHE, F-75005, Paris, France\\
Email:~\email{m.zaro@nikhef.nl}}

\abstract{
We perform a  study of lepton-pair production in association with bottom quarks at the LHC
based on the predictions obtained at next-to-leading order in QCD,
both at fixed order and matched with a QCD parton shower.
We consider a comprehensive set of observables and estimate the associated theoretical uncertainties by studying the dependence on the perturbative QCD scales (renormalisation, factorisation and shower) and by comparing different parton-shower models (\pythia and {\sc Herwig++}) and matching schemes (\amcnlo and {\sc POWHEG}).
Based on these results, we propose a simple procedure to include bottom-quark effects in neutral-current Drell-Yan production, going
beyond the standard massless approximation. Focusing on the inclusive lepton-pair transverse-momentum distribution $\ptz$, we quantify the impact of such effects
on the tuning of the simulation of charged-current Drell-Yan observables and the $W$-boson mass determination.}

\preprint{DESY 18-024, CP3-18-16, NIKHEF/2018-008, TIF-UNIMI 2018-2}

\keywords{Drell-Yan, heavy quarks, W boson mass, Monte Carlo simulations}
\begin{document}

\clearpage{}\section{Introduction}
\label{sec:intro}
The production of a pair of high-transverse-momentum leptons in hadron-hadron collisions
is one of the historical testing grounds of perturbative Quantum Chromodynamics (QCD). At the lowest order (Born approximation), it proceeds through the parton level amplitude $q\bar q \to Z/\gamma^* \to\ell^+\ell^-$, which once folded with parton distributions, gives the (first order) prediction for the inclusive rate, the so called Drell-Yan (DY) process. As shown a long time ago, higher-order QCD corrections~\cite{Altarelli:1979ub} are important and need to be included to improve both the precision and accuracy of the calculation.  Predictions for more exclusive final states can also be calculated in perturbative QCD,  including for example QCD jets or heavy quarks (bottom o top quarks).

The theoretical interest in this process is matched (or even surpassed) by the experimental one: dilepton pairs in high-energy collisions have been always considered golden final states for Standard Model measurements as  well as for new physics searches. In the long and impressive list of experimental results which feature an $\ell^+\ell^-$ final state at hadron colliders,
the measurement of the inclusive lepton-pair transverse-momentum distribution,
conventionally dubbed $\ptz$, in neutral-current (NC) DY,
has now reached an impressive level of accuracy at the LHC.
Using 8 TeV measurements, ATLAS \cite{Aad:2015auj} and CMS \cite{Khachatryan:2016nbe}  have attained
a total experimental uncertainty below the 0.5\% level
in a large interval of transverse-momentum values, ranging between 2 and 50 GeV.
These achievements represent a formidable challenge for the theoretical predictions which need to combine approximate results obtained  with different techniques (fixed higher-order corrections vs resummation to all orders of logarithmically-enhanced terms) matched  together,
to perform a sensible test of the Standard Model (SM).

As mentioned above, the DY processes start at Leading Order (LO)
as a purely electroweak (EW) scattering, $q\bar q \to Z/\gamma^* \to \ell^+\ell^-$.
The radiative corrections are exactly known up to
${\cal O}(\alpha_s^2)$
~\cite{Hamberg:1990np,Anastasiou:2003ds,Anastasiou:2003yy,Melnikov:2006di,Melnikov:2006kv}
in the strong-interaction coupling,
while the ${\cal O}(\alpha_s^3)$ threshold corrections
have been presented in Refs.~\cite{Ahmed:2014cla,Ahmed:2014uya}
for the inclusive cross section and for the rapidity distribution
of the dilepton pair, respectively. The corrections up to ${\cal O}(\alpha)$
\cite{Wackeroth:1996hz,Baur:1998kt,Baur:2001ze,Dittmaier:2001ay}
in the EW coupling are available. The $\ptz$ spectrum, at large transverse momenta, is known with
next-to-next-to-leading (NNLO) QCD accuracy
\cite{Boughezal:2015ded,Boughezal:2016isb,Ridder:2016nkl,Gehrmann-DeRidder:2016jns,Gehrmann-DeRidder:2017mvr}.

The approximate inclusion of initial-state logarithmically-enhanced corrections
to all perturbative orders is necessary to perform a meaningful comparison with differential distributions of the leptons and is known up to next-to-next-to-leading
logarithmic (NNLL) QCD accuracy \cite{Collins:1984kg,Catani:2000vq}
with respect to $\log(p_\perp^V/m_V)$,
where $p_\perp^V$ is the lepton-pair transverse momentum and $m_V$ is
the relevant gauge boson mass ($V=W,Z$);
these corrections have been implemented in simulation codes such as
\resbos\, \cite{Balazs:1997xd} or
\dyqt/\dyres\, \cite{Bozzi:2010xn,Catani:2015vma}.

The problem of merging fixed-order and all-order results,
avoiding double counting, has been separately discussed in the context of QCD
\cite{Balazs:1997xd,Bozzi:2008bb,Frixione:2002ik,Nason:2004rx,Hoeche:2014aia,Karlberg:2014qua,Alioli:2015toa}  and in the EW \cite{CarloniCalame:2006zq,CarloniCalame:2007cd,Placzek:2003zg} computations. QCD and EW results have to be combined together to obtain a
realistic description of the DY final states:
general-purpose Shower Monte Carlo programs,
such as \pythia \cite{Sjostrand:2007gs, Sjostrand:2014zea},
\herwig \cite{Bahr:2008pv} or \sherpa \cite{Gleisberg:2008ta},
include the possibility of multiple photon, gluon and quark emissions
via a combined application of QCD and QED Parton Shower (PS),
formally retaining only LL accuracy
in the respective logarithmic expansions.
The combined matching with exact matrix elements of QCD and QED PS,
respecting the NLO-QCD and NLO-EW accuracy on the quantities inclusive
with respect to additional radiation has been presented in \cite{Bernaciak:2012hj,Barze:2012tt,Barze:2013yca,CarloniCalame:2016ouw}.
For a systematic comparison of the tools that simulate the DY processes
including higher-order radiative corrections see Ref.~\cite{Alioli:2016fum}.

Given the very precise experimental results available for
all the relevant observables in the NC-DY process,
it is necessary to carefully quantify all possible sources of uncertainties, including those coming from sets of radiative corrections which are formally subdominant in the perturbative expansions
in the strong and electromagnetic couplings. These higher-order corrections include contributions from
subprocesses with additional coloured particles in the final state. Among them, the production of a lepton pair in association with a bottom quark pair is of special interest. In this case the presence of (at least) two well-separated perturbative scales, $m_b$ and $m_Z$, where  $\Lambda_{\rm QCD}\ll m_b\ll m_Z$,  can potentially lead to large perturbative logarithmic corrections whose impact needs to be carefully assessed on a observable-by-observable basis.  Starting from the pragmatic point of view that $b$ quarks can be found as partons in the proton, it can be easily checked that at the LHC $b \bar b \to Z$ provides a small but non negligible fraction of the total cross section for inclusive lepton-pair production. As this contribution affects both the normalisation and the shape of the kinematic distributions, a careful analysis is required that can also estimate bottom mass effects.

A first estimate of  the relative importance of contributions from different flavours of quarks in DY processes can be obtained by computing the individual contributions of quarks
to the total cross section for NC-DY in the so-called five-flavour scheme (5FS), i.e. in terms of five massless active quarks. Results are shown in Table~\ref{tab:flavourdecomposition}. While this decomposition is not physical per se as it is ambiguous beyond NLO, it allows to appreciate the precision needed in the predictions of NC-DY production through heavy quark flavours.
\begin{table}[!h]
  \renewcommand{\arraystretch}{1.2}
\begin{center}
\begin{tabular}{@{}|c|c|c|@{}}
\toprule
\textbf{initial-state quark} &  \textbf{cross section (pb)} & \textbf{\%} \\
  \midrule
$d$ &  $277.98 \pm 0.14$ &  37.4 \\
$u$ &  $245.54 \pm 0.13$ & 33.0 \\
$s$ &  $127.90 \pm 0.09$ &  17.2 \\
$c$ &  $63.86\pm 0.07$ &  8.6 \\
$b$ &  $28.31 \pm 0.05$ &  3.8 \\
\midrule
\textbf{total} &  $743.61\pm 0.22$ &   100.0\\
\bottomrule
\end{tabular}
\end{center}
\caption{\label{tab:flavourdecomposition}
  Flavour decomposition of the total cross section within the acceptance cuts described in Section \ref{sec:setup},  computed at NLO accuracy with five active massless quarks in the proton.
}
\end{table}
\begin{figure}[!h]
  \centering
  \includegraphics[width=0.6\textwidth,angle=0]{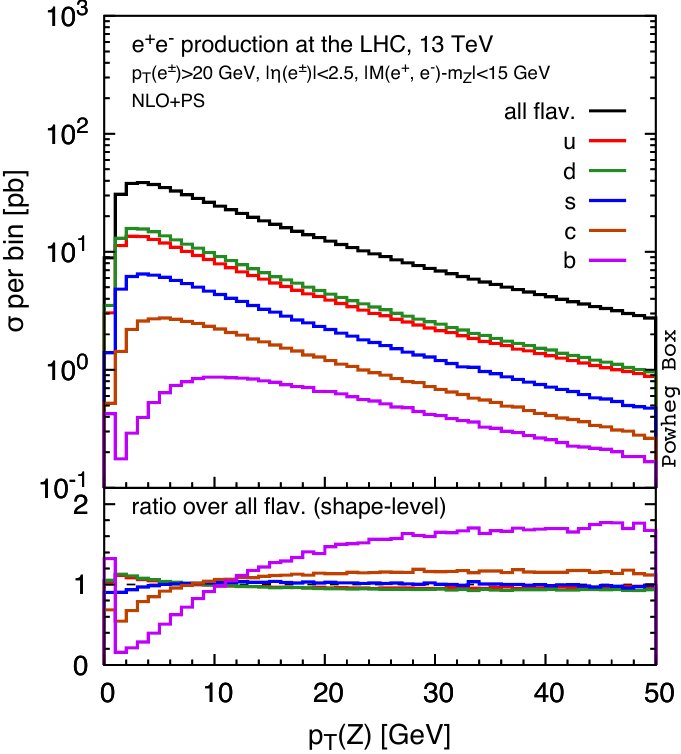}
\caption{\label{fig:flavourdecomposition}
  Flavour decomposition of the $\ptz$ distribution computed with five active
  massless quarks in the proton.
}
\end{figure}

Although in this case all the active flavours in the proton are described as massless,
in the case of heavy quarks the effect of their mass $m_Q$ is introduced in an initial condition
that controls the evolution equations of the respective parton densities;
the latter start to be non-zero at an energy scale of ${\cal O}(m_Q)$.
These boundary conditions, combined with all the other constraints satisfied by the proton PDFs,
result a heavy quark PDFs behaviour
which is typically quite different from those of light quarks,
leading to significant differences in observables like the $\ptz$ distribution.
In Figure~\ref{fig:flavourdecomposition} we appreciate the shape of the various
contributions initiated by different quark flavours, which display a harder spectrum in
the case of heavy quarks. We conclude that given the present experimental uncertainty,
the bottom-quark contribution to the $\ptz$ distribution
deserves a dedicated study of the residual uncertainties due to the treatment of the bottom mass effects.
With the latter, we mean an improved description of the bottom-quark kinematics, including mass-dependent terms, at NLO-QCD.

The $\ptz$ distribution provides access to a large range of scales and therefore it offers a stringent test of perturbative QCD in different regimes:
in the low-momentum region it is sensitive to non-perturbative QCD contributions
and possibly to the flavour structure of the proton \cite{Konychev:2005iy}.
The precise knowledge of this part of the $\ptz$ spectrum
makes it possible to calibrate the non-perturbative models that describe
the partonic transverse degrees of freedom inside the proton.
Such models, implemented e.g.~in QCD PS,
are then used to simulate other scattering processes, and their uncertainties
propagate in the prediction of the new observables.
A striking example is the determination of the $W$-boson mass $\mw$
\cite{Aaltonen:2013vwa,D0:2013jba,Aaboud:2017svj},
which relies on the $\ptz$ input to obtain an accurate simulation of
the $W$-boson transverse-momentum spectrum
and in turn of the leptonic final state.
Eventually, the extraction of $\mw$ displays a strong sensitivity to the modelling assumptions
for the low-momentum part of the $\ptw$ spectrum.
In this context, it is important to remark that
the heavy-quark contribution is different in CC- and NC-DY,
because of the different initial-state flavour structure,
following from electric charge conservation and Cabibbo-Kobayashi-Maskawa (CKM) mixing.
More specifically, the bottom-quark effects which are present in NC-DY
are marginal in CC-DY as the bottom-quark density appears only
in the CKM-suppressed $c\bar b$ initiated subprocess.
If (part of) the perturbative effects are non-universal and flavour dependent,
as it is the case for the bottom-quark contributions,
then a non-perturbative model based on the fit of NC-DY data
could include in its parameterisation these effects
and erroneously propagate them also to processes like CC-DY,
where instead they are absent or marginal.
In summary, improving the accuracy and precision of the heavy-quark contributions to the inclusive $Z$-boson production, is relevant:
$i)$ to reduce the amount of information
which has to be encoded in a model that describes the low-momentum part of the gauge-boson transverse-momentum spectrum;
$ii)$ to capture some non-universal flavour-dependent contributions,
which distinguish massless and massive quarks,
leaving for the non-perturbative model, to a greater extent, only universal, flavour-independent effects.

Understanding heavy-quark contributions to lepton-pair production
benefits also from the analysis of exclusive final states where the leptons are associated to a pair of bottom-antibottom quarks, which are explicitly tagged in terms of either $b-$jets or $B$ hadrons. The presence of additional energy scales, such as the masses and transverse momenta of the measured $b$ quarks, imposes non-trivial constraints on the structure of the radiative corrections that have to be included in the simulations to obtain accurate predictions. Understanding these final states is also propedeutic to that of other heavy systems, e.g., a Higgs boson or a $t\bar t$ pair, accompanied by a $b\bar b$ pair.

The production of $\ell^+\ell^-b\bar b$, with the inclusion of NLO-QCD corrections
has been discussed in  Refs.~\cite{Campbell:2003dd,Campbell:2005zv,Maltoni:2005wd,FebresCordero:2008ci,Cordero:2009kv}, and more recently in Ref.~\cite{Baranov:2017tig}, for final states with at least one or with two tagged $b$-quark jets, in the so called four-flavour scheme (4FS), namely using a parameterisation of the proton structure in terms of only four active quarks and considering bottom quarks in the final state as massive.
The matching of fixed-order matrix elements with a Parton Shower has been implemented in Ref.~\cite{Frederix:2011qg} in the \amcnlo framework \cite{Alwall:2014hca} and in Ref.~\cite{Krauss:2016orf} in the \sherpa framework.

Given on the one hand
the very high level of precision necessary to obtain sensible results
in the description of $\ptz$ and eventually in the determination of $\mw$,
and, on the other hand, the link between exclusive and inclusive final states characterised by the presence of heavy quarks, we deem necessary to scrutinise the theoretical uncertainties affecting the prediction of the observables for $\ell^+\ell^-b\bar b$ final states.
To this aim, we present a systematic comparison of two different schemes
of matching between fixed order results with a QCD PS\footnote{
For a similar study in the case of Higgs production in gluon fusion, cfr. Ref.~\cite{Bagnaschi:2015bop}.
  }, namely the \amcnlo and the \powheg ones;
we expose the impact of different treatments for the QCD PS phase space assignment
and we present the phenomenological results obtained with two QCD PS models, namely \pythia and \herwignospace.

To summarise we $i)$ thoroughly compare the implementations of the production of a lepton pair in association with a $b\bar b$ pair in the 4FS between two available Monte Carlo
event generators, with a systematic analysis of all the relevant QCD theoretical uncertainties;
$ii)$ consider the effects of including  bottom-quark-mass  contributions on the
inclusive transverse-momentum spectrum of the lepton pair;
$iii)$ estimate the impact that such contributions may have on the determination of the $W$-boson mass.

The paper is structured as follows. In Section~\ref{sec:setup}
we describe the setup employed for the numerical simulations;
in Section~\ref{sec:zbb} we study
lepton-pair production in association with bottom quarks in the 4FS,
we compare the implementations in the \amcnlo and \powheg frameworks
and discuss several sources of theoretical uncertainties for inclusive observables.  We defer to Appendix~\ref{sec:appendix} an extensive comparison of more exclusive observables.
In Section~\ref{sec:ptz}, in order to evaluate the effects of the bottom-quark mass, we consistently combine the 4FS prediction for $\ell^+\ell^-b\bar b$ with the usual 5FS inclusive lepton-pair calculation and study the transverse-momentum distribution $\ptz$. In Section~\ref{sec:impact} we consider the impact of bottom-quark mass effects on CC-DY observables and on the determination of the $W$-boson mass. We draw our conclusions in Section~\ref{sec:conclusions}.

\clearpage{}

\clearpage{}
\section{Setup of the simulations}
\label{sec:setup}
In this work we study the processes
\bea
&& pp\to \ell^+\ell^-+X,\label{eq:proc-NC5F}\\
&& pp\to \ell^+\ell^-+b\bar b+X, \label{eq:proc-NC4F}\\
&& pp\to \ell^+\nu_\ell+X,\label{eq:proc-CC}
\eea
for one leptonic family, in a setup typical of the LHC,
with $\sqrt{S}=13$ TeV.

Unless stated otherwise, the simulations have been run at NLO+PS accuracy
with the codes
\amcnlo
(all the processes have been generated within the same computational framework)
and \powhegboxnospace.
Both codes have been interfaced with the same QCD-PS programs,
namely \pythia (version 8.215, Monash tune)
\cite{Sjostrand:2014zea,Skands:2014pea}
and \herwig (version 2.7.1) \cite{Bahr:2008pv,Bellm:2013hwb}.
We did not include any QED effect via QED PS.
The simulation of the underlying event is not performed.
For the proton parton-density parameterisation
we use the NNPDF 3.0 NLO PDFs with $\as(\mz)=0.118$,
with the same flavour-number scheme as for the hard process \cite{Ball:2014uwa}.
The SM parameters are set to the following values~\cite{Patrignani:2016xqp,Aaltonen:2013iut}:
\bea
&&\alpha = 1/132.507,
\quad
G_\mu =1.16639 \cdot 10^{-5} \, \gev^{-2},
\quad
m_b = 4.7\,\gev,
\quad
m_t = 173\,\gev,
\nonumber
\\
&&
\mz=91.188\,\gev,
\quad
\Gamma_\smallz=2.4414\, \gev,
\quad
\mw=80.385\,\gev,
\quad
\Gamma_\smallw=2.085\, \gev,
\nonumber\\
&&
|V_{ud}| \, = \, |V_{cs}| = 0.975,
\quad
|V_{us}| \, = \, |V_{cd}| = 0.222,
\quad
|V_{tb}| = 1,
\nonumber \\
&&
|V_{ub}| \, = \, |V_{cb}| =\,  |V_{td}| \, = \, |V_{ts}| \, = 0.
\eea
It is understood that the quoted value of $m_b$ is employed only in the 4FS process,
Eq.~\ref{eq:proc-NC4F}. For the central value of the renormalisation and factorisation scales,
we use for all samples the lepton-pair transverse mass divided by four:
\be
\mu = \frac{1}{4}\sqrt{M^2(\ell^+, \ell^-) + (\ptz)^2}\,.
\label{eq:transversemass}
\ee
For the 5FS NC-DY, this choice was advocated in Ref.~\cite{Lim:2016wjo}.
The only exception to what stated above is represented by the samples
for charged-current Drell-Yan used in Section~\ref{sec:template},
where the transverse mass of
the (reconstructed) $W$ boson is used:
\be
    \mu_{\rm CC-DY} = \sqrt{M^2(\ell^+, \nu) + (\ptw)^2}\,.
\label{eq:transversemassCC}
\ee
In Eq.~\ref{eq:transversemass} (\ref{eq:transversemassCC})
$M^2$ and $\ptz$ ($\ptw$) are respectively
the squared invariant mass and the transverse momentum
of the lepton pair (lepton-neutrino pair).

In the simulation of processes \ref{eq:proc-NC5F} and \ref{eq:proc-NC4F},
a generation cut $M(\ell^+, \ell^-) > 30\, \gev$ is applied
in order to avoid the singularity related to the photon contribution.
At the analysis level, for the
processes \ref{eq:proc-NC5F} and \ref{eq:proc-NC4F}, we apply a cut on the transverse momentum of each lepton,
$p_\perp (\ell^\pm) > 20\, \gev$, and on their pseudorapidity, $\eta(\ell^\pm)<2.5$,
together with an invariant-mass cut around the $Z$ peak, $\left|M(\ell^+,\ell^-) -m_Z\right|<15\,\gev$. In process \ref{eq:proc-CC} we impose
a cut on the charged-lepton transverse momentum and on the missing transverse energy (the transverse momentum of the neutrino),
$p_\perp(\ell^+) > 20\, \gev, p_\perp^{\rm miss} > 20\,\gev$,
and again a pseudorapidity cut $|\eta(\ell^+)| < 2.5$ for the charged lepton.\\

\clearpage{}

\clearpage{}\section{Lepton-pair production in association with bottom quarks in the 4FS}
\label{sec:zbb}
In this section we study the process $pp\to \ell^+\ell^- b \bar b+X$
in the 4FS.
The bottom quarks, absent in the proton PDFs,
are treated as massive final-state hard partons.
In Section~\ref{sec:matchingschemes} we compare
the formulation of two matching recipes to combine fixed- and all-order results
with NLO-QCD accuracy, with special attention to the details of the inclusion
of multiple parton radiation and to the perturbative sources of uncertainty.
We then discuss phenomenological results, obtained with the setup outlined in Section~\ref{sec:setup}:
in Section~\ref{sec:zbbscale} we first determine a typical scale
which characterises the process and then in sections~\ref{sec:zbbpt} and \ref{sec:ptzfour}
we compare respectively the results of the various matched schemes
for the transverse momentum of the $\ell^+\ell^- b\bar b$ system and
the effect of higher-order corrections and of the matching for the lepton-pair transverse momentum.
The interested reader can find more details on other differential observables in the Appendix~\ref{sec:appendix}.

\subsection{\amcnlo and \powhegbox implementations}
\label{sec:matchingschemes}
In this section we compare two different matching schemes,
namely those implemented in the \amcnlo\ and \powhegbox Monte Carlo event
generators. The aim is to disentangle genuine bottom-quark effects from those due to a different treatment of higher-order emissions in the two approaches.

The simulation of scattering processes in hadron collisions
requires not only the inclusion of fixed-order corrections
in order to obtain a reliable estimate of the overall normalisation
of the cross sections,
but also the inclusion of multiple parton emissions at all orders
in order to achieve a realistic description of the shape of the distributions.
The possibility
of simultaneously preserving the NLO accuracy for all the observables
that are regular when the radiative corrections are included,
together with the description of multiple parton emissions,
is achieved by matching fixed- and all-order results.
Different matching schemes have been proposed in the literature, and here we focus on \mcnlonospace~\cite{Frixione:2002ik}
and \powhegnospace~\cite{Nason:2004rx};
they share the same fixed-order accuracy, but differ for the inclusion
of subsets of higher-order terms.
The latter are beyond the accuracy of the calculation with respect to
the coupling constant expansion
and are formally subleading in a logarithmic expansion in powers
of $\log(\ptv/\mv)$,
where $\ptv$ represents a generic transverse-momentum variable that yields
a singularity of the amplitude
in the limit $\ptv\to 0$, and $\mv$ is the invariant mass of the
system whose transverse momentum is described by $\ptv$;
although subleading,
these terms can nevertheless have a sizeable numerical impact on
the predictions,
in particular for those observables that have only the lowest order accuracy.

The matching of fixed- and all-orders corrections
should avoid double counting between the two contributions and
respect the ordering of the emissions of QCD partons,
in order to preserve the logarithmic accuracy of the results.
In a Monte Carlo approach,
the hardest QCD parton,
with respect to the radiation ordering parameter $t$,
plays a special role,
for it receives the exact matrix element corrections of the fixed-order
calculation.
The subsequent emissions are instead generated by the QCD-PS
and the associated phase-space volume is part of the matching prescription.

A generic scattering process,
whose lowest order (LO) is characterised by the presence of
$k$ final state particles, receives radiative corrections due to the emission
of $n$ additional partons.
In the \mcnlo approach
an event is generated according to the following
steps:
\begin{enumerate}
\item
  The event weight is split in two contributions,
  called standard ($ \mathbb S$-events) and hard ($ \mathbb H$-events), which describe final states with respectively
  $k$ and $k+1$ final state particles;
  both standard and hard terms are matched with a QCD-PS
  that generates $n$ additional parton emissions using:
  $i)$ the standard Sudakov form factor computed in the collinear approximation;
  $ii)$ an approximated phase-space measure;
  $iii)$ an upper limit for the hardness of the emission set by a scale $\qsh$
  called shower scale.

\item
  $ \mathbb H$-events  account for the exact real matrix-element corrections
  describing the first real emission,
  evaluated in the full phase space with exact integration measure.
  The double counting in the generation of the hardest parton between the PS
  and the exact matrix element is avoided with an appropriate counterterm.

\item
  $ \mathbb S$-events account for all the terms entering a NLO cross sections (Born, virtual, counterterms, etc.) except for
  the real-emission matrix elements and the corresponding counterterms.

\item
  The shower scale $\qsh$ associated to each $\mathbb S$-event is extracted from a probability distribution.
  The latter parametrically depends, event-by-event, on a reference scale, which we denote with the symbol $\mu_{sh}$ and which is computed considering the $\mathbb S$-event kinematics.
  For the corresponding $\mathbb H$-event, the maximum of the allowed values by the same distribution is used.
  The details of this procedure and the functional form of the distribution of are given in Section 2.4.4 of Ref.~\cite{Alwall:2014hca}.
\end{enumerate}

In the \powheg approach
an event with $n$ additional partons is generated according
to the following steps:
\begin{enumerate}
\item
  Each LO configuration is rescaled by a factor, usually denoted by $\tilde{B}$,
  that accounts for virtual corrections and the integral over the first real emission.
  This rescaling guarantees the full NLO accuracy for inclusive quantities.
\item
  The expression of the \powheg Sudakov form factor depends on the splitting,
  in the full real-emission matrix elements $R$,
  between the singular $R_s$ and a remaining regular part $R_f$,
  controlled by a scale $h$ according to:
  $R=R_s+R_f; R_s\equiv f(h,t) R,\,\,R_f=(1-f(h,t)) R$
  where the damping factor $f(h,t)$ depends on the radiation variable $t$,
  it goes to 1 in the collinear limit $t\to 0$ and vanishes for large $t$.
  The scale $h$ defines the region where the Sudakov suppression is active
  and the effects of multiple parton emissions are systematically included.

\item
  The probability of the first emission is evaluated using:
  $i)$ the \powheg Sudakov form factor;
  $ii)$ the exact radiation phase space;
  $iii)$ the exact matrix elements for the real emission.
\item
  The fact that the first parton with emission variable $t=\bar t$ is by
  construction the hardest is obtained:
  $i)$ computing the product of the Sudakov form factor for an emission
  with $\bar t$ with the corresponding real-emission matrix element;
  $ii)$ limiting the QCD-PS phase space at the value $\bar t$ of the emission
  variable $t$.
\item
  The QCD-PS populates the available phase space,
  assigning to the ordering variable $t$ a maximum value equal
  to the shower scale $\qsh$ of the event;
  this value is by default $t=\bar t$
  (where $\bar t$ varies on a event by event basis);
  however, a redefinition of the $\qsh$ value,
  different than $\bar t$, is allowed in the generation of the events based
  on the non singular part $R_f$ of the real-emission matrix element
  (remnant events), without spoiling the accuracy of the calculation;
  once $\qsh$ is assigned, the generation of $n-1$ additional emissions
  proceeds in the PS approximation
  for the branching probability and integration measure.

\end{enumerate}

These two approaches share the NLO-QCD accuracy in the prediction of the
total inclusive cross section,
and differ by the inclusion of terms of higher order in the perturbative
expansion in powers of $\as$.
As already said,
the latter are formally subleading with respect to the enhancement due to
$\log(\ptv/\mv)$ factors, but can nevertheless be numerically sizeable,
depending on the phase space region under study.

We note that in general when PS programs take in short-distance events,
a comparison is performed between the shower scale  $\qsh$ provided in the event (in the event-record field {\tt SCALUP}) and the corresponding scale that would be associated by the code based on its own phase space evaluation.
Since all the PS emissions must be ordered with respect to the hardness parameter $t$,
the smallest value between the two is eventually used in the QCD-PS.

We focus now our attention on the actual distribution of $\qsh$ in the event samples
produced in the two Monte Carlo frameworks, i.e.~the distribution of the values of the {\tt SCALUP} field of the event records.

\begin{figure}[!h]
    \centering
    \includegraphics[width=0.47\textwidth,angle=0,clip=true,trim={0.cm 5.5cm 0.cm 1cm}]{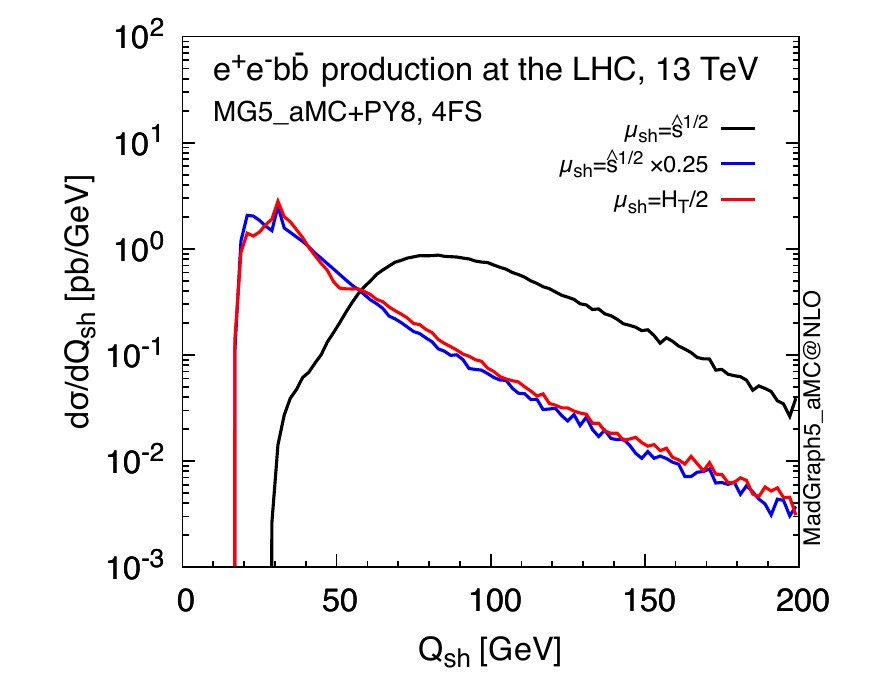}
    \includegraphics[width=0.47\textwidth,angle=0,clip=true,trim={0.cm 5.5cm 0.cm 1cm}]{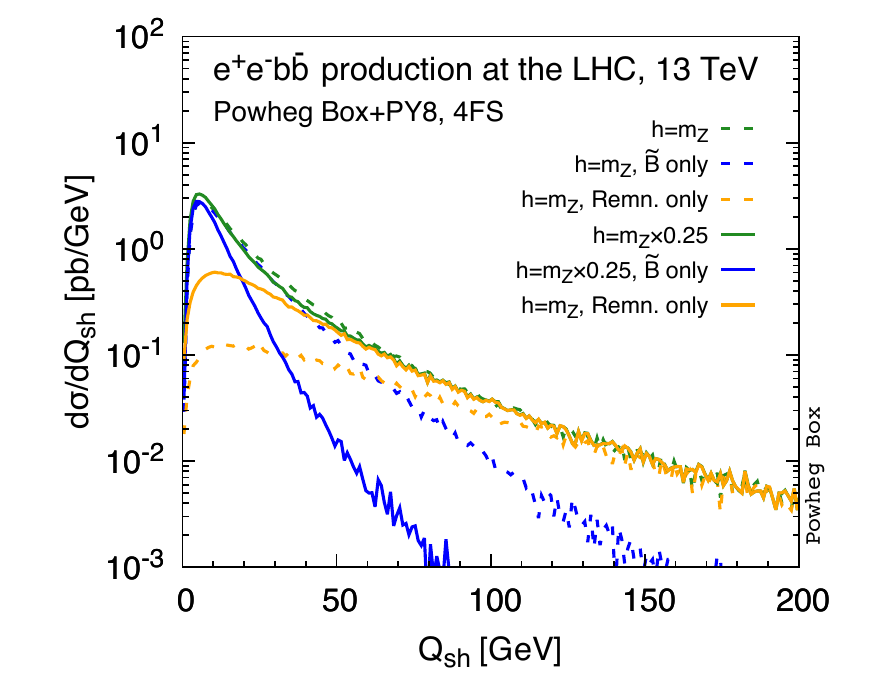}
    \caption{
      Probability distributions of the shower scale $\qsh$ (i.e.~{\tt SCALUP}) in various event samples generated by \amcnlo\ (left) and by the \powhegbox (right).
\label{fig:showerscale}
}
\end{figure}

In Figure~\ref{fig:showerscale} (left plot) we show
the histograms obtained with \amcnlo \\ for different choices of $\mu_{sh}$, in the 4FS simulation.\footnote{
    The choice $\mu_{sh}=\sqrt {\hat s}$ has been the default in \amcnlo up to version 2.5.2. From version 2.5.3, $\mu_{sh}=H_T/2$ is the new
    default. In these newer versions, it
    is still possible to use $\mu_{sh}=\sqrt {\hat s}$, by setting the {\tt i\_scale=0} in the subroutine {\tt assign\_ref\_scale}
    inside {\tt montecarlocounter.f}.
  }
For reference, we also show the same distribution in the 5FS simulation.

In Figure~\ref{fig:showerscale} (right plot) we show the distributions obtained in the \powhegbox 4FS simulation, for the events describing the singular part and the regular reminder of the real matrix element ($\tilde B$ and remnant events respectively), for different values of the damping factor $h$.
In the \powhegbox default setup, $\qsh$ coincides with the transverse momentum of the first emission.
As said, it is possible to preserve the logarithmic accuracy of the calculation
with a different choice of $\qsh$ for the remnant events.\footnote{
For example, we tested a different option in Figure~\ref{fig:nbjetbjet}
where we also considered the distributions of Figure~\ref{fig:showerscale} divided by a factor two.}
We recall that the generation probability of a remnant event depends also on
the scale $h$ introduced in the \powheg Sudakov form factor, as it can be seen from the plot.

From the comparison of the two plots, one may appreciate how different the distributions of $\qsh$ are
in the event samples generate by the two frameworks. This fact and the different structure of the matching procedure itself, impact the final results of the simulations (after showering),
formally beyond the claimed accuracy but still in a phenomenologically relevant way.
We also note that the final numerical results, after showering, depend on the PS ordering variable, so that different QCD-PS models may yield different results even if using the same sample,  $\qsh$ distribution.

To summarise, the formulation of the matching  and
the interface with the QCD PS are closely entangled;
their ambiguities and prescriptions represent an important source of
theoretical uncertainty, beside the canonical ones,
related e.g.~to the choice of the renormalisation and factorisation scales.
These additional uncertainties are relevant in the study of the shape
of the kinematic distributions, in particular of those sensitive to the
details of real radiation.

\subsection{Identification of the reference energy scale for
  lepton-pair production in association with a $b$-quark pair }
\label{sec:zbbscale}
In Ref.~\cite{Lim:2016wjo},
the production of a Higgs boson in association with a massive $b\bar b$ pair
is considered and,
following the discussion of Ref.~\cite{Maltoni:2012pa},
a universal logarithmic factor
$L\equiv\log\left({\cal Q}^2(z)/\mb^2\right)$,
associated to each $g\to b\bar b$ splitting is identified.
We adapt this approach to the case under study of the subprocess
$g/q(p_1)\, g/\bar q(p_2)\to \ell^+(q_+) \ell^-(q_-) b(k_1) \bar b(k_2)$
and obtain that the universal corrections have the form
\be
L=\log
\left(
\frac{M^2(\ell^+,\ell^-)}{\mb^2}\,\frac{(1-z_i)^2}{z_i}
\right)
\quad
\textrm{with}
\quad
z_i=\frac{M^2(\ell^+,\ell^-)}{s_i},
\quad
s_i=(q_+ + q_- + k_i)^2\,,
\ee
where $M^2(\ell^+,\ell^-)$ is the squared lepton-pair invariant mass.
The effective scale that characterises the process is
\be
\overline{M} \equiv M(\ell^+,\ell^-) \frac{(1-z_i)}{\sqrt{z_i}}\, .
\label{eq:Mbar}
\ee
\begin{figure}[!h]
\centering
\includegraphics[width=0.60\textwidth,angle=0,clip=true,trim={0.65cm 7.8cm 0.6cm 0.5cm}]{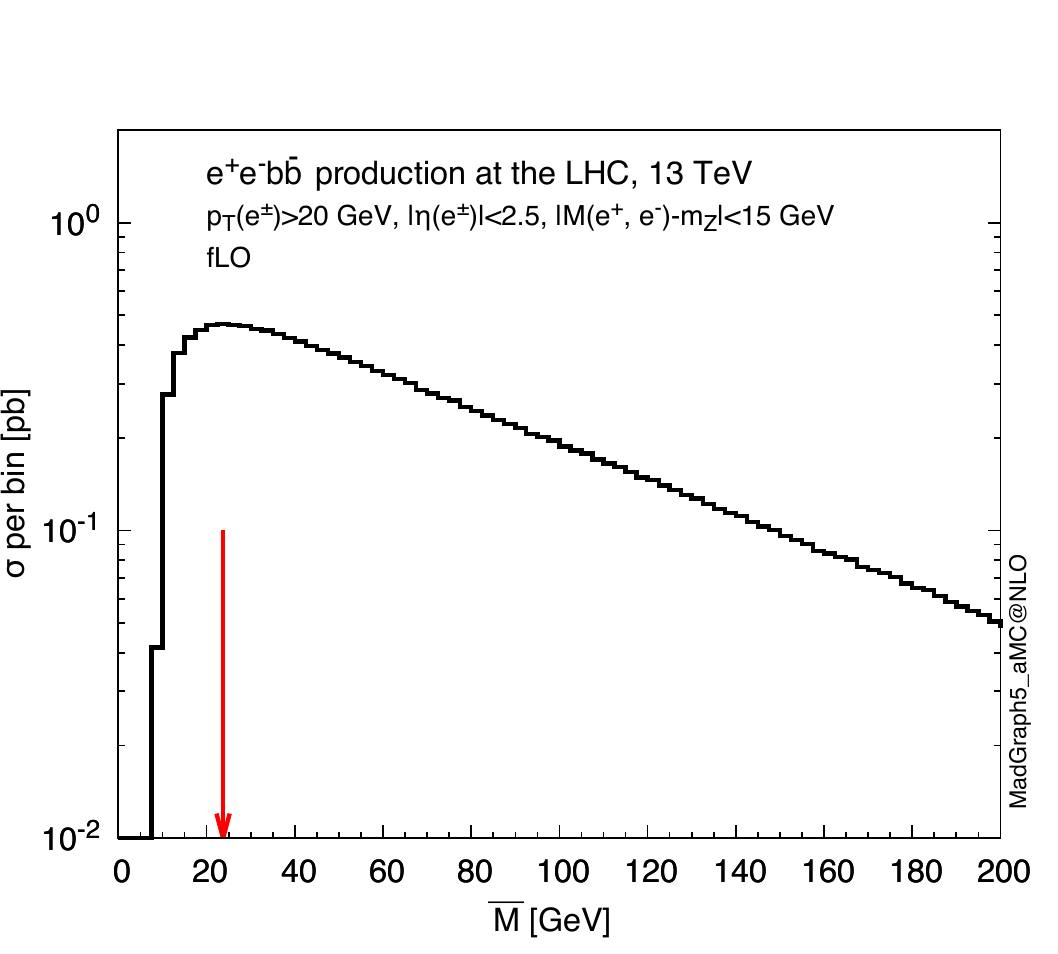}
\caption{\label{fig:effective-scales-Mbar}
  Event distribution, in $\ell^+\ell^-b\bar b$ production,
  with respect to the variable $\overline{M}$
  defined in Eq.~\protect\ref{eq:Mbar}. The red arrow corresponds to the peak of the distribution.}
\end{figure}
In Figure~\ref{fig:effective-scales-Mbar}
we plot the distribution $d\sigma / d\overline{M}$
and observe the presence of a peak at $\overline{M}\sim 25\,\gev $.
We interpret this value as one of the typical energy scales
that characterise the process
and justify, following Ref.\cite{Lim:2016wjo}, our choices described in Section
\ref{sec:setup} for the renormalisation and factorisation scales in the 4FS.

\subsection{The $\ell^+\ell^-b\bar b$ transverse-momentum distribution}
\label{sec:zbbpt}
\begin{figure}[!h]
\centering
\includegraphics[width=0.48\textwidth,angle=0,clip=true,trim={0.65cm 6.7cm 0.cm 0.4cm}]{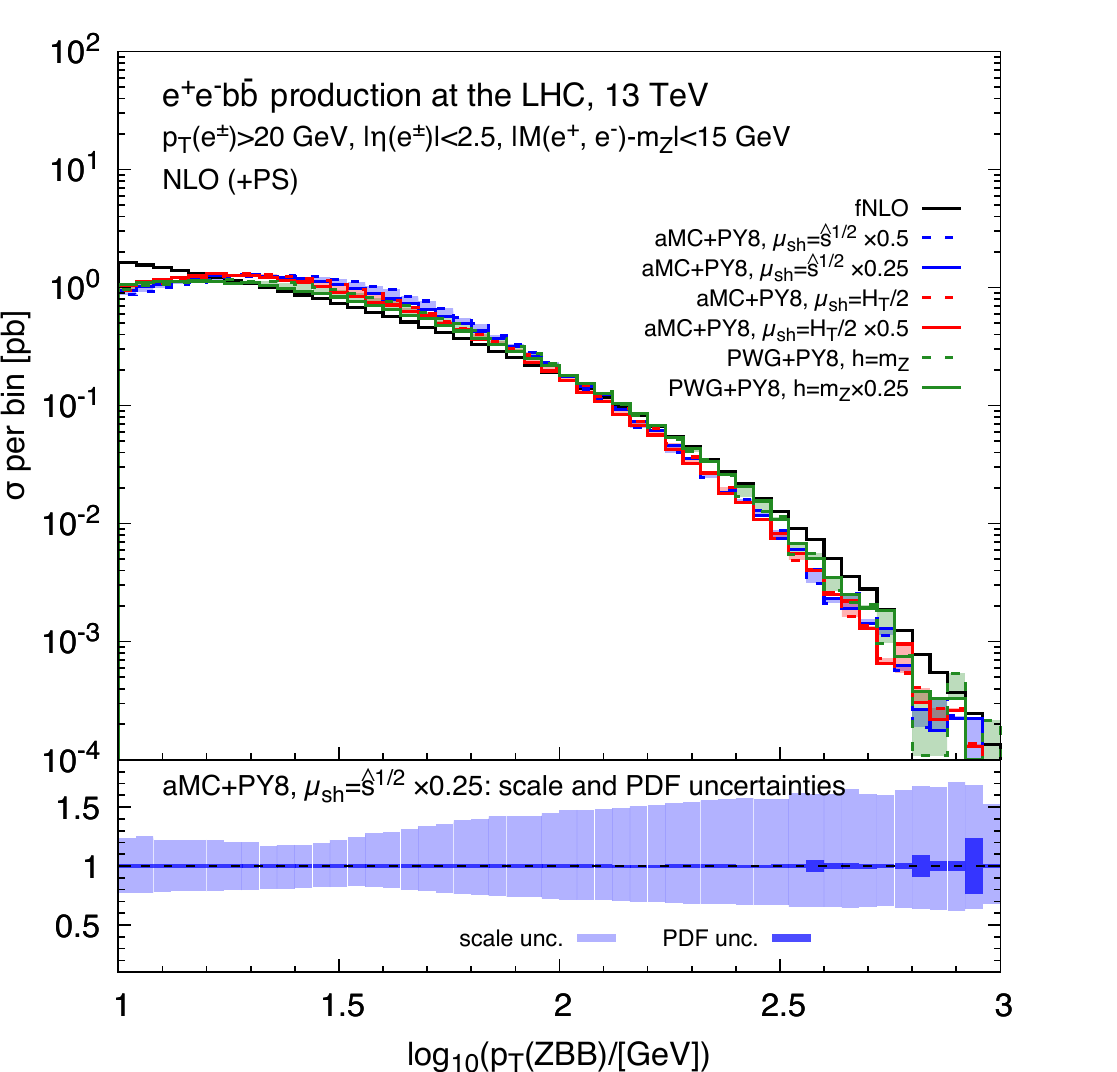}
\includegraphics[width=0.48\textwidth,angle=0,clip=true,trim={0.65cm 6.7cm 0.cm 0.4cm}]{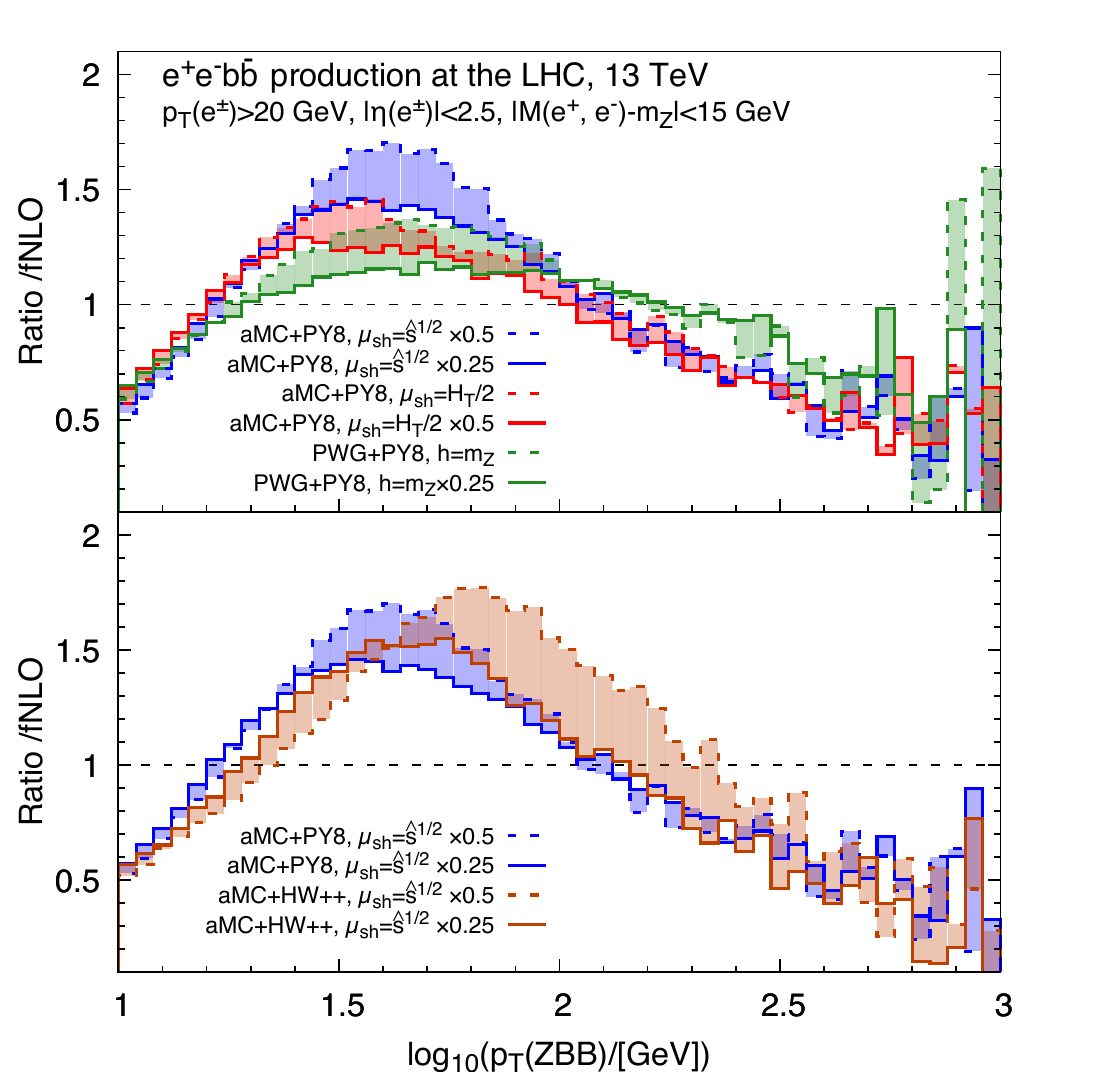}
\caption{\label{fig:ptzbbQ}
  Transverse momentum distribution of the $\ell^+\ell^-B\bar B$ final state,
  computed in the 4FS.
  In the plot we represent the  $\log(p_\perp(\ell^+\ell^-B\bar B))$ distribution.
  The left plot shows the distribution in different approximations;
  the lower inset shows the size of the corresponding PDF,
  factorisation and renormalisation scale uncertainty bands,
  with respect to the NLO prediction.
  The right plot shows the relative effect of different approximations,
  with respect to the fixed-order NLO results:
  different matching schemes (\amcnlo vs \powhegboxnospace) with the corresponding
  matching-parameter uncertainty band, $\mu_{sh} = \sqrt{\hat s}/2 , \sqrt{\hat s}/4$ or
  h = $\mz/4, \mz$ (lower inset);
  different Parton Shower models (\pythia vs \herwignospace).
}
\end{figure}

In this section we consider  the
transverse-momentum distribution of the $\ell^+\ell^-b\bar b$ system
and present numerical results in different approximations in
Figure~\ref{fig:ptzbbQ}.
This quantity is of technical interest,
because it makes it possible to study  the impact of QCD radiation
on this system,
with interesting features due to the presence of coloured particles in the final
state, whose emissions contribute to the recoil of $\ell^+\ell^-b\bar b$.

In the left plot of Figure~\ref{fig:ptzbbQ}
we show the distribution of the logarithm of the
transverse=momentum distribution of the $\ell^+\ell^-B\bar B$
system\footnote{
  In the evaluation of the distribution we tag and analyse
  the $B$ mesons present in the final state.}
computed with different combinations of generators and of
scales\footnote{The same colour codes,
combinations of codes and approximations
are valid also in Figures
\ref{fig:ptZ4F-inclusive},
\ref{fig:nbjetbjet},
\ref{fig:ptZ1jets},
\ref{fig:ptZ2jets},
\ref{fig:minvbjetpair},
\ref{fig:minvBBpair0jets},
\ref{fig:minvBBpair1jets},
\ref{fig:minvBBpair2jets},
\ref{fig:deltarBBpair0jets},
\ref{fig:deltarBBpair1jets},
\ref{fig:deltarBBpair2jets},
\ref{fig:deltarbjetpair},
\ref{fig:ptbjet1},
\ref{fig:ptbjet2},
\ref{fig:etabjet1},
\ref{fig:etabjet2}.
}:
in black we present the divergent distribution computed in the fixed order
NLO calculation
(we stress that this quantity is only LO accurate in this calculation);
in blue we show results obtained with \amcnlo and \pythia,
using as reference shower scale, $\mu_{sh}$, the partonic variable
$\sqrt{\hat s}/2$ (dashed) or $\sqrt{\hat s}/4$ (solid);
in red we show results obtained with \amcnlo and \pythia,
using as $\mu_{sh}$ the sum of the final-state transverse masses
$H_T/2$ (dashed) or $H_T/4$ (solid);
in green we show results obtained with \powheg and \pythia,
setting the value of the scale $h$ in the damping factor equal to
$\mz$ (dashed) or $\mz/4$ (solid).

In the upper inset of the right plot of Figure~\ref{fig:ptzbbQ},
we compare the different combinations of tools and scales of the left plot
with the fixed-order NLO-QCD results. As a function of the transverse momentum of the system
we observe: the Sudakov suppression at low momenta;
the redistribution of the events due to the unitarity of the
PS and of the matching procedure, yielding an increase of the distribution at
intermediate momenta;
a decrease of the distributions compared to the fixed order results
occurs at large momenta, for both \amcnlo and the \powhegbox and irrespective of
the choices of the PS parameters.

In the lower inset of the left plot we show the PDF and
renormalisation/factorisation scale uncertainties,
which are at the $\pm 20$\% level for transverse momenta smaller than 120 GeV,
but increase and reach the $\pm 45$\% level for transverse momenta close to 1 TeV,
and are dominated by the scale uncertainties.

In the lower inset of the right plot we compare again
with the fixed-order NLO-QCD curve
the result obtained with \amcnlo combined with \pythia (blue)
and with \herwig (brown) using for $\mu_{sh}$
the partonic variable $\sqrt{\hat s}/2$ (dashed) or $\sqrt{\hat s}/4$ (solid).
We observe similar trends of the two QCD-PS models
but a quantitative significant difference in the comparison of the bands
obtained with a variation of the shower scale in the same range,
in particular for the size of the bands corresponding to the $\sqrt{\hat s}/2$ scale choice.

\subsection{The $\ptz$ distribution, inclusive over $b$-quark contributions,
  in different approximations}
\label{sec:ptzfour}
In this section we study
the transverse-momentum distribution $\ptz$ of the lepton pair,
in presence of a $b\bar b$ pair in the final state,
inclusive over the $b$-quark contributions.

The process $pp\to \ell^+\ell^- b\bar b$ is studied in the 4FS
in different perturbative approximations,
namely at LO, at fixed NLO-QCD, including QCD-PS effects matched with
the LO or with the NLO-QCD results.
At variance with the 5FS case,
where the $\ptz$ distribution is divergent at fixed-order ${\cal O}(\as)$
when $\ptz\to 0$,
this observable in the 4FS is regular in the same limit at fixed order
and {\it a fortiori} after matching with a QCD PS.
The regular behaviour of $\ptz$ in the 4FS is due to the bottom-quark mass,
which acts as a regulator for the singularity
associated with the limit $\ptz\to 0$.

At NLO-QCD the $\ptz$ distribution is sensitive to large
logarithmic corrections due to QCD initial-state radiation,
mostly from the $gg$-initiated subprocess\footnote{
A similar statement is also present in Ref.~\cite{Frederix:2011qg} (cfr. Figure 6).
}.
The origin of these large effects can be understood by considering
the two mechanisms that yield a transverse momentum of the lepton pair:
$i)$ the LO distribution of the $\gamma^*/Z$ boson in a three-body final state
and $ii)$ the recoil against QCD radiation of the $\ell^+\ell^-b\bar b$ system.
While the former is regular in the whole phase space,
the latter is sensitive to the presence of collinear divergences
due to initial-state radiation.
In fact, the transverse-momentum distribution of the $\ell^+\ell^-b\bar b$ system
is divergent at fixed order for vanishing transverse momentum and
requires the resummation of logarithmically-enhanced terms to all orders
to become regular.
After the resummation, the transverse-momentum distribution
of the $\ell^+\ell^-b\bar b$ system is still sensitive to logarithmically-enhanced
corrections, which contribute in turn to the second of the two mechanisms that yields the $\ptz$ distribution,
explaining why the prediction of the latter requires not only a fixed-order calculation
but also the matching with multiple parton emissions at all orders via QCD-PS.
\begin{figure}[!h]
\centering
\includegraphics[width=70mm,angle=0,trim={0cm 4.5cm 0.6cm 0.5cm}]{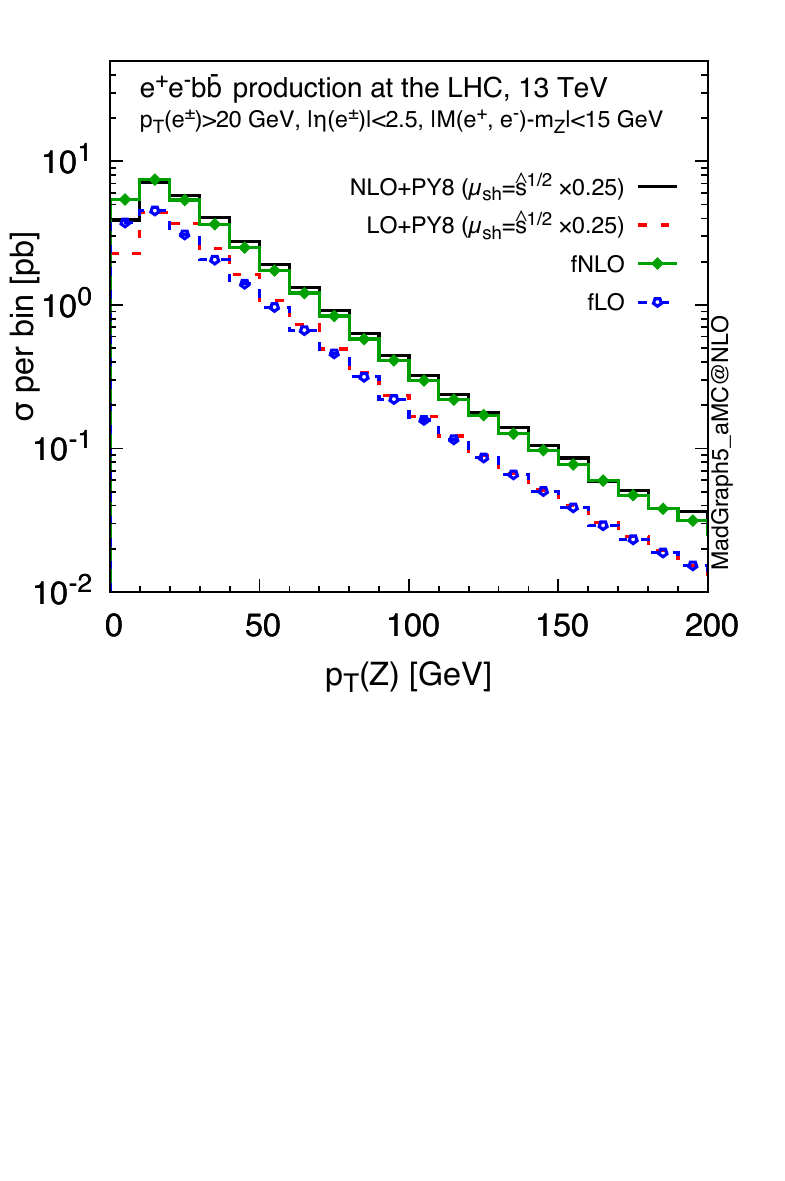}~~~
\includegraphics[width=70mm,angle=0,trim={0cm 4.5cm 0.6cm 0.5cm}]{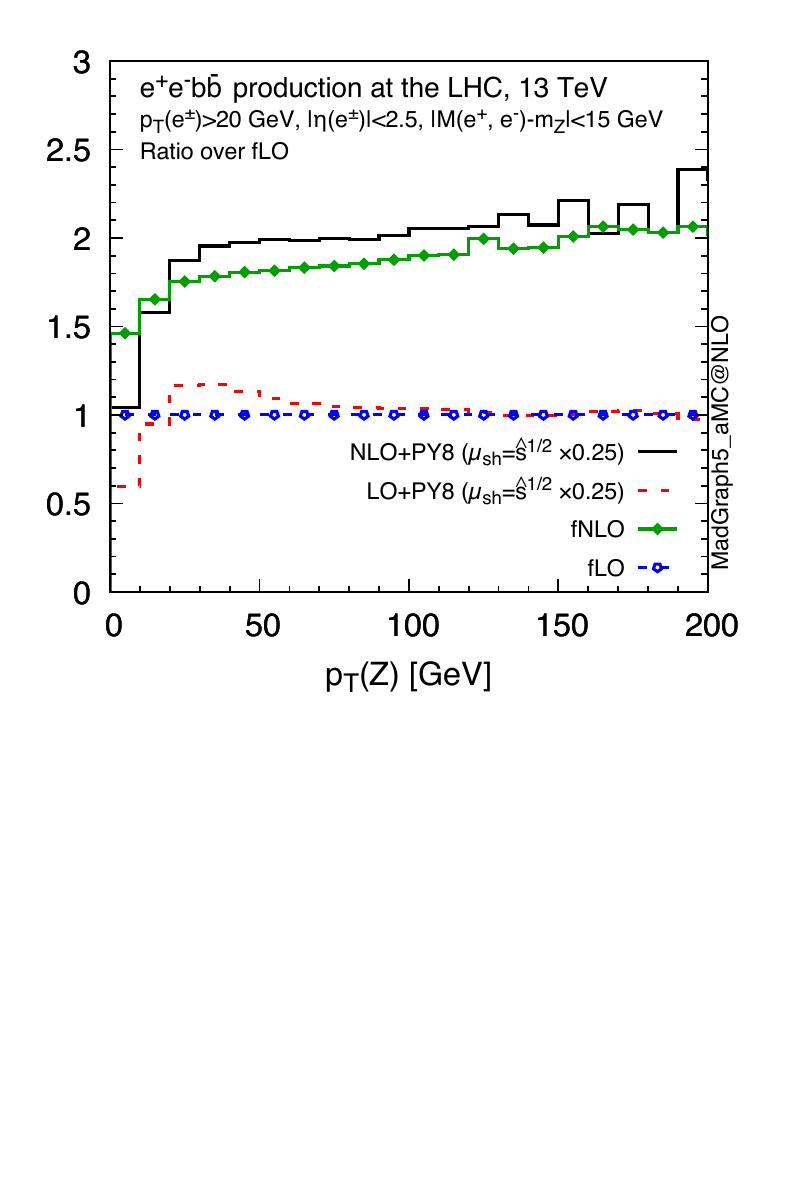}
\caption{\label{fig:ptZ4F-LO-NLO-matched}
$\ptz$ distribution in 4FS inclusive over the $b$-quarks contribution:
  comparison of predictions in different perturbative approximations:
  LO, NLO, LO matched with the \pythia QCD-PS, NLO matched with a QCD-PS
  (left panel) and relative impact of higher-order corrections relative to
  the LO prediction (right panel).  }
\end{figure}

In Figure~\ref{fig:ptZ4F-LO-NLO-matched} (left panel)
we compare the 4FS distributions in different perturbative approximations:
at fixed-order LO and NLO and, after the matching of \amcnlo \\ with the \pythia
QCD-PS, with LO+PS and NLO+PS accuracy.
We use the inputs described in Section \ref{sec:setup} and
set $\mush=\sqrt{\hat s}/4$ as the reference shower scale.
In Figure~\ref{fig:ptZ4F-LO-NLO-matched} (right panel)
we show the relative impact of the various approximations relative to
the LO results.
The NLO corrections (green) yield a large \kfactor of ${\cal O}(70\%)$,
flat almost the whole $\ptz$,
with the exception of the low transverse-momentum region,
where the corrections are smaller, of ${\cal O}(50\%)$.
The action of a QCD-PS on top of the LO distributions
strongly modifies the shape of the distribution,
with a corrections which is negative and reaches -40\% at very low $\ptz$
values, vanishes at $\ptz\sim 25 $ GeV,
then increases and has a maximum of ${\cal O}(+20\%)$ at $\ptz\sim 35$ GeV,
decreases and eventually vanishes for larger values of $\ptz$.
After matching the NLO results with the QCD-PS,
the relative impact of the latter with respect to the fixed NLO results
is similar to the difference between LO+PS and LO for low $\ptz$,
while for large $\ptz$ a positive correction of ${\cal O}(+20\%)$ remains.
\begin{figure}[!h]
\centering
\includegraphics[width=0.48\textwidth,angle=0,clip=true,trim={0.65cm 6.7cm 0.cm 0.4cm}]{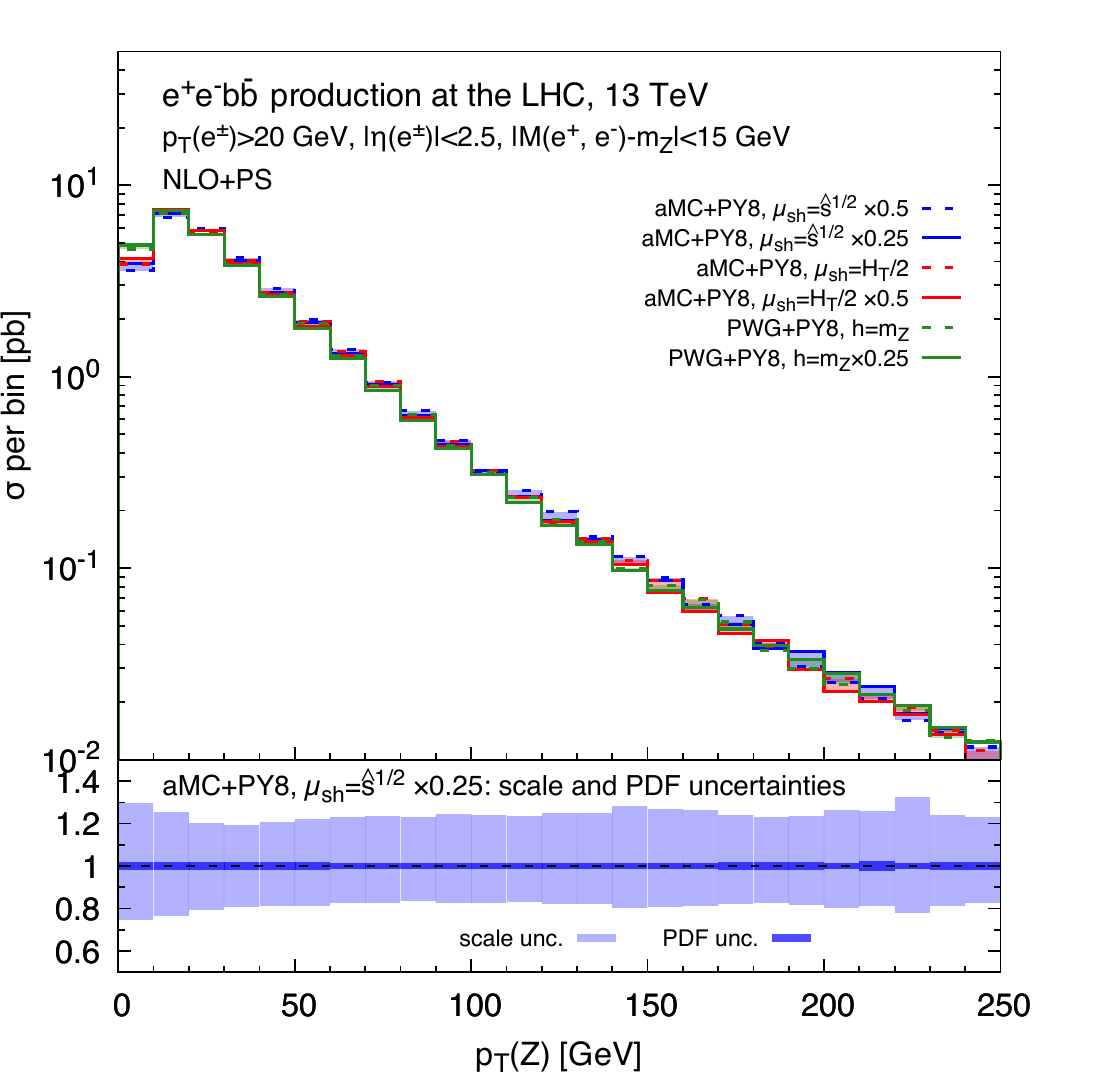}
\includegraphics[width=0.48\textwidth,angle=0,clip=true,trim={0.65cm 6.7cm 0.cm 0.4cm}]{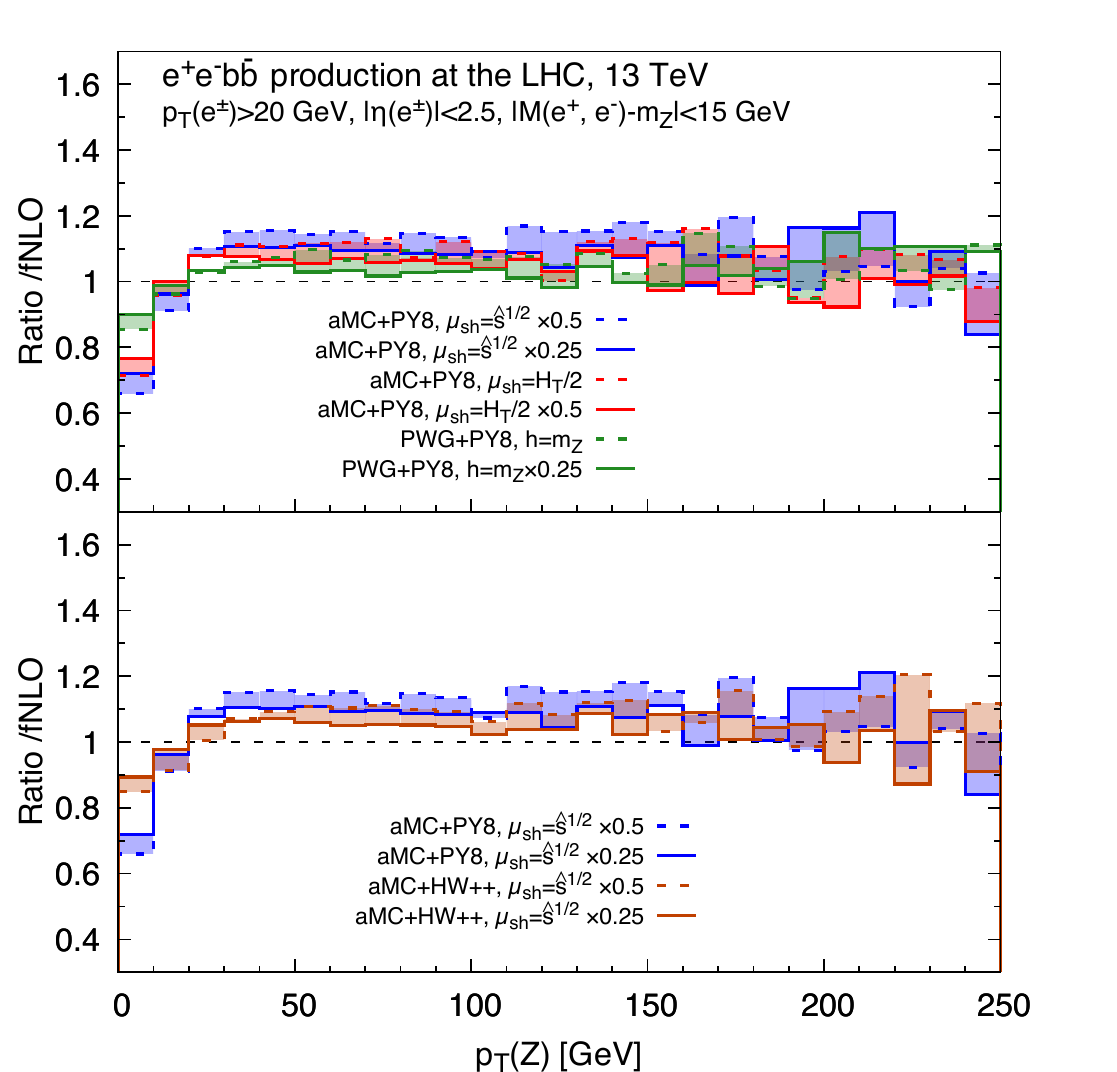}
\caption{\label{fig:ptZ4F-inclusive}
$\ptz$ distribution in the 4FS inclusive over the $b$-quark contribution.
Same approximations and colour codes as in Figure \protect\ref{fig:ptzbbQ}.
}
\end{figure}

In Figure~\ref{fig:ptZ4F-inclusive} we study different sources of
theoretical uncertainty,
using the same colour code and comparing the same approximations
as in Figure~\ref{fig:ptzbbQ}.
In the left plot, upper inset,
we compare the predictions obtained with \amcnlo and \powhegboxnospace,
both interfaced with \pythianospace,
using different variables and values for $\mush$ and $h$.
We observe a global compatibility between the different options:
if we consider the envelope of the different bands
as an estimate of the matching and shower uncertainties,
we conclude that they are at the ${\cal O}(\pm 7\%)$ level,
with the exception of the first bin, as long as $\ptz<100$ GeV, and slightly increase for larger values of $\ptz$.
We observe a common trend of the corrections due to multiple parton emissions,
which are negative down to $-30\%$, with respect to the fixed NLO prediction,
for $\ptz<20$ GeV and positive up to $+15\%$ for larger values.
From the upper inset of the right plot of Figure~\ref{fig:ptZ4F-inclusive}
we observe that in the low-$\ptz$ region  the \powhegbox corrections are
slightly smaller in size than those of \amcnlonospace.

In the left plot, lower inset,
we show the uncertainty bands associated to scale variations
(the renormalisation and factorisation scales are varied independently
in the interval $[\mu/2, 2\mu]$) and to PDFs.
As it can be seen, scale variations provide an uncertainty of ${\cal O}(\pm 20\%)$
which is quite independent on the value of $\ptz$,
and represent by far the dominant source of uncertainty.
PDF uncertainties are much smaller, below $3\%$.
In the right plot, lower inset,
we compare the PS \pythia and \herwignospace, both matched to \amcnlo and
with the same reference shower scale $\mush$.
We observe that accidentally the combination of \amcnlo with \herwig
yields results which are in size and shape similar to those obtained with
the \powhegbox and \pythianospace.

\begin{figure}[!h]
\centering
\includegraphics[width=0.48\textwidth,angle=0,clip=true,trim={0.65cm 8.7cm 0.cm 0.4cm}]{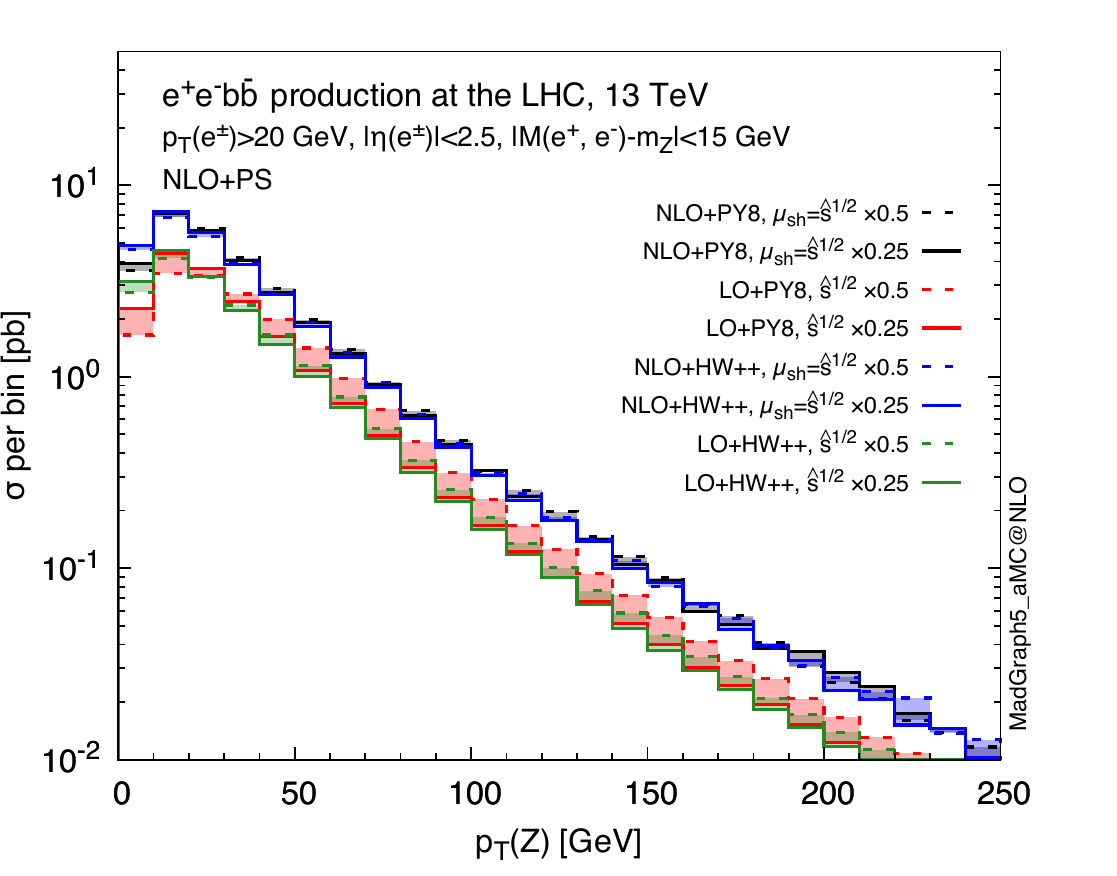}
\includegraphics[width=0.48\textwidth,angle=0,clip=true,trim={0.65cm 8.7cm 0.cm 0.4cm}]{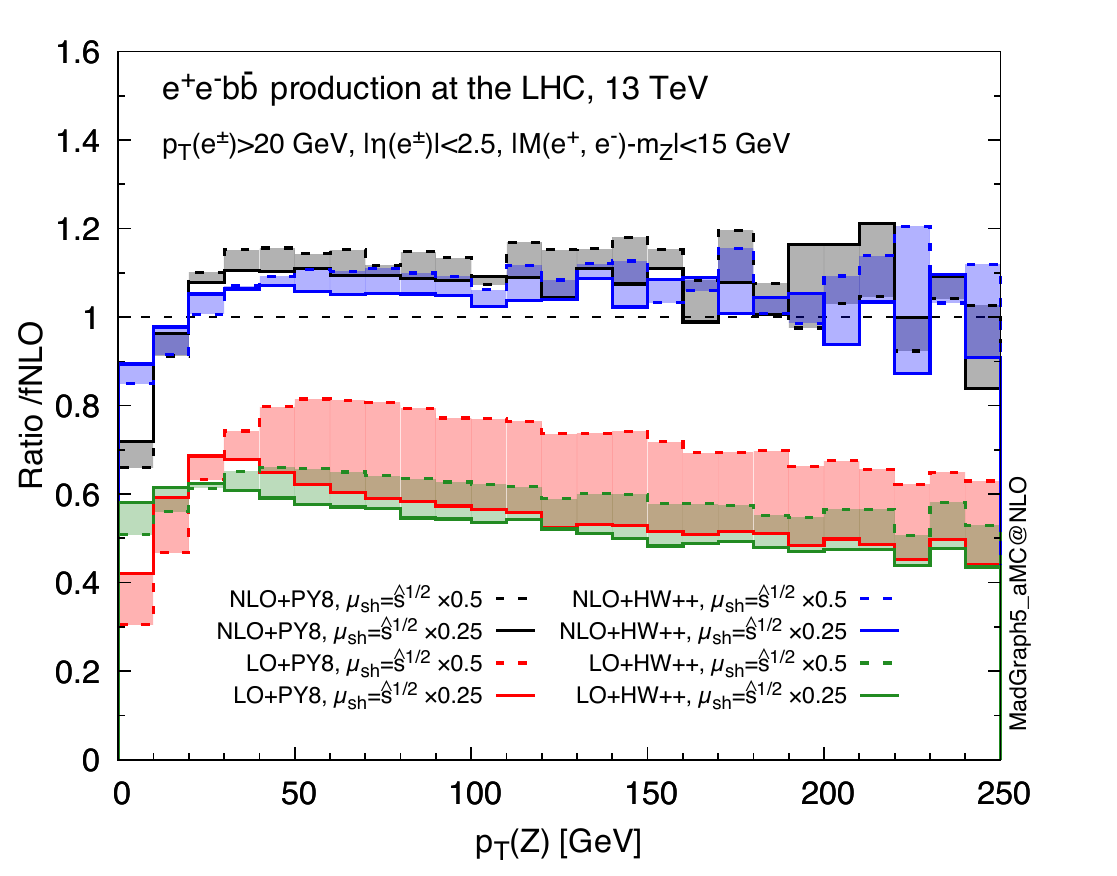}
\caption{\label{fig:ptZ4F-LOPS-NLOPS}
$\ptz$ distribution in the 4FS inclusive over the $b$ quarks contribution.
Comparison of the results obtained with LO+PS and NLO+PS accuracy
with \pythia and \herwignospace,
for a given choice of the matching parameter.
}
\end{figure}

In Figure~\ref{fig:ptZ4F-LOPS-NLOPS} we compare the predictions of \amcnlo
at LO+PS and NLO+PS accuracy obtained with \pythia and \herwig as QCD-PS
(left plot),
with either $\mush = \sqrt{\hat s}/4$ or $\mush = \sqrt{\hat s}/2$;
we show the relative difference with respect to the NLO fixed-order prediction
in the right panel.
The differences of order $\pm 15\%$ at LO+PS (green and red bands)
are reduced down to $\pm 7\%$ at NLO+PS,
because the first real emission is described, in the latter case,
by the exact matrix element and the PS differences appear
from the second emission.

\clearpage{}

\clearpage{}\section{Inclusive lepton-pair transverse-momentum distribution}
\label{sec:ptz}
In this section we discuss the prediction of
the inclusive lepton-pair transverse-momentum distribution
and propose a formulation that includes a refined treatment of the bottom-quark contributions,
exploiting the advantages of both 4FS and 5FS formulations
of the lepton-pair production process.

\subsection{Four- {\it vs} five-flavour schemes}
\label{sec:fourvsfive}

\subsubsection{Generalities}
Two procedures are commonly followed to calculate
high-energy processes characterised by a hard scale $Q$,
that involve the production of heavy quarks such as the bottom quark.

In the so called ``massive'' or four-flavour scheme (4FS)
the heavy quarks do not contribute to the proton wave-function
because the value of their mass, larger than the one of the proton,
makes their creation in pairs possible only in high-energy interactions.
In this scheme the active degrees of freedom are $n_f$ light quarks,
while the heavy quarks are decoupled and do not contribute to the running of
the strong coupling constant nor to the PDF evolution; in particular
a bottom-quark PDF is absent.
The validity of this approach is guaranteed when the hard scale $Q$ of the
process is comparable to the heavy-quark mass $\mb$.
The latter acts as a natural cut-off in the case of additional collinear
emissions.

In the case when a hierarchy between the heavy-quark mass and the hard scale
of the process is present ($\mb\ll Q$)
it is possible that large corrections enhanced by $\log\frac{ Q^2}{\mb^2}$
appear in the cross sections, spoiling the convergence of the perturbative
expansion, while powers of the ratio $\mb/Q$ are naturally suppressed.
The initial-state logarithmic corrections can be resummed to all orders via
the Altarelli-Parisi equations and reabsorbed in the definition of a
bottom-quark proton PDF, while in the final-state case it is possible
to introduce appropriate fragmentation functions.
The bottom quark belongs then to the light quarks present in the proton ($n_f=5$)
and contributes to the running of the strong coupling constant.
This approach is called ``massless'' or five-flavour scheme (5FS).


The advantages of the 5FS are related to the lower multiplicity of scattering
particles: the simplicity of the final-state structure makes it possible to include
higher-order radiative corrections more easily than in the corresponding 4FS
processes. In addition, the presence of a bottom PDF in the proton resums to all orders
initial-state collinear logarithms due to gluon emissions.
As of today, the final-state higher multiplicity in the 4FS forbids
the inclusion of corrections beyond NLO-QCD.
On the other hand, the exact description of the massive-quark kinematics is already present at LO and can be analysed in detail upon inclusion of the NLO corrections and also after matching with a QCD PS. In addition, as argued in Section \ref{sec:zbbscale} based on the results of Ref.~\cite{Lim:2016wjo},
possibly large logarithms $\log\frac{Q^2}{\mb^2}$ are in fact suppressed by phase space effects, and the effective scale ${\cal Q}^2$ is parametrically lower than the vector boson mass. One therefore expects that in this region bottom mass effects to be more relevant than the collinear logarithms and the 4FS could be preferred over the 5FS.
Another motivation for employing the 4FS scheme is provided by the inclusive gauge-boson transverse-momentum distribution, which has a peak in the interval between three and eight GeV, namely at values comparable to the bottom-quark mass. The simulation of the lepton-pair transverse momentum distribution in shower Monte Carlo codes based on the 5FS description requires, in the case of the bottom-induced partonic contributions, that the emission of real radiation stops at a transverse momentum scale of ${\cal O}(m_b)$, that the $b$ parton be put on its mass shell and that the hadronisation  of $b$ quark into $B$ hadron takes place. This  is typically handled by an ad hoc procedure in the QCD PS, which features intrinsic ambiguities.  In the 4FS instead, the lepton-pair transverse momentum distribution receives an exact matrix-element description including the ${\cal O}(\alpha_s)$ corrections in the full range from zero GeV up to the kinematic limit.

\subsubsection{Bottom-quark contributions to DY in 4FS and 5FS}

We are interested in combining the advantages of the 4FS and 5FS approaches,
in order to improve the description of the bottom-quark effects in the lepton-pair transverse-momentum distribution.\footnote{The formulation of five-flavour schemes retaining power-suppressed mass-effects at some level of accuracy in inclusive or semi-inclusive observables has a long history, e.g., see Section 2 of Ref. \cite{Maltoni:2012pa}.  For a recent proposal in the context of fully-exclusive predictions see Ref.~\cite{Krauss:2017wmx}.}

The merging of 4FS and 5FS results is in principle possible provided that
double counting is avoided. To this aim, equivalent
terms that contribute in the two schemes need to be identified, then subtracted
from the 5FS description of the process and added back as evaluated in the 4FS.
The rationale behind this combination is the possibility
of exploiting the improved description
offered by the 4FS of the heavy-quark contribution
to observables like the gauge-boson transverse momentum
at low-/intermediate-momentum values.
\begin{figure}[!h]
\centering
\includegraphics[width=75mm,angle=0, trim= 0cm 5cm 0cm 0cm]{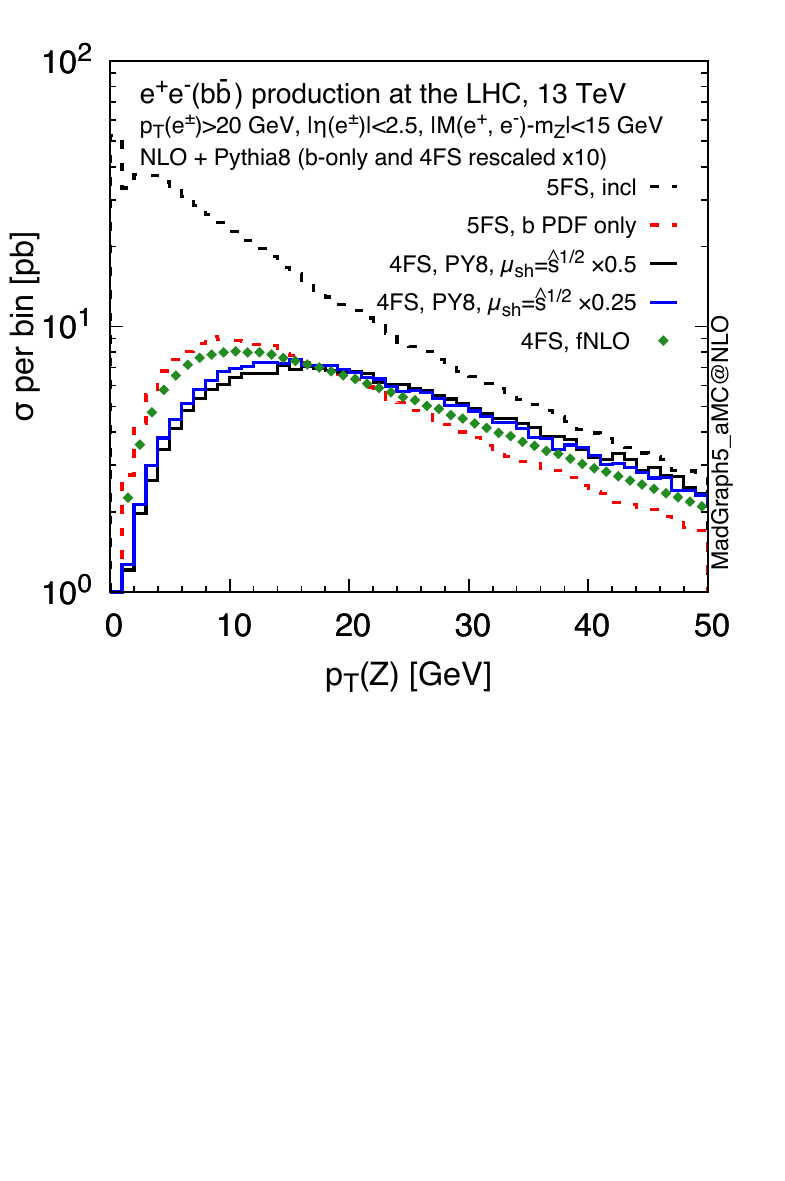}
\caption{\label{fig:ptz-fourvsfive}
Comparison of the $\ptz$ distribution in the plain 5FS
with the contribution associated to the bottom quark,
the latter evaluated in different schemes and approximations.}
\end{figure}

In the 5FS, at tree level, the DY process occurs through quark-antiquark annihilation,
the partonic cross section starts at ${\cal O}(G_\mu^2)$
and bottom-initiated subprocesses are already present.
Since we are interested in the bottom contribution,
we remark that this density, generated inside the proton
by a radiative mechanism, is proportional to $\as$ and it contains,
via Altarelli-Parisi evolution, the resummation to all orders in $\as$
of terms enhanced by a factor $\log(\mu_F/\mb)$.

In Figure~\ref{fig:ptz-fourvsfive}
we show, with NLO+PS accuracy,
in black dashed the complete $\ptz$ distribution in the 5FS
and
in red dashed the contribution given by the subprocesses initiated
by at least one bottom PDF.
The size of the latter is consistent with the overall contribution
of ${\cal O}(4\%)$
to the total cross section,
but the peak of the distribution is at a larger value than the one of the
all-flavour $\ptz$ distribution (10 GeV vs.~3 GeV).

After the matching of exact NLO matrix elements with a QCD-PS
that simulates parton radiation to all orders,
we have to consider the possibility that the emitted gluons split
into $b\bar b$ pairs, which appear as final-state hard partons;
such terms are of ${\cal O}(\as^2\,G_\mu^2)$
(when the initial state contains only light quarks) or higher.
Since it is not possible to make a distinction between initial- and final-state
bottom contributions,
we are lead to define the bottom contribution to DY in the 5FS
as the one given by all the events that contain at least one $B$ hadron
in the final state (generated in the hadronisation phase of the QCD-PS).
We recall that in the 5FS the cross section is evaluated with five active flavours
contributing to the strong coupling-constant running,
inducing a bottom contribution also in the subprocesses initiated
by light quarks and gluons;
the latter are not tagged by the $B$ hadron selection.

In the 4FS,
the bottom quark in the proton is by definition absent;
lepton-pair production in association with a $b\bar b$ pair
starts at ${\cal O}(\as^2 G_\mu^2)$,
with the strong coupling-constant running with four active flavours.
This LO cross section is exact in the description of the kinematics
of the massive $b\bar b$ pair.
In a NLO-QCD accurate calculations, also terms of ${\cal O}(\as^3 G_\mu^2)$
are exactly included.
In this scheme, heavy-quarks contributions to the $\as$ running
are decoupled and included in the renormalisation condition.
After matching with a QCD PS, additional $b\bar b$ pairs
might be created, although with suppressed rate,
starting from ${\cal O}(\as^4 G_\mu^2)$.

In Figure~\ref{fig:ptz-fourvsfive}
we show in green dotted the $\ptz$ distribution in the 4FS
inclusive over the $b$ quarks, at NLO QCD,
while in blue and in black solid we present the results with NLO+PS accuracy,
for two different choices of the reference shower scale.
The sizeable impact of the matching with a QCD-PS can be appreciated at glance.

\subsection{Merging 4FS and 5FS results: bottom-quark effects on the $\ptz$ distribution}
\label{sec:collage}
As discussed in Section \ref{sec:fourvsfive},
the improvement over the plain 5FS description can be obtained
by the subtraction of the bottom-related contributions and their
replacement with the 4FS results.

\begin{figure}[!h]
\includegraphics[width=70mm,angle=0,clip=true,trim=0.65cm 0.cm 0.cm 0.4cm]{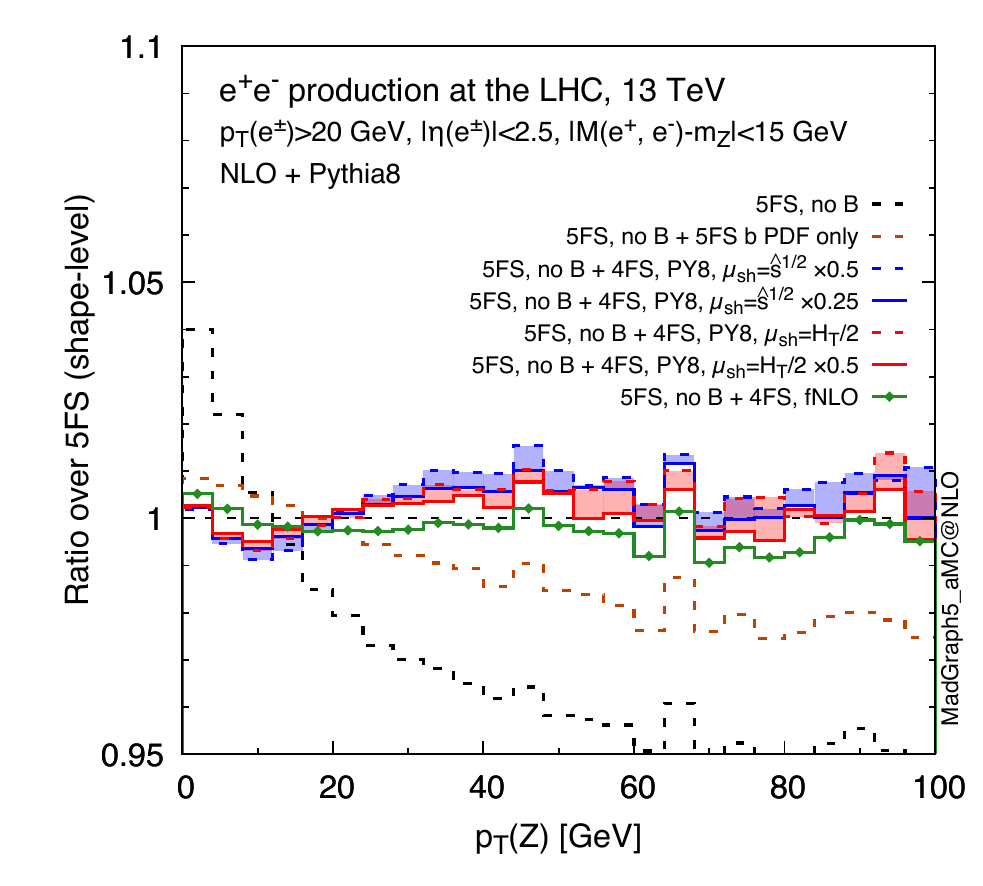}
\includegraphics[width=70mm,angle=0,clip=true,trim=0.65cm 0.0cm 0.cm 0.4cm]{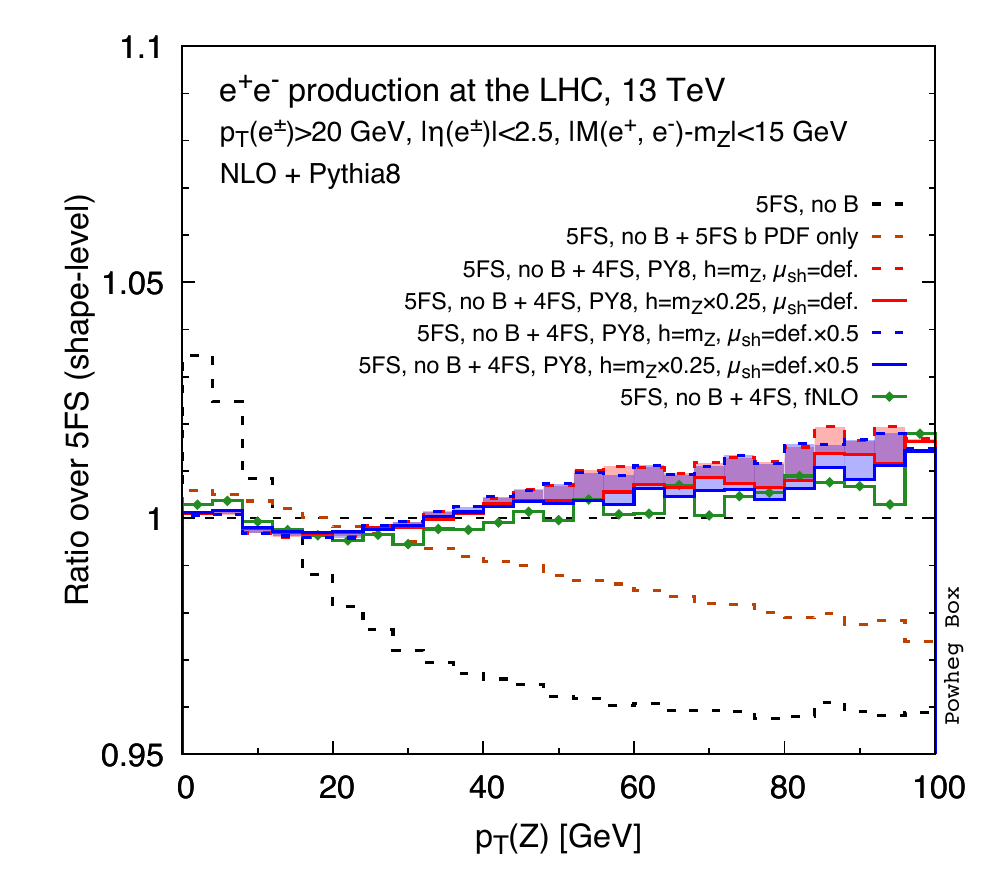}
\caption{\label{fig:ratios}
  Ratio of the $\ptz$ distribution in the refined approximation including bottom-quark mass effects
  over the plain 5FS.}
\end{figure}

We define two physical distributions, namely the production of a lepton pair
strictly without $B$ hadrons (our $B$-vetoed 5FS calculation,
that we label \fivebvetonospace)
and the production of a lepton-pair accompanied by at least one $B$ hadron
(our 4FS results), which are complementary with respect to the additional
particles beside the lepton pair.\footnote{
  A similar procedure has been proposed in~Ref.~\cite{Moretti:2015vaa} in order
  to have an improved description of $t\bar t b \bar b$ in $t\bar t$+jets
  samples.}
The orthogonality of the two quantities allows us to take their sum
and to consider it  as our alternative prediction
for any DY observable,
in particular for the lepton-pair transverse-momentum distribution,
with respect to the treatment of the bottom-quark effects
\be
\frac{d\sigma^{\rm mass}}{d\ptz}
=
\frac{d\sigma^{\fivebveto}}{d\ptz}
\,+\,
\frac{d\sigma^{\rm 4FS}}{d\ptz}\,.
\label{eq:best}
\ee
The impact of our combination is illustrated by the ratio
${\cal R}(\ptz)$
of the shape of our alternative combination for the $\ptz$ distribution
over the corresponding results obtained in the plain 5FS, defined as
\be
{\cal R}(\ptz)
=
\left( \left.
\frac{1}{\sigma_{\rm fid}^{\rm mass}}\frac{d\sigma^{\rm mass}}{d\ptz}
\right|_{\tt tune X}  \right)\cdot
\left( \left. \frac{1}{\sigma_{\rm fid}^{\rm 5FS}}\frac{d\sigma^{\rm 5FS}}{d\ptz}  \right|_{\tt tune X} \right)^{-1}
\label{eq:ratio}
~.
\ee
The ratio in Eq.~\ref{eq:ratio} is defined for a generic Parton Shower model, which we label with {\tt tuneX}.
In Figure~\ref{fig:ratios} we show the function ${\cal R}(\ptz)$,
computed using,
in all the terms that enter in its definition,
the same matching scheme (\amcnlo in the left plot, \powhegbox in the right plot)
and QCD PS model (\pythianospace).
We argue that the ratio deviates from one
because of the different content of perturbative terms associated
to the treatment of the bottom quark,
and also for the choice of the Parton Shower phase space.
The different approximations shown in Figure~\ref{fig:ratios} correspond to:
 5FS computation with a veto on all the $B$ hadrons  (black-dashed);
 5FS computation with a veto only on the $B$ hadrons originating from a
 final-state $g\to b\bar b$ splitting (brown-dashed);
 $B$-vetoed 5FS description combined with the fixed NLO 4FS prediction
(green solid with symbols);
 $B$-vetoed 5FS description combined with the 4FS prediction at NLO+PS
(solid or dashed, red or blue lines). In the last case, the colour code is specific to the matching scheme used.
For \amcnlo, it is the same as for Figs.~\ref{fig:ptzbbQ} and~\ref{fig:ptZ4F-inclusive}:
blue (red) lines correspond to use
$\mush = \sqrt{\hat s} \times f$ ($\mush = H_T \times f$),
where $f=0.5$ (dashed) or $f=0.25$ (solid).
For the \powhegboxnospace, solid (dashed) lines correspond to
using $h=m_Z\times 0.25$ ($h=m_Z$),
while red (blue) lines correspond to using a default (reduced by $1/2$)
$\qsh$ in the ``remnant'' events.

We see that the effect due to bottom-quark effects
on the $\ptz$ distribution hardly exceeds $\pm1\%$.
In both the \amcnlo and the \powhegbox cases, the shape of these effects is such that
the $\ptz$ distribution gets depleted below 20 GeV, while for larger values
the improved prediction is slightly harder;
the \amcnlo \\results tend to flatten again at around 50 GeV,
while the \powhegbox ones keep a positive slope until the end of
the explored $\ptz$ range.
For what concerns the effects due to the shower parameters,
the choice of $\mush$ has always a visible effect on the shape of
the \amcnlo predictions,
and reflects the pattern of the $\qsh$ probability distribution.
In the \powhegboxnospace, the main effect is due to the variation of $h$,
while the variation of the prescription for $\qsh$
in the ``remnant'' events has no effect in practice.

\begin{figure}[!h]
\centering
\includegraphics[width=0.48\textwidth,angle=0,clip=true,trim=0.2cm 3.5cm 0.cm 0.4cm]{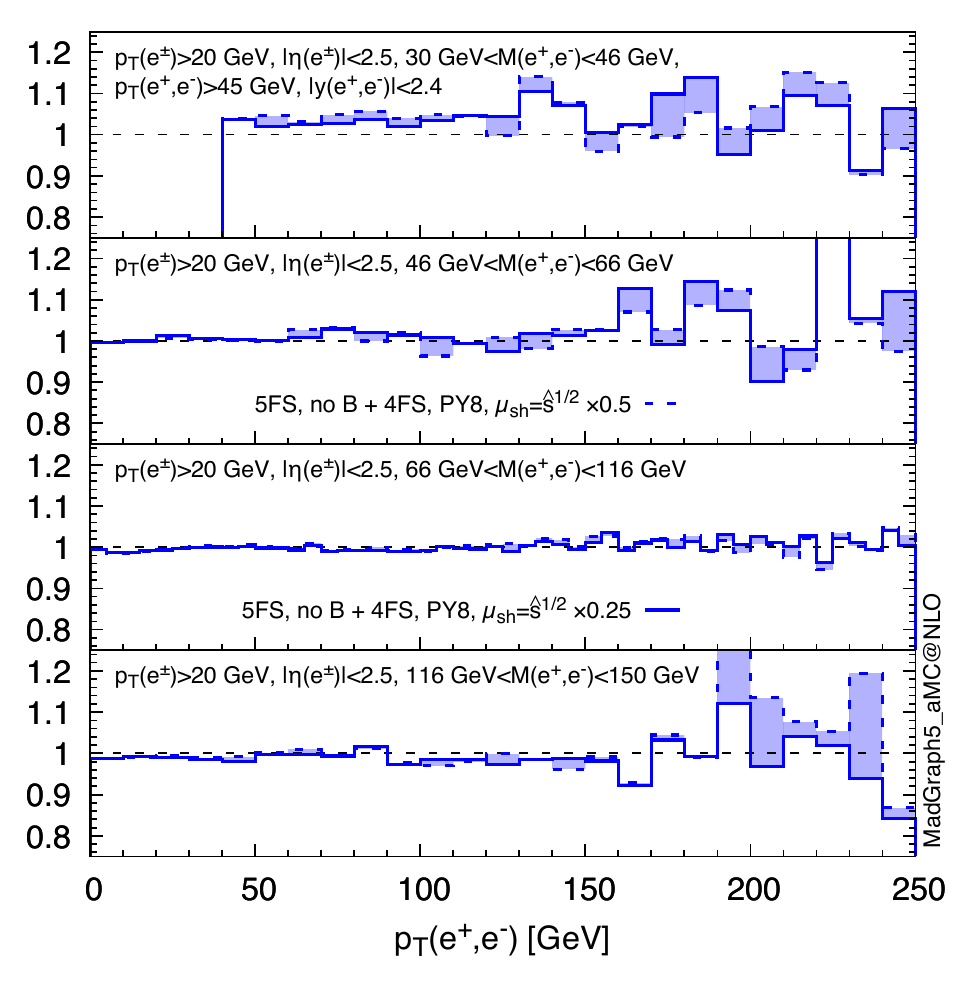}
\includegraphics[width=0.48\textwidth,angle=0,clip=true,trim=0.2cm 3.5cm 0.cm 0.4cm]{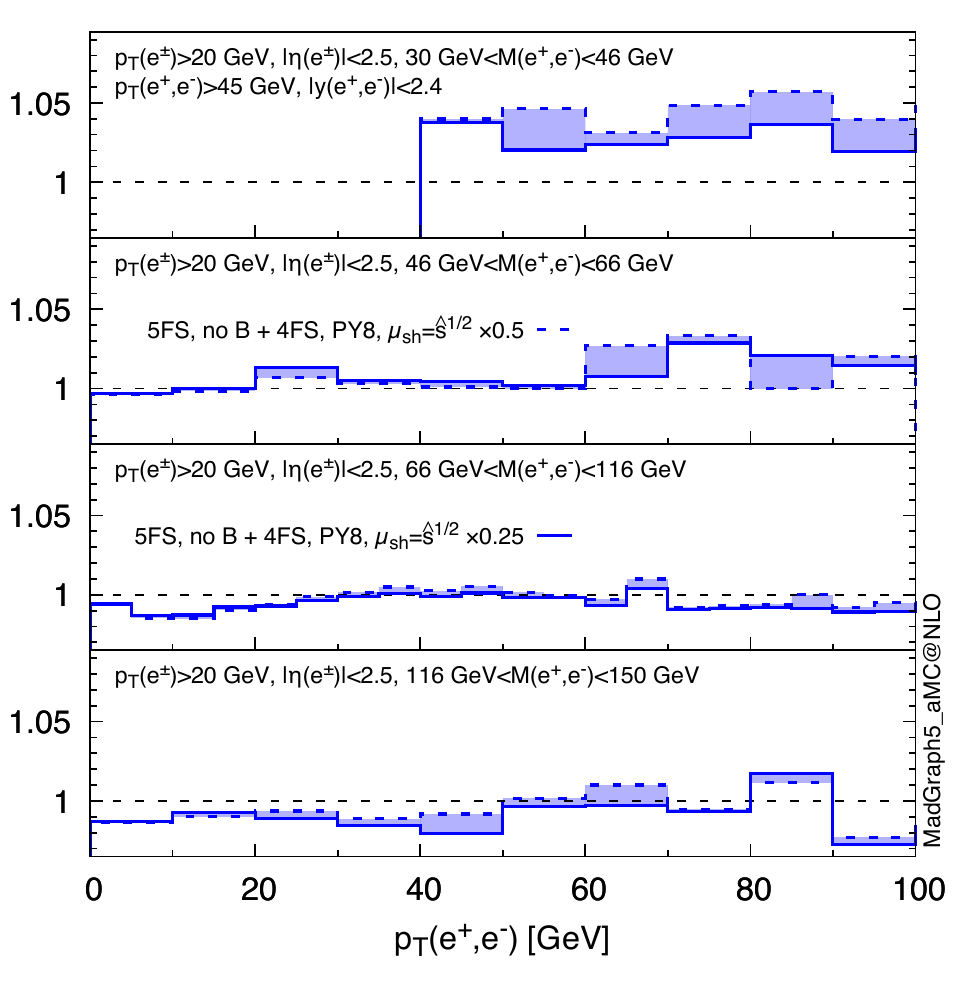}
\caption{\label{fig:invmass}
  Ratio of the $\ptz$ distribution in the refined approximation including the bottom-quark mass
  effects over the plain 5FS,
in different bins of the lepton-pair invariant mass.
The plot on the left is a zoom on the low-$\ptz$ region
of the plot on the right.}
\end{figure}

\begin{figure}[!h]
\centering
\includegraphics[width=0.48\textwidth,angle=0,clip=true,trim=0.2cm 2.cm 0.cm 0.4cm]{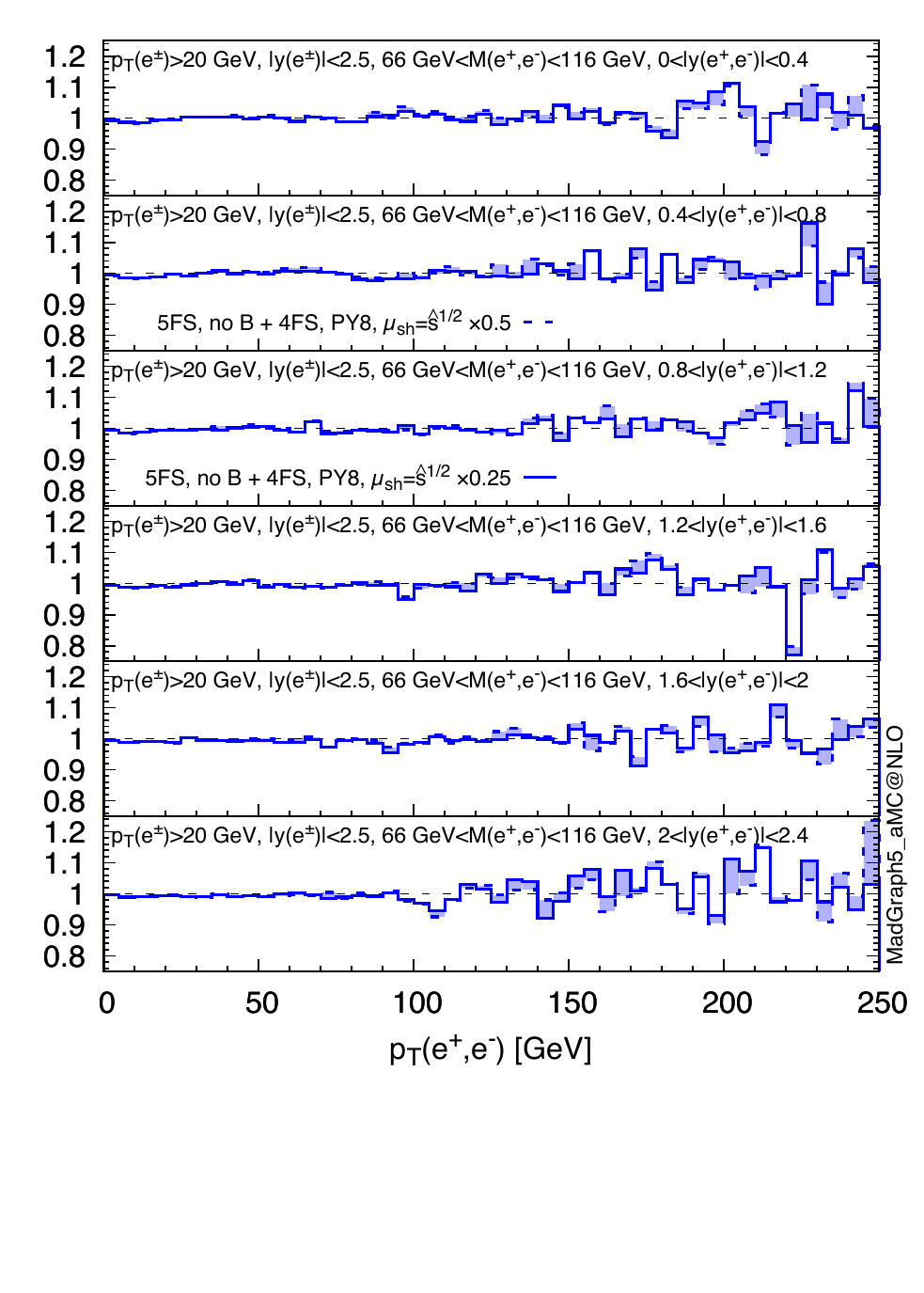}
\includegraphics[width=0.48\textwidth,angle=0,clip=true,trim=0.2cm 2.cm 0.cm 0.4cm]{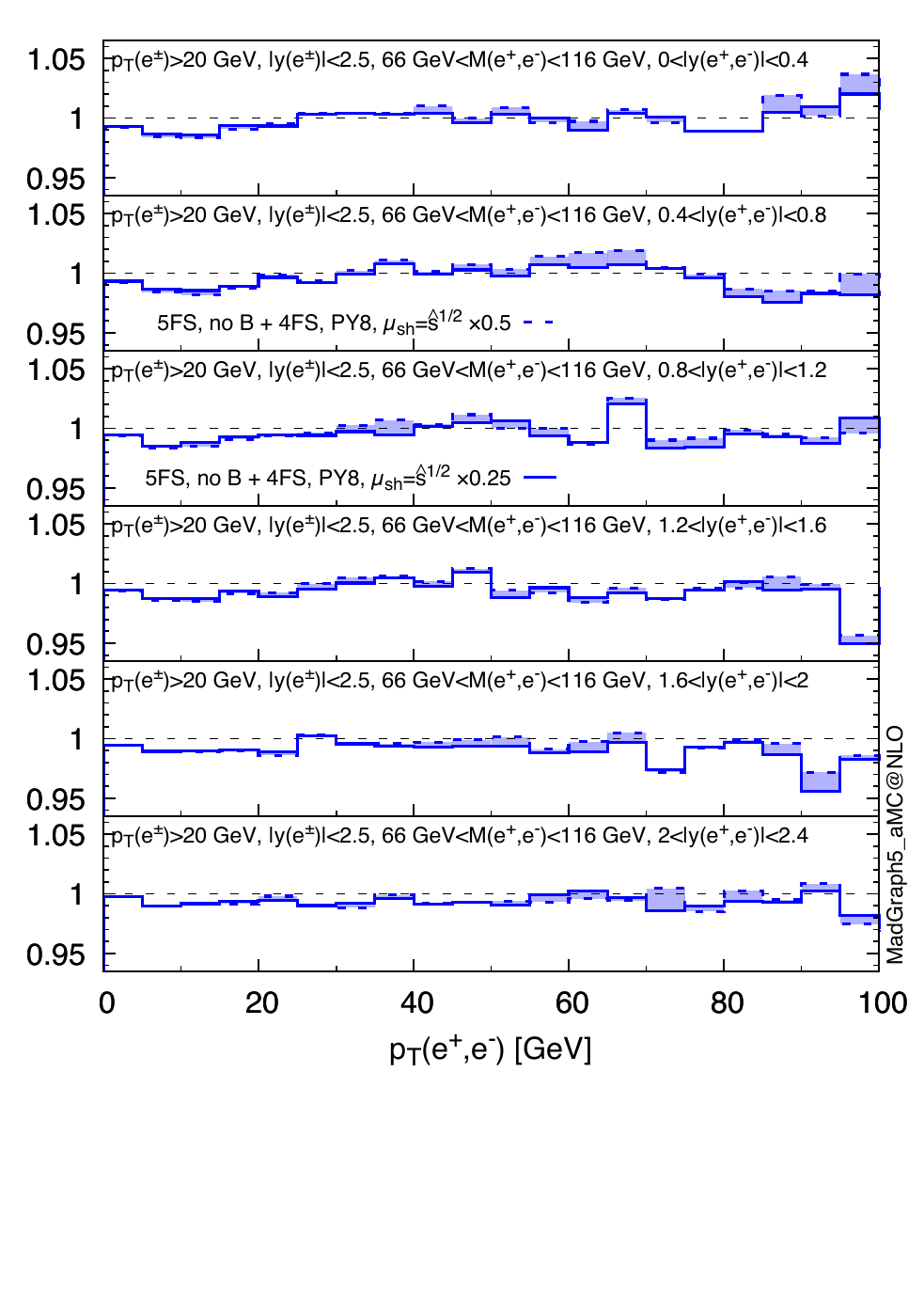}
\caption{\label{fig:pseudo}
Ratio of the $\ptz$ distribution in the refined approximation including the bottom-quark mass effects over the plain 5FS,
in different bins of the lepton-pair rapidity.
The plot on the left is a zoom on the low-$\ptz$ region
of the plot on the right.}
\end{figure}

One may wonder whether the effects of the improved prediction can be enhanced
in some region of the lepton-pair phase space:
in fact, if the dominant effects enter through terms of
${\cal O}\left( m_b/M(\ell^+,\ell^-), \log(\mb/M(\ell^+,\ell^-))\, \right)$,
with $M(\ell^+,\ell^-)$ being the lepton-pair invariant mass,
one may expect a dependence of these effects with respect to
the $\gamma^*/Z$ virtuality.
A study in this direction is further motivated by Ref.~\cite{Aad:2015auj}
(see in particular Figures 14 and 15 therein),
where no single generator gives a satisfactory description
of the $\ptz$ spectrum across different lepton-pair invariant-mass
or rapidity bins,
with the data-theory disagreement at the level of several tens of percent.
In Figures~\ref{fig:invmass} and~\ref{fig:pseudo}
we show the ratio ${\cal R}(\ptz)$
defined in Eq.~\ref{eq:ratio} and
already studied in Figure~\ref{fig:ratios},
now analysed in different bins of lepton-pair
invariant mass and rapidity, respectively.
The binning corresponds to the one adopted in Ref.~\cite{Aad:2015auj}
(with the exception of the invariant-mass bin below 30 GeV,
which we do not consider).
For the sake of simplicity, we only show \amcnlo predictions,
and, among these,
only those with $\mush \sim \sqrt{\hat s}$
(which are those giving the largest effect in Figure~\ref{fig:ratios}).
However, we have also performed the same analysis with the POWHEG-BOX and found similar results.
As a function of the lepton-pair invariant mass,
one indeed observes a trend in the corrections:
they are flat and positive (+3-4\%) in the bin with the smallest invariant
mass ($30\, \gev < M(\ell^+,\ell^-) < 46\, \gev$),
while going at higher invariant masses the corrections are
smaller and become negative(-1\%) in the full $\ptz$ range
of the largest invariant-mass bin ($116\, \gev < M(\ell^+,\ell^-) < 150\, \gev$).
As expected, the bin around the $Z$ peak has a shape which closely follows
the one considered in the inclusive case shown in Figure~\ref{fig:ratios}.

The effect is identical to that of the inclusive case for rapidity up to $|y(\ell^+,\ell^-)| < 1.6$
At larger values, the effect of the improved prediction becomes flatter and negative (-1\%).
Given the size and the shape of these effects,
we conclude that the data-generator differences found in Ref.~\cite{Aad:2015auj}
cannot be attributed to heavy-quark effects.
The new NNLO computations which have been recently published
for $Z$+jet production
\cite{Boughezal:2015ded,Boughezal:2016isb,Ridder:2016nkl,Gehrmann-DeRidder:2016jns,Gehrmann-DeRidder:2017mvr}
hint that these discrepancies may be due to missing higher-order QCD effects.

\clearpage{}

\clearpage{}\section{Interplay between NC-DY and CC-DY}
\label{sec:impact}

Effects due to the strong interaction in a non-perturbative regime,
which cannot be evaluated from first principles,
affect several observables studied at hadron colliders.

The NC-DY,
thanks to the full kinematic reconstruction of the lepton pair,
allows us to perform a precise tuning of the models describing
non-perturbative contributions to the $\ptz$ distribution,
at small values of the transverse momentum.
Under the assumption that these long-distance physics effects are universal,
it is possible to use these models to predict other observables.

In CC-DY the neutrino transverse momentum is inferred from the study
of the recoil of the whole hadronic system that accompanies the lepton pair,
but a precision measurement of $\ptw$ is not possible at the level
necessary for the $\mw$ determination;
furthermore, a stand-alone determination of the non-perturbative effects
relevant to describe $\ptw$ at small transverse momenta is not possible.
For the above reasons,
the parameters fitted from NC-DY are used in the simulation of CC-DY,
to predict the $\ptw$ distribution.

An imperfect evaluation of the non-perturbative parameters in the NC-DY fit
will propagate to CC-DY and in turn affect the $\mw$ determination.
The different heavy-quark flavor content of the initial state in NC-DY
with respect to CC-DY suggests that the non-perturbative parameters
fitted in NC-DY might not be fully universal as one would wish
in view of a high-precision simulation of CC-DY,
where a bottom quark in the initial state is in fact absent.
In Section~\ref{sec:collage} we have proposed a way to include
in the description of the $\ptz$ distribution
an explicit treatment of the non-universal elements
peculiar of the bottom quarks, in particular due to mass corrections.
We can employ this method and re-fit the non-perturbative parameters,
so that such a fit can be sensitive only to effects which are (more) universal,
in the sense that they are common to light and heavy quarks.

\subsection{Transferring the bottom-quark effects to the simulation
  of charged-current Drell-Yan}
\label{sec:transfer}

In this section we try to estimate how our alternative prediction of the inclusive
$\ptz$ distribution,
which accurately includes the bottom quarks contributions,
will affect the prediction of the $\ptw$ distribution and,
in turn, the $\mw$ determination.

We do not rely on the experimental data,
but rather try to explore the role of the bottom quark with a simplified
approach based only on the available simulation tools.
We make the following assumptions:
$i)$ it is possible to tune the parameters of the QCD-PS
to perfectly describe the shape of the $\ptz$ data
in the 5FS (we call this setup {\tt tune1});
$ii)$ given the smallness of the bottom-quark effects,
it is possible to find a second combination of the QCD-PS parameters
to perfectly describe the shape of the $\ptz$ data
also when we use our alternative prediction Eq.~\ref{eq:best}
for the perturbative cross section (we call this setup {\tt tune2});
$iii)$ the parameters of the QCD-PS describing non-perturbative effects
are universal,
i.e.~flavor independent, and constant, i.e.~energy-scale independent.
We assume that our perturbative description provides the bulk of
the prediction and non-perturbative effects are just a correction
that compensates for the different perturbative approximations.
As a consequence of $iii)$, the non-perturbative parameters contribute
to the description of the gauge-boson transverse-momentum distribution,
irrespective of the boson, $W$ or $Z$, as a function only of the
transverse-momentum value.

If {\tt tune1} and {\tt tune2} provide the same exact description of the
shape of the data in the fiducial region (fid) defined by the acceptance cuts,
we can write
\be
\frac{1}{\sigma_{\rm fid}^{\rm exp}}
\frac{d\sigma^{\rm exp}}{d\ptz}
\,=\,
\frac{1}{\sigma_{\rm fid}^{\rm 5FS}}
\left.\frac{d\sigma^{\rm 5FS}}{d\ptz} \right|_{{\tt tune1}}
\,=\,
\left.
\frac{1}{\sigma_{\rm fid}^{\rm mass}}
\frac{d\sigma^{\rm mass}}{d\ptz} \right|_{{\tt tune2}}
\,=\,
{\cal R}(\ptz)
\frac{1}{\sigma_{\rm fid}^{\rm 5FS}}
\left.\frac{d\sigma^{\rm 5FS}}{d\ptz} \right|_{{\tt tune2}}\,,
\label{eq:equalities}
\ee
where the last equality follows from Eq.~\ref{eq:ratio}
and we use the labels (exp, 5FS, mass) to indicate the experimental data,
the plain 5FS massless simulation and our alternative predictions.
From these equalities we read that
the function ${\cal R}$ expresses
the difference in the predictions of the shapes computed
with the same 5FS massless partonic cross section,
using {\tt tune1} or {\tt tune2}
\be
\frac{1}{\sigma_{\rm fid}^{\rm 5FS}}
\left.\frac{d\sigma^{\rm 5FS}}{d\ptz} \right|_{{\tt tune2}}
\,=\,
\frac{1}{{\cal R}(\ptz) }
\frac{1}{\sigma_{\rm fid}^{\rm 5FS}}
\left.\frac{d\sigma^{\rm 5FS}}{d\ptz} \right|_{{\tt tune1}}\,.
\ee
In summary, the function ${\cal R}$ represents the impact of the
improved perturbative treatment of bottom-quark effects;
alternatively, if these effects can be perfectly absorbed in a QCD-PS tune,
it describes the difference of the predictions obtained in the plain 5FS,
using either the plain 5FS tune or the tune derived from the
improved partonic cross section.

In our study, we would like to simulate CC-DY using {\tt tune2},
i.e. with a Parton Shower that has been tuned to account
for the bottom-quark effects,
and compare these predictions with the standard ones based on {\tt tune1}.
Since {\tt tune2} is not yet available,
we can mimic the CC-DY results corresponding to this tune in the following way:
we work with the plain 5FS code interfaced to a {\tt tune1} QCD-PS
and
we reweigh by $\left( 1/{\cal R}(\ptw) \right)$ each event according to its
lepton-pair transverse momentum $\ptw$.
This last combination allows us to assess the impact in the CC-DY simulation
of an improved treatment of the bottom-quark effects in the NC-DY fit.
The reweighting of $\ptw$ then propagates to all the other leptonic
observables used in the $\mw$ determination and leads eventually
to a shift in the measured $\mw$ value.

\subsection{Template-fit determination of $\mw$}
\label{sec:template}

The procedure of template fit  to a distribution of experimental data
consists in the comparison with the data of several theoretical distributions,
the {\it templates}, obtained varying the fit parameter, in our example $\mw$.
The template that maximises the agreement with the data selects the preferred,
i.e.~the measured value of the fit parameter.
\begin{figure}[h!]
\includegraphics[width=75mm,angle=0]{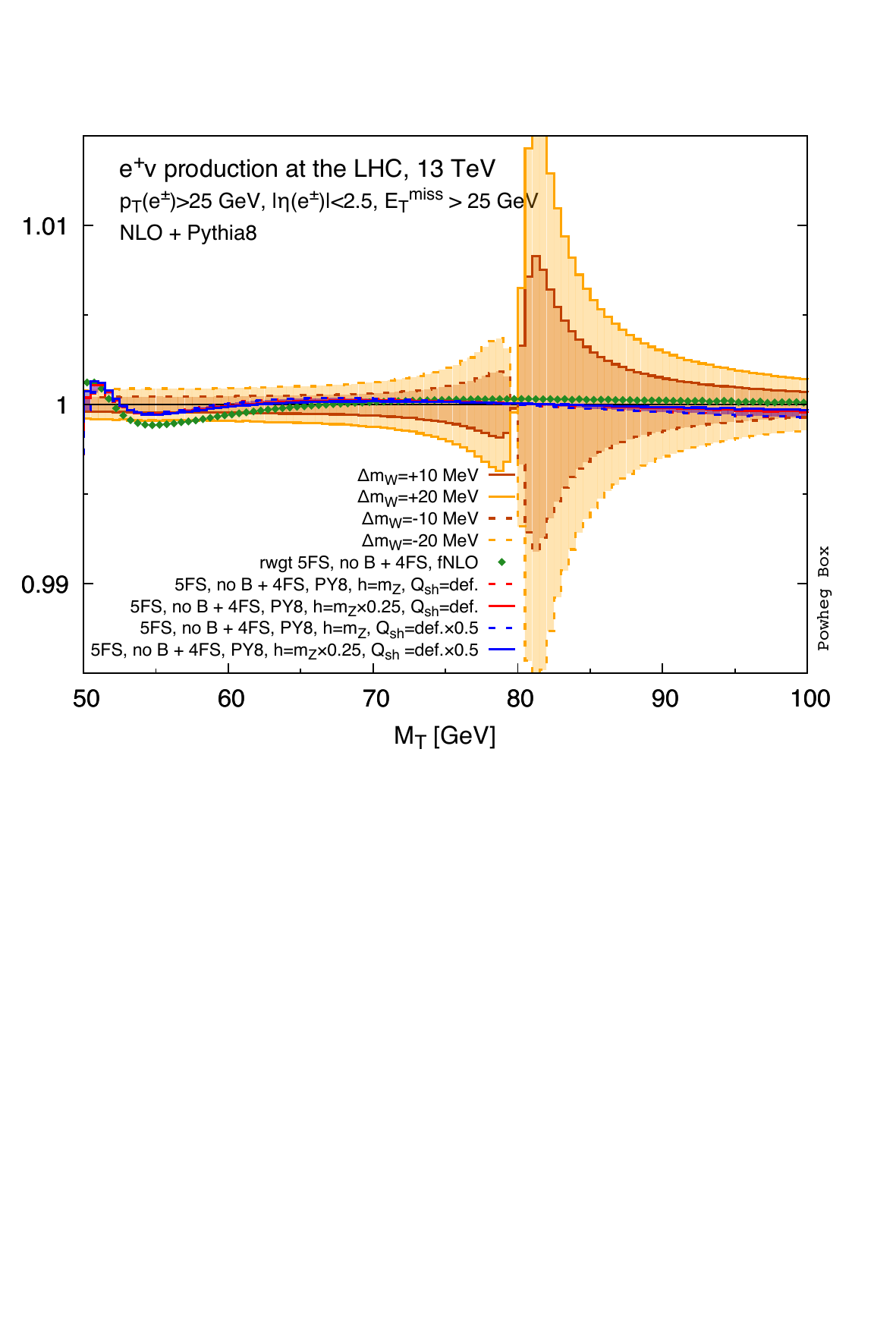}
\includegraphics[width=75mm,angle=0]{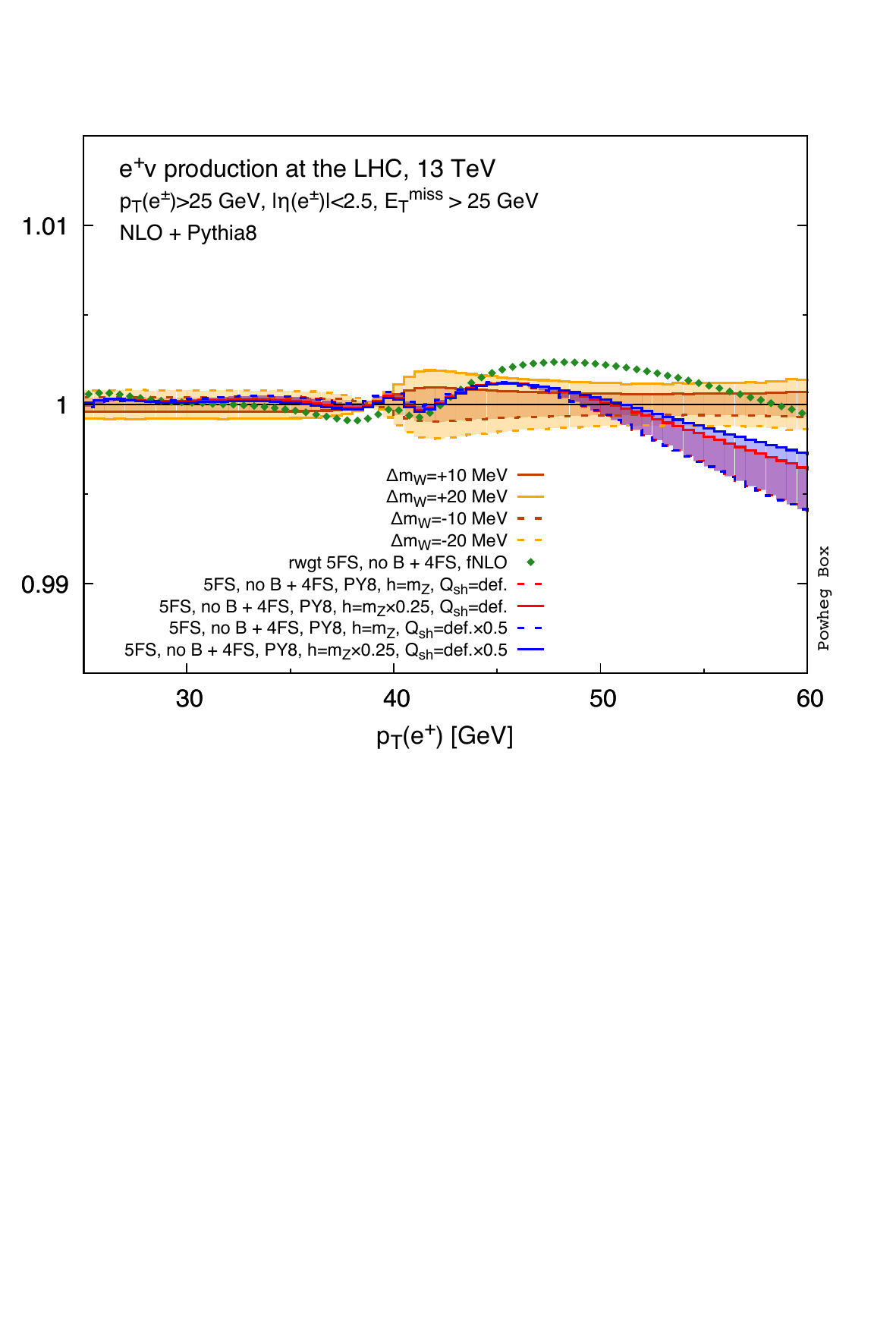}\\[-5cm]
\includegraphics[width=75mm,angle=0]{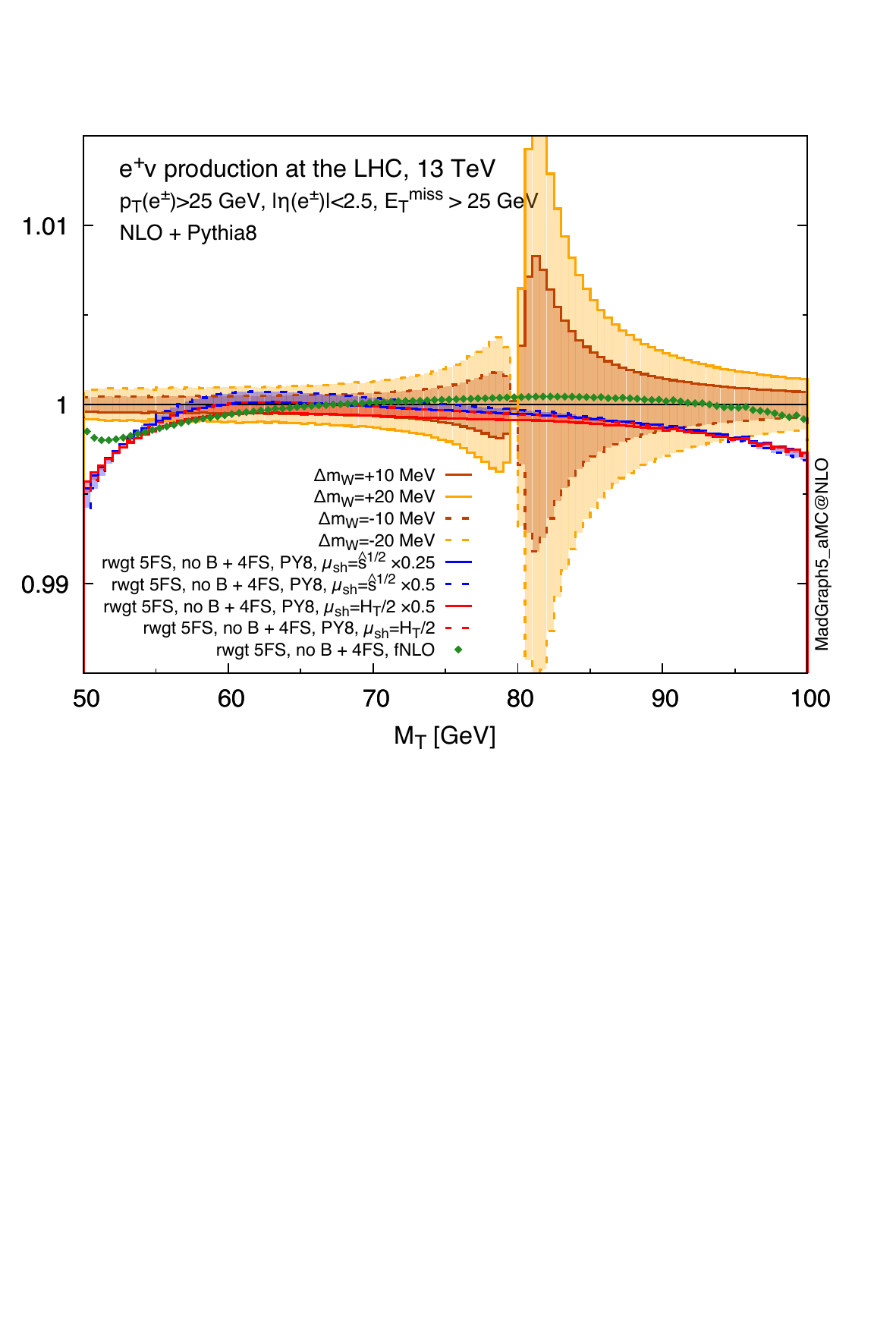}
\includegraphics[width=75mm,angle=0]{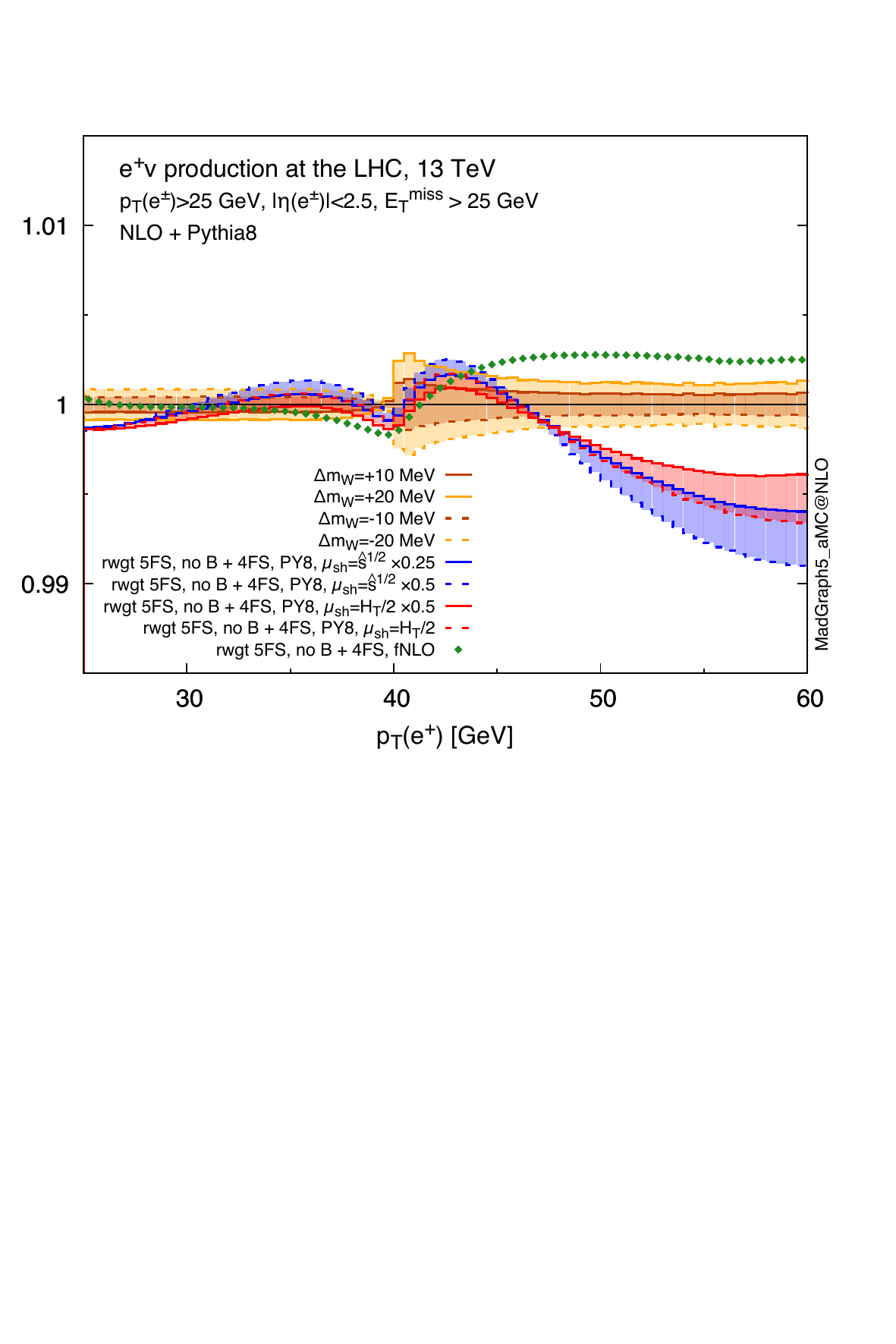}\\[-5cm]
\caption{\label{fig:templates}
  Comparison of templates generated for different $\mw$ values in the CC-DY
  in the 5FS (orange shaded areas). In the top row we show the \powhegbox results, while in the bottom row we present the equivalent curves from \amcnlonospace.
  In blue, red and green we display the curves obtained by using the reweighting function ${\cal R}$, for several different setups.
}
\end{figure}
In the present study we do not directly compare the theoretical distributions
with the data.
We choose one set of input parameters as reference and prepare the templates
accordingly, letting $\mw$ vary in a given range.
We then simulate the distribution with a second set of inputs,
keeping $\mw$ at a fixed nominal value $\mwc$. We fit this distribution,
that we call {\it pseudodata}, with the templates based on the first set of
inputs.
The preferred value $\mwj$ is in general different than $\mwc$,
because the fitting procedure tries to accommodate the distortion induced
by the second set of inputs with a shift of $\mw$.
The difference $\mwj-\mwc$ is an estimate of the difference between
the two results that would be observed,
if one would use templates based on these two sets of inputs
to fit the same experimental data.

In the present study we use all our central choices for the input parameters,
as described in Section \ref{sec:setup} to compute the templates
for the CC-DY lepton transverse momentum and lepton-pair transverse mass
in the plain 5FS using {\tt tune1}, i.e.~our default \pythia tune.
We generate the distributions corresponding to different values of $\mw$
with the reweighting technique described in
refs.\cite{Bozzi:2011ww,Bozzi:2015hha}.
For illustration we show in Figure~\ref{fig:templates},
for the transverse mass of the $W$ boson as well as
for the transverse momentum of the charged lepton,
the comparison of templates computed with different $\mw$ values
in a range up to $\pm 20$ MeV about the central PDG value:
we observe a spread of the curves  at the Jacobian peak
in correspondence of the $W$ resonance.
The same figure also shows the effect
of the $\ptw$ reweighting on these observables,
according to our alternative predictions
as shown in Figure~\ref{fig:ratios}. These curves represent the
pseudodata that will be employed in the fit, which are obtained
again in the plain 5FS with default \pythia tune,
PDG values for the masses and
the $\ptw$-dependent reweighting by ${\cal R}(\ptw)$,
described in Section \ref{sec:transfer},
to simulate the impact of the improved bottom-quark treatment.
For both templates and pseudodata we consider the shape of the distributions:
we define a range of values of the fitted variable around its Jacobian peak
and normalise the distribution to the corresponding integral.
In this way we enhance the sensitivity of the template fit procedure
to the precise position of the peak.

\begin{figure}[h!]
\includegraphics[width=75mm,angle=0]{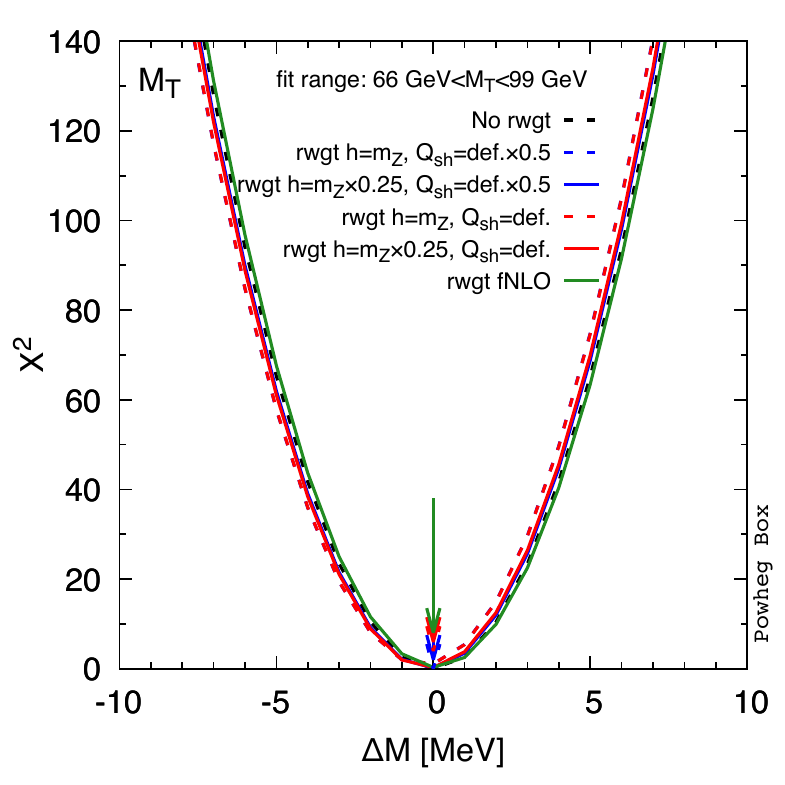}
\includegraphics[width=75mm,angle=0]{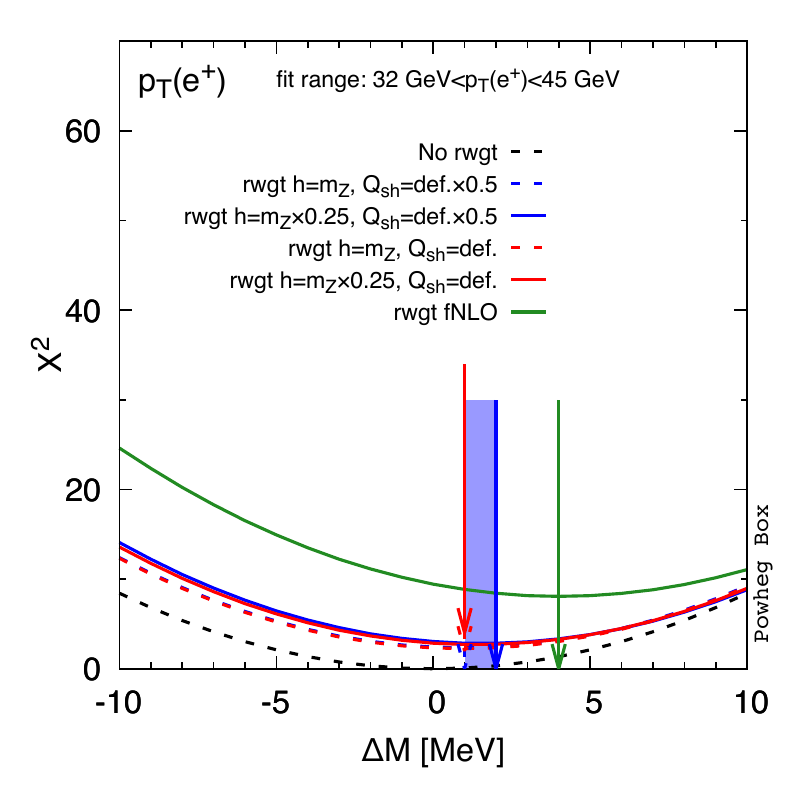}\\
\includegraphics[width=75mm,angle=0]{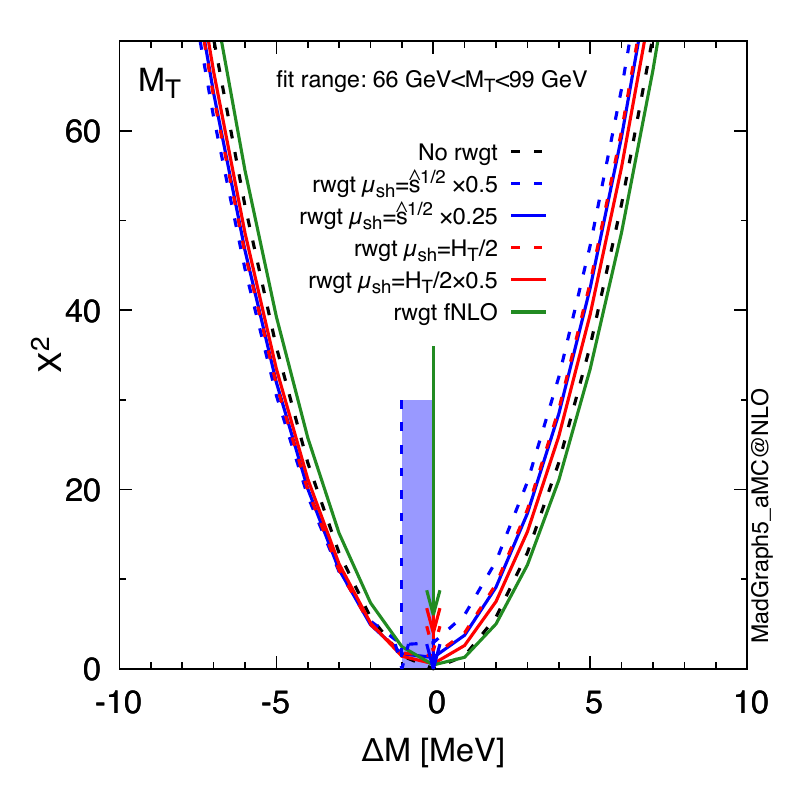}
\includegraphics[width=75mm,angle=0]{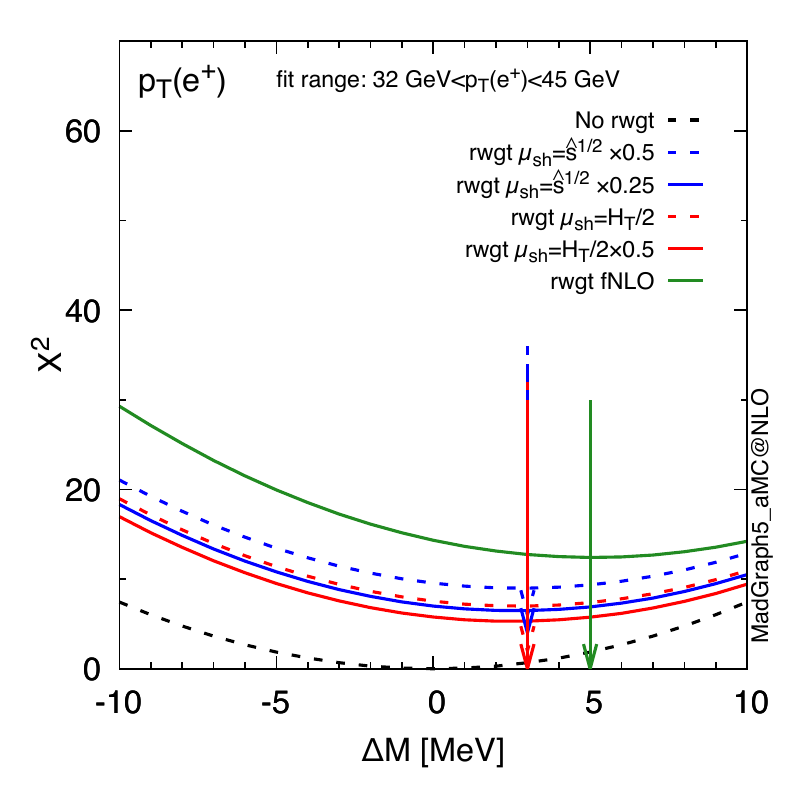}
\caption{\label{fig:chiduemw}
Result of the template fit to distributions that include the improved bottom-quark effects in different QCD approximations.
}
\end{figure}

\begin{figure}[h!]
\centering
\includegraphics[width=145mm,angle=0]{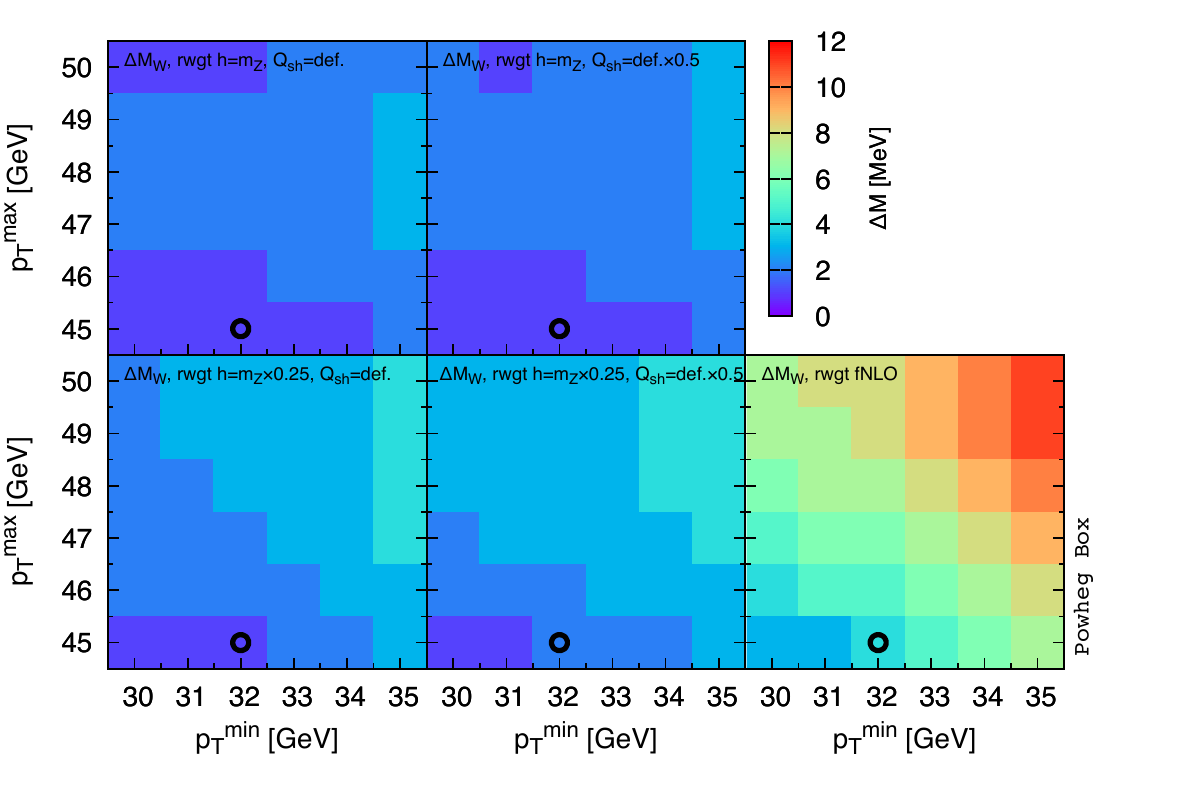}
\includegraphics[width=145mm,angle=0]{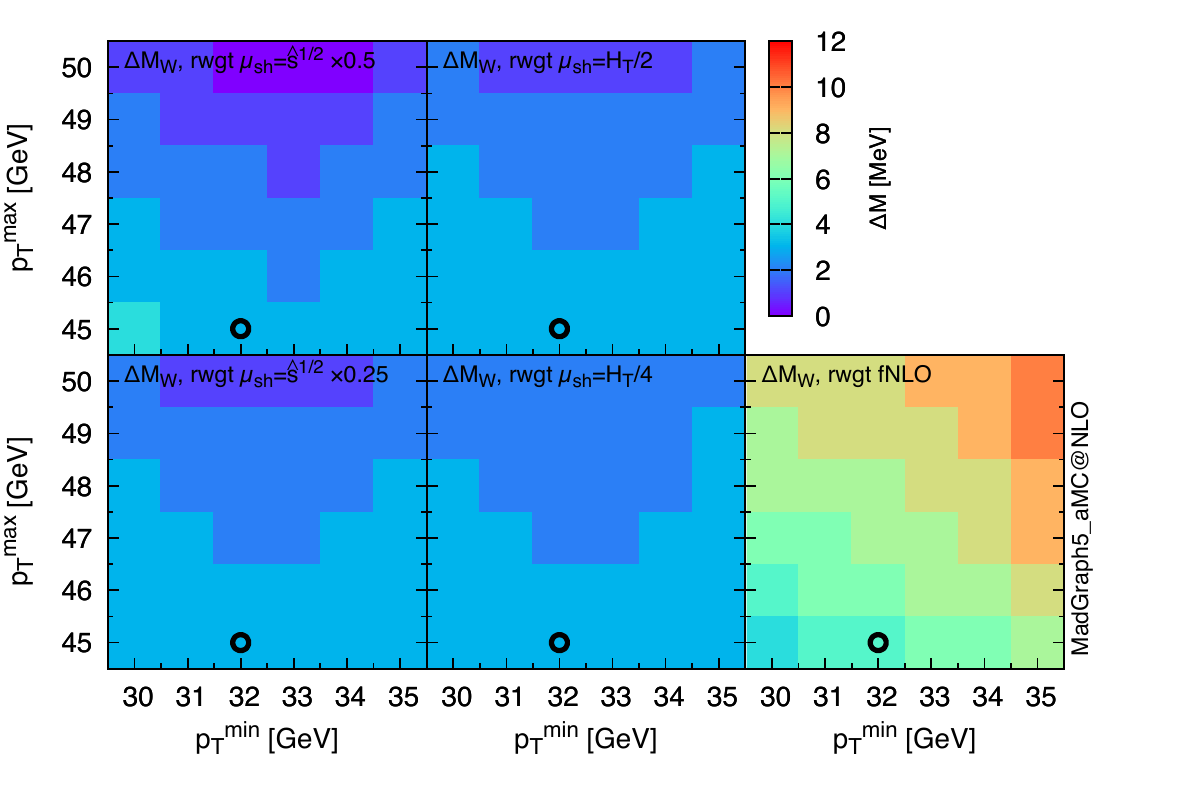}
\caption{\label{fig:fitwindow}
  Dependence on the fit window of the template-fit results,
  in the case of the lepton transverse-momentum distribution.
  The black marks correspond to the fit range used in Figure
  \protect\ref{fig:chiduemw}.
}
\end{figure}

The level of agreement between templates and pseudodata
can be assessed with the least squares method.
The standard definition of a $\chi^2$ indicator,
\be
\chi^2_j(\vec{\cal O}^{\rm data})
=
\sum_{k\in \rm bins}
\frac{({\cal O}^{\rm data}_k -  {\cal O}^{j,\rm template}_k)^2 }{\sigma_k^2}\,,
\label{eq:chidueuncorr}
\ee
assumes that all the bins are uncorrelated and that each contributes
according to its statistical error, represented by $\sigma_k^2$.
For each template $j$ we compute $\chi^2_j$;
as a function of $j$ we should obtain a parabola
whose minimum indicates the preferred value of the fit parameter.
We perform two independent fits on the lepton transverse momentum
and on the $W$-boson transverse mass,
which is defined, starting from the transverse momentum of the charged lepton
and of the neutrino\footnote{
  We identify the neutrino transverse momentum and
  the missing transverse energy, i.e.~$p_\perp(\nu)\equiv p_\perp^{\rm miss} $},
as
\be
M_T(\ell^+,\nu)\equiv
\sqrt{2p_\perp(\ell^+)p_\perp(\nu)(1-\cos \phi_{\ell^+\nu})}\,,
\ee
where $\phi_{\ell^+\nu}$ is the azimuthal angle between the two leptons.
The fit is performed in the following ranges,
which correspond to the ones employed by ATLAS~\cite{Aaboud:2017svj}:
\be
  32\,\gev < p_\perp(\ell^+) < 45\,\gev,\qquad
  66\,\gev < M_\perp(\ell^+,\nu) < 99\,\gev\,.\qquad
\ee
The granularity of the $m_W$ scan is of $1\, \mev$.\\
In Figure~\ref{fig:chiduemw} we show the $\chi^2$ parabolas
and the shift induced by reweighting the $\ptw$ distribution
with our alternative $\ptz$ description.
The left column of the figure refers to the transverse mass,
the right column to the lepton transverse momentum.
Plots in the top row are obtained with the \powhegboxnospace,
those in the bottom row with \amcnlonospace.
As far as the transverse mass is concerned,
all induced shifts are compatible with zero.
In fact this observable is known to be insensitive to the details
of the $\ptw$ modelling \cite{Ellis:1991qj}.
When the lepton transverse momentum is considered, the mass shifts are
of the order $\Delta_{m_W}\sim 4-5\, \mev$
when the 5FS is improved with the fixed-order 4FS prediction,
and of $\Delta_{m_W}\sim 1-2\, \mev$ or  $\Delta_{m_W}\sim 3\, \mev$
for the predictions improved with the NLO+PS 4FS,
in the \powhegbox and \amcnlo respectively.
We take the results based on the fixed-order NLO 4FS calculation
as a technical benchmark,
while we consider the results obtained matching fixed- and all-orders
calculations,
discussed in detail in Section \ref{sec:ptzfour},
as more accurate in the description of these transverse-momentum
distributions.
In conclusion, our estimate of the $\mw$ mass shift due to $b$-quark effects,
in the measurement from the lepton transverse-momentum distribution,
is in general smaller than 5 MeV.
We conclude our section by investigating how the extracted $W$-mass shift
depends on the details of the fitting window,
in particular on its boundaries.
To do so, we compute the shift when the window boundaries
$[p_\perp^{\rm min},p_\perp^{\rm max}]$ vary in the ranges
$30\, \gev < p_\perp^{\rm min} < 35\,\gev$ and
$45\, \gev < p_\perp^{\rm max} < 50\,\gev$.
The resulting values of the mass shift are shown in Figure~\ref{fig:fitwindow}
and can be justified with a comparison with Figure~\ref{fig:templates},
where the reweighted distributions and the templates are compared.
We should consider, for a given template $j$,
which bins contribute the most to the $\chi^2_j$ value
and this can be guessed at glance by checking when a reweighted distribution
follows and when it deviates from the template.
Above the Jacobian peak the reweighting procedure yields a shape
qualitatively different than any of the templates,
so that the inclusion of the bins where the deviation is more pronounced
may affect the position of the minimum of the $\chi^2$ parabola.
This problem is particularly evident when considering the fixed-order results
(green dots in Figure~\ref{fig:templates}),
which correspond to the lower right plot of Figure~\ref{fig:fitwindow}.

\clearpage{}\section{Conclusions}
\label{sec:conclusions}

The high luminosity of the LHC together with the stunning performances of
the detectors (ATLAS, CMS, and also LHCb)
have turned Drell-Yan processes into high-precision arenas where to test our
understanding of the fundamental interactions on the one hand and
to perform the most precise measurements of
the parameters of the SM, on the other hand.
At this level of precision one needs to control not only
higher-order perturbative effects, QCD as well as EW,
but also less obvious ones, such as non-perturbative or parametric effects.
An interesting example, discussed in this work,
is given by the contribution from bottom quarks to Drell-Yan processes
which so far has been considered in the massless approximation.
In fact, the associated production of a lepton pair together
with a $b\bar b$ pair is a rather complicated process,
featuring several (if not all the) aspects that make the description
of final states involving $b$ quarks an
interesting challenge for theorists as well as for experimentalists.

In this paper we have considered $\ell^+\ell^-b\bar b$ production in the 4FS
at the LHC with the main goal of assessing the accuracy and precision
currently achievable of  $\ell^+\ell^-$ observables inclusive over
the bottom quarks. To this aim we have employed state-of-the-art Monte Carlo
tools accurate at NLO in QCD and matched to parton showers.
We have shown that predictions from different NLO MC tools for quantities
that are inclusive with respect to the bottom quark
in the $\ell^+\ell^- b\bar b$ final state are in agreement within the expected
uncertainties.
We have employed a simple prescription which makes it possible to consistently include
the contributions from massive bottom quarks into the inclusive DY production calculated in the 5FS, and studied their effects together with the associated uncertainties.
In so doing we have been able to estimate that the residual uncertainties
have a small but visible ($\Delta\mw < 5$ MeV)
impact on the $W$-mass extraction.
The stability of this prescription,
with respect to the inclusion of higher-order QCD corrections,
could be further explored with the help of codes which make it possible to match QCD-PS with NNLO-QCD accurate predictions for Drell-Yan
processes~\cite{Hoeche:2014aia, Karlberg:2014qua, Alioli:2015toa}.
We have also performed an extensive study in the 4FS of the observables
that are exclusive on the bottom quarks.
This analysis, which is documented in Appendix \ref{sec:appendix},
reveals differences  between formally equivalent methods
that are larger than the (estimated) associated uncertainties,
at least in some cases.
A thorough comparison of many distributions has allowed us to identify
the regions in phase space where the differences arise.
Assessing the origin of such discrepancies in the specific
case of $\ell^+ \ell^- b\bar b$ and providing a resolution
will be an important task for the SM and BSM programme of the LHC
(see e.g., the measurement of $HZ$-associated production
at small transverse momentum and the search for dark matter
in the missing-transverse energy +$b$-jet final states are two examples
directly related to $\ell^+ \ell^- b\bar b$).
A deeper understanding of the treatment of the bottom quark contributions
can be crucial also for the precision prediction of very important
final states like $t\bar t b\bar b$.
This, however, needs a dedicated effort which goes beyond the scope
of our work and it is left for future investigations.

\section{Acknowledgements}
We would like to thank
Maarten Boonekamp, Stefano Camarda, Rikkert Frederix, Stefano Frixione,
 Luca Perrozzi, Paolo Torrielli, for many stimulating discussions.
MZ is grateful to the Physics Department of the University of Milan for
the hospitality in many different occasions.
This work has received funding from the European Union's Horizon 2020 research
and innovation programme as part of the Marie Sklodowska-Curie Innovative
Training Network MCnetITN3 (grant agreement no. 722104).
EB is supported by the Collaborative Research Center SFB676 of the DFG, ``Particles, Strings and
the early Universe''.
AV is supported by
the Executive Research Agency (REA) of the European Commission
under the Grant Agreement PITN-GA-2012-316704  (HiggsTools)
and by
the  European  Research  Council  under  the  Grant
Agreement 740006NNNPDFERC-2016-ADG/ERC-2016-ADG.
MZ is supported
 by the Netherlands National Organisation for Scientific
 Research (NWO), by the European Union's Horizon 2020 research and
innovation programme under the Marie Sklodovska-Curie grant
agreement No 660171 and in part by the ILP LABEX (ANR-10-LABX-63),
in turn supported by French state funds managed by the ANR
within the ``Investissements d'Avenir'' programme
under reference ANR-11-IDEX-0004-02.

\clearpage{}

\appendix
\clearpage{}\section{Appendix: Differential observables in $\ell^+ \ell^- b \bar b$ production}
\label{sec:appendix}
In this appendix we compare results obtained with different PS and/or matching schemes for various
differential observables
in $pp\to \ell^+\ell^-b\bar b$ production, possibly
distinguishing different signatures
depending on the number of tagged $b$-jets.\\

We use the setup described in Section~\ref{sec:setup}.
After parton shower and hadronisation, hadrons are clustered into jets using
the anti-$k_T$ algorithm \cite{Cacciari:2008gp} as implemented in
{\sc FastJet} \cite{Cacciari:2005hq,Cacciari:2011ma}, using a radius
parameter $R=0.4$. Jets are required to satisfy the following conditions
\begin{equation}
    p_\perp(j) > 30\,\gev\,,\qquad \left|\eta(j)\right| < 2.5\,.
\end{equation}
A jet is considered as a $B$-tagged jet if at least one $B$-flavoured hadron is
found among its constituents. For fixed-order predictions we apply the same jet-clustering algorithm to QCD
partons (gluons and quarks, including the $b$),
and we consider a jet as $B$-tagged if at least one $b$ quark appears
among its constituents.
In both cases we assume a 100\% $B$-tagging efficiency and zero mis-tagging rate.\\

The point of this comparison is to stress the fact that the differences that emerge by employing
different matching approaches and QCD PS models
(as one can appreciate in figures \ref{fig:ptzbbQ}, and \ref{fig:nbjetbjet}-\ref{fig:etabjet2}),
make it apparent that higher-order terms with respect to the
$\as$ expansion, subleading in the counting of logarithmic enhancing factors,
can nevertheless be numerically sizeable.\\
We consider the width of the envelope of the different uncertainty bands
presented in these figures
as a conservative quantity useful to characterise our level of understanding
of the observable under consideration and of the accuracy of our simulations.
When we observe a similar shape in the correction factors expressing the impact of all the terms beyond NLO-QCD,
we tend to consider the envelope a reliable conservative estimate of the residual uncertainties;
when this is the case, in all the plots considered the envelope
has a width typically of ${\cal O}(\pm 20\%)$ with respect to its mid point, a value that also represent the typical uncertainty from scale variations in most kinematic
configurations (scale and PDF uncertainties are shown for all differential observables).
When instead we observe different trends in the corrections,
rather than quoting a very large uncertainty,
we can only argue that the comparison is signalling the presence of a quantity
whose description is very sensitive to the details of the radiation
and deserves further analytical and numerical investigation. The colour code employed in all
figures in this appendix is the same as in Figure~\ref{fig:ptzbbQ}.

\subsection{Jet multiplicities}
\begin{figure}[!h]
\centering
\includegraphics[width=0.48\textwidth,angle=0,clip=true,trim={0.65cm 6.7cm 0.cm 0.4cm}]{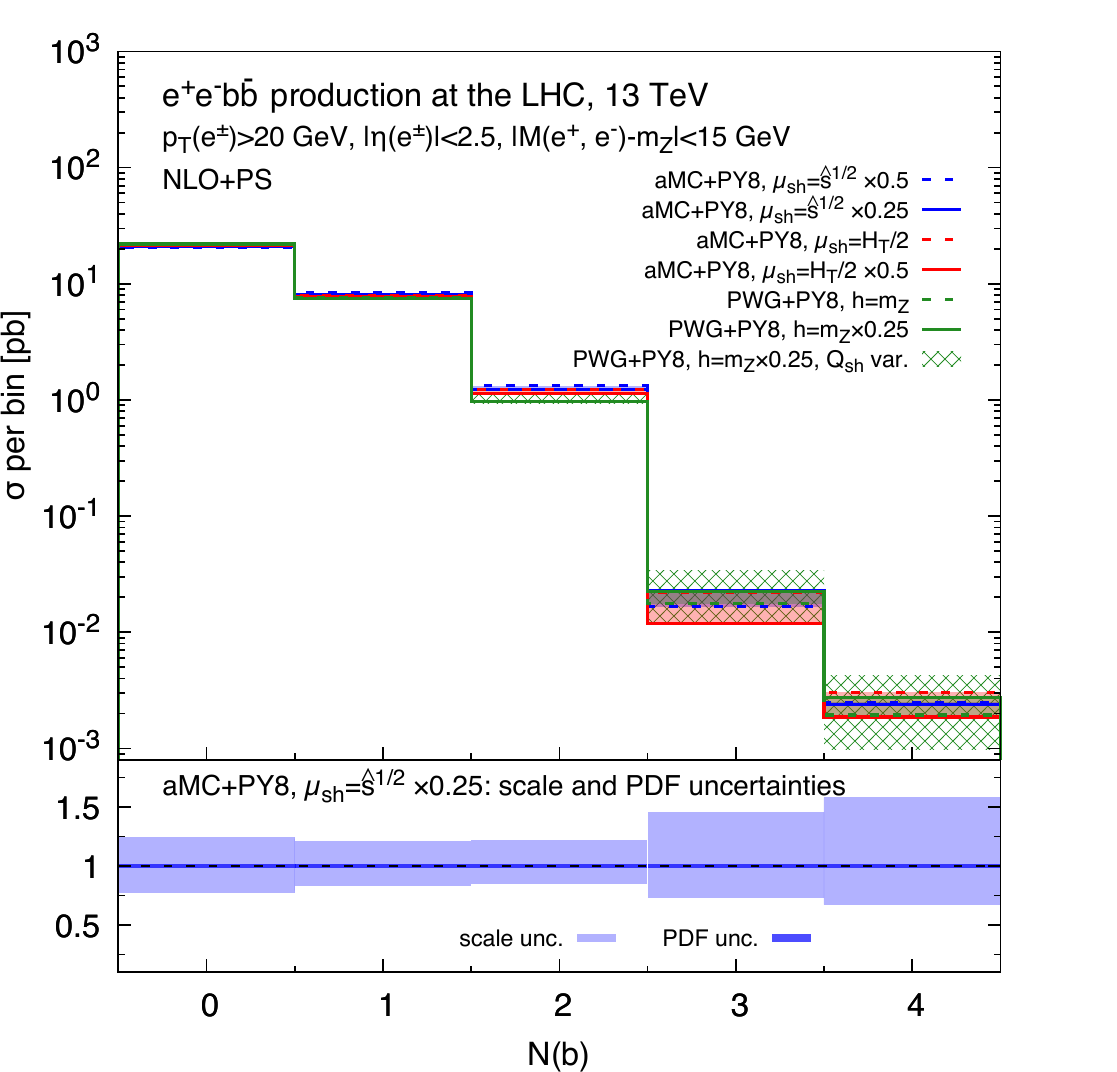}
\includegraphics[width=0.48\textwidth,angle=0,clip=true,trim={0.65cm 6.7cm 0.cm 0.4cm}]{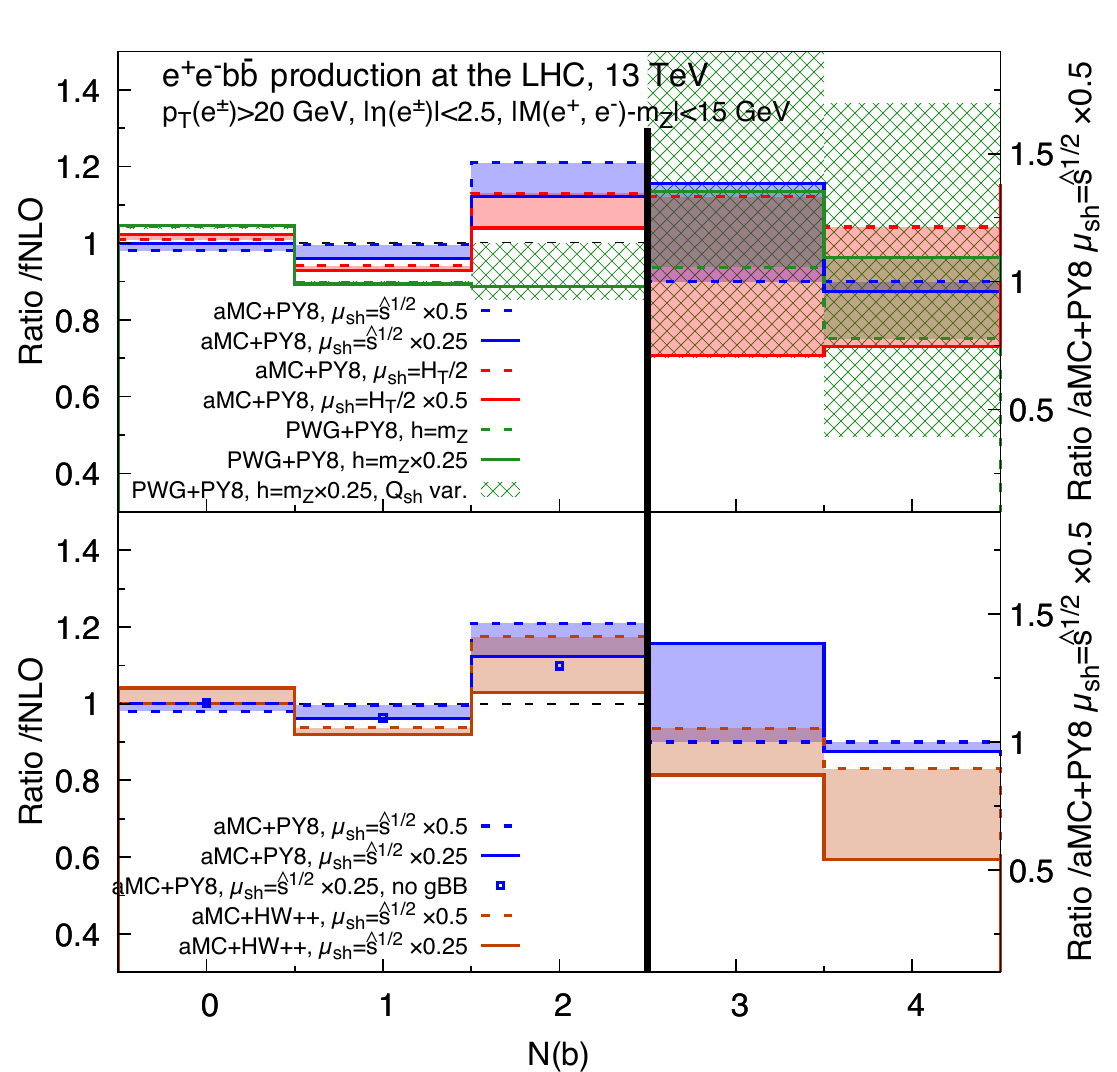}
\caption{\label{fig:nbjetbjet}
  The $b$-jet multiplicities.
  Same colour codes and approximations of figure \protect\ref{fig:ptzbbQ}.}
\end{figure}
The first observable we investigate is the number of reconstructed $b$ jets, shown in
 Figure~\ref{fig:nbjetbjet}.
With respect to the normal layout of the figures, for this specific observable we also show as
a green-patterned band the uncertainty related to the variation of $\qsh$
in the ``remnant'' events of the \powhegbox samples, as described in Section~\ref{sec:setup}.
Higher-order QCD corrections play a non trivial role in the jet reconstruction,
yielding in turn sizeable effects.
The $b$-jet multiplicity is thus the first quantity that has to be discussed,
for a correct interpretation also of the other observables.
The largest bin is the one with zero $b$-jets,
because the production of $b$ quarks is due to
the collinear splitting of the incoming gluons,
so that the transverse momentum of the jet that includes the $b$ quark
does not fulfil the jet definition.
The number of events with 1 or 2 $b$ jets depends on the transverse momentum
distribution of the final state $b$ and $\bar b$ at NLO-QCD.
Higher-order corrections beyond NLO-QCD, simulated with a QCD-PS,
yield a redistribution of the events.
We observe that in \amcnlo there is a moderate stability
of the 0-jet and 1-jet bins
(changes do not exceed the $\pm 5\%$ level)
and an increase of the 2-jets bin with respect to the fixed NLO prediction.
The precise description of the effects in the first two bins and their overall
stability depend on the details of the QCD-PS and PS phase space adopted.
The increase of the third bin
is due to a migration of events from the 1-jet to the 2-jets bin.
Even if the absolute number of events that migrate is not large,
the percentage effect is large, of ${\cal O}(+20\%)$,
because of the steeply-falling shape of the distribution.
The hard recoil of the $\ell^+\ell^-b\bar b$ system,
compared to the fixed-order prediction,
in \amcnlo at intermediate transverse momentum values
(see Figure~\ref{fig:ptzbbQ}) may explain the larger number of events
with both $b$ quarks passing the $b$-jet requirements.

In the \powhegbox case we observe an increase of the 0-jet bin and
a corresponding reduction of the rates with 1 and 2 $b$ jets
(the observables with a genuine NLO accuracy),
independently of the value of the $h$ scale in the damping parameter,
if the default prescription for $\qsh$
(i.e.~the transverse momentum of the first, hardest emission)
is used.
Variations in the shower scale of the ``remnant'' events
give instead effects comparable with those from
shower-scale variations in \amcnlo, in the 2 $b$-jets bin.
The latter is the most sensitive to changes in the treatment of
the ``remnant'' events, characterised by a large transverse momentum of the
first parton.
Although the variation of $\qsh$ will not be shown
for the other observables presented in this section,
the reader should keep in mind that the $h$ variation in the \powhegbox
may give only a partial estimate of the theoretical uncertainties,
and that other sources of uncertainty exist.

Events with 3 or 4 $b$ jets are due to additional splittings
via the QCD-PS and are affected by large parametric uncertainties.

\subsection{$\ptz$ with extra tagged $b$ jets}
\begin{figure}[!h]
\centering
\includegraphics[width=0.48\textwidth,angle=0,clip=true,trim={0.65cm 6.7cm 0.cm 0.4cm}]{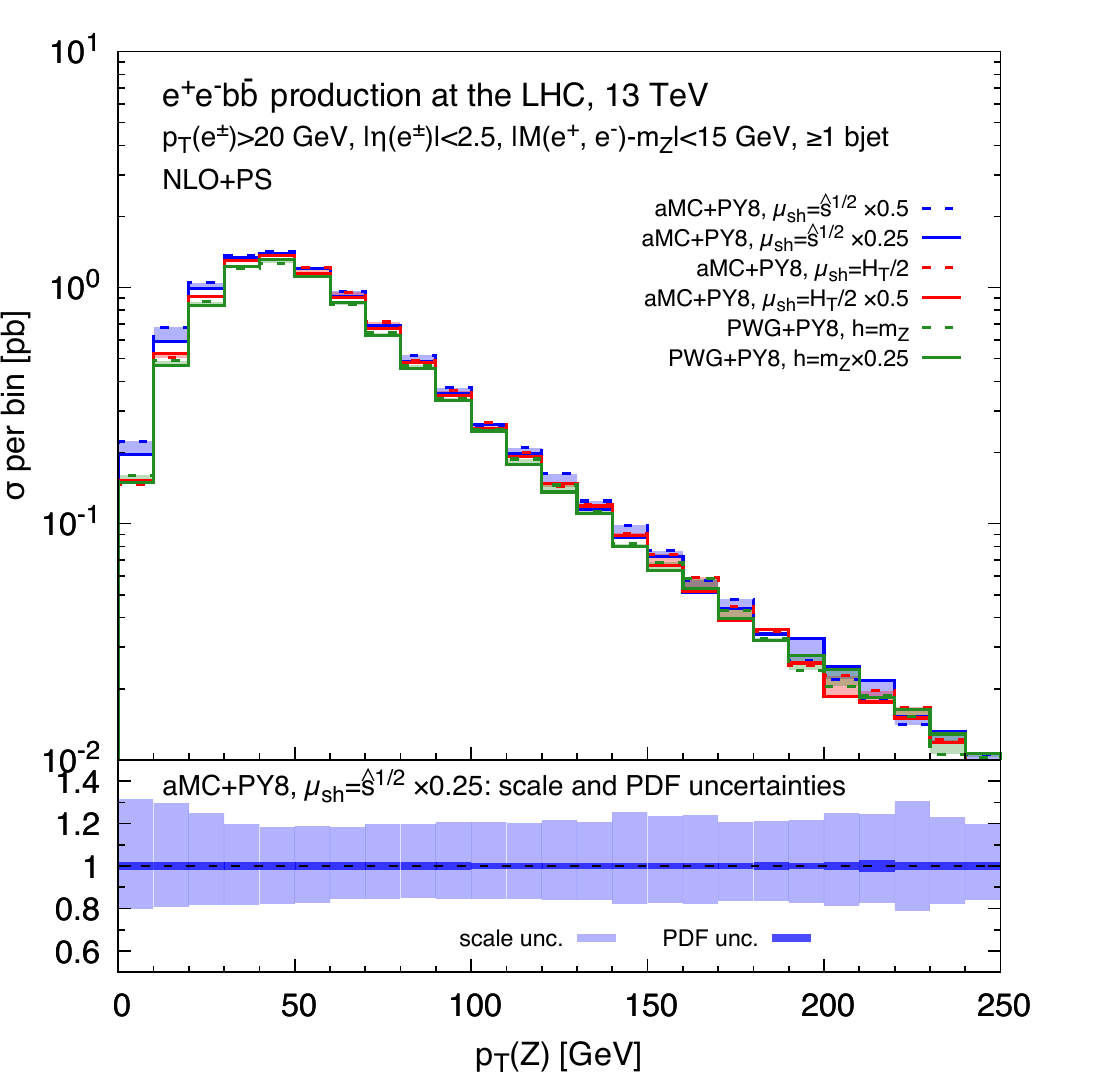}
\includegraphics[width=0.48\textwidth,angle=0,clip=true,trim={0.65cm 6.7cm 0.cm 0.4cm}]{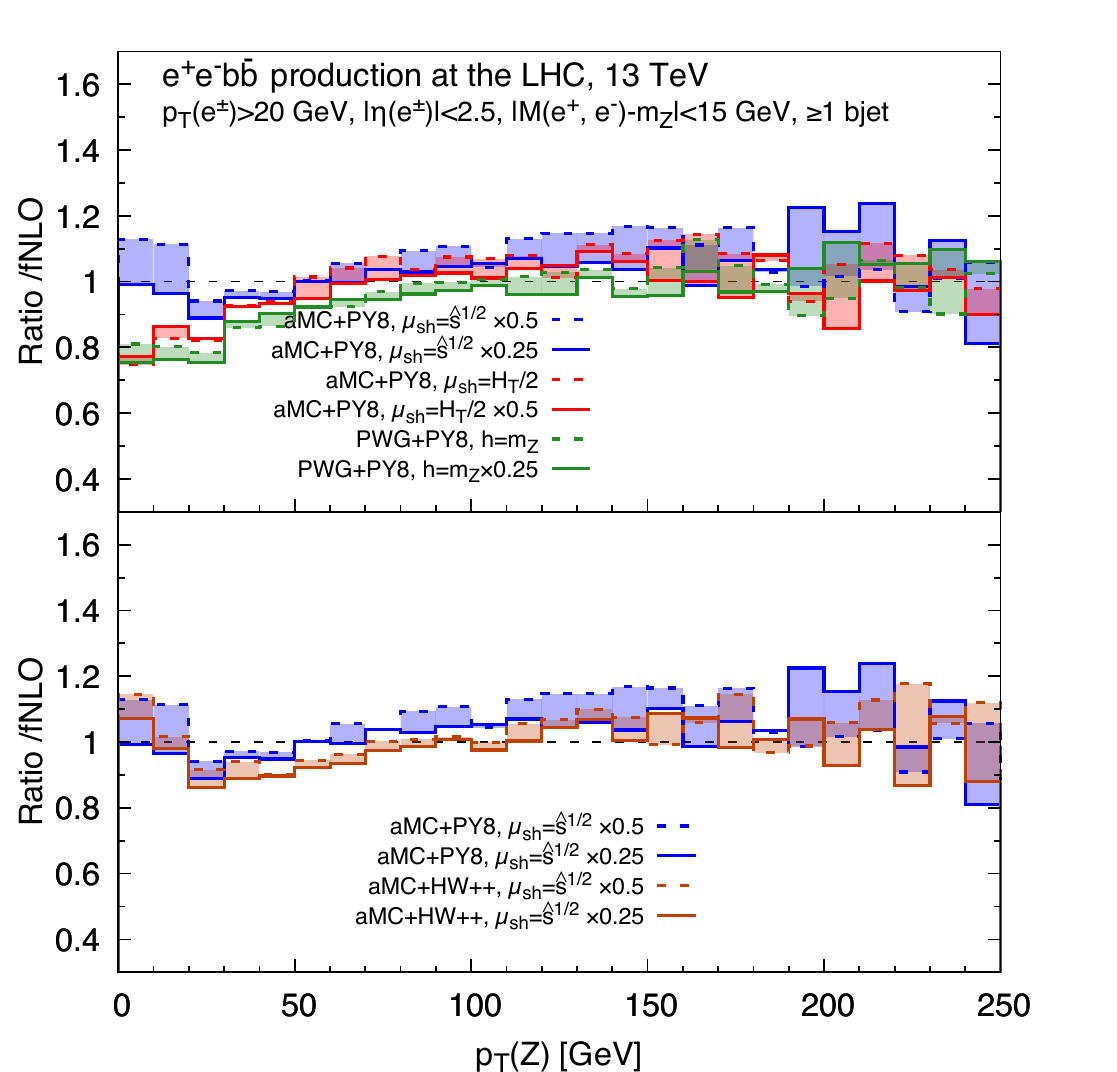}
\caption{\label{fig:ptZ1jets}
$\ptz$ distribution in association with at least 1 $b$ jet.
}
\end{figure}
\begin{figure}[!h]
\centering
\includegraphics[width=0.48\textwidth,angle=0,clip=true,trim={0.65cm 6.7cm 0.cm 0.4cm}]{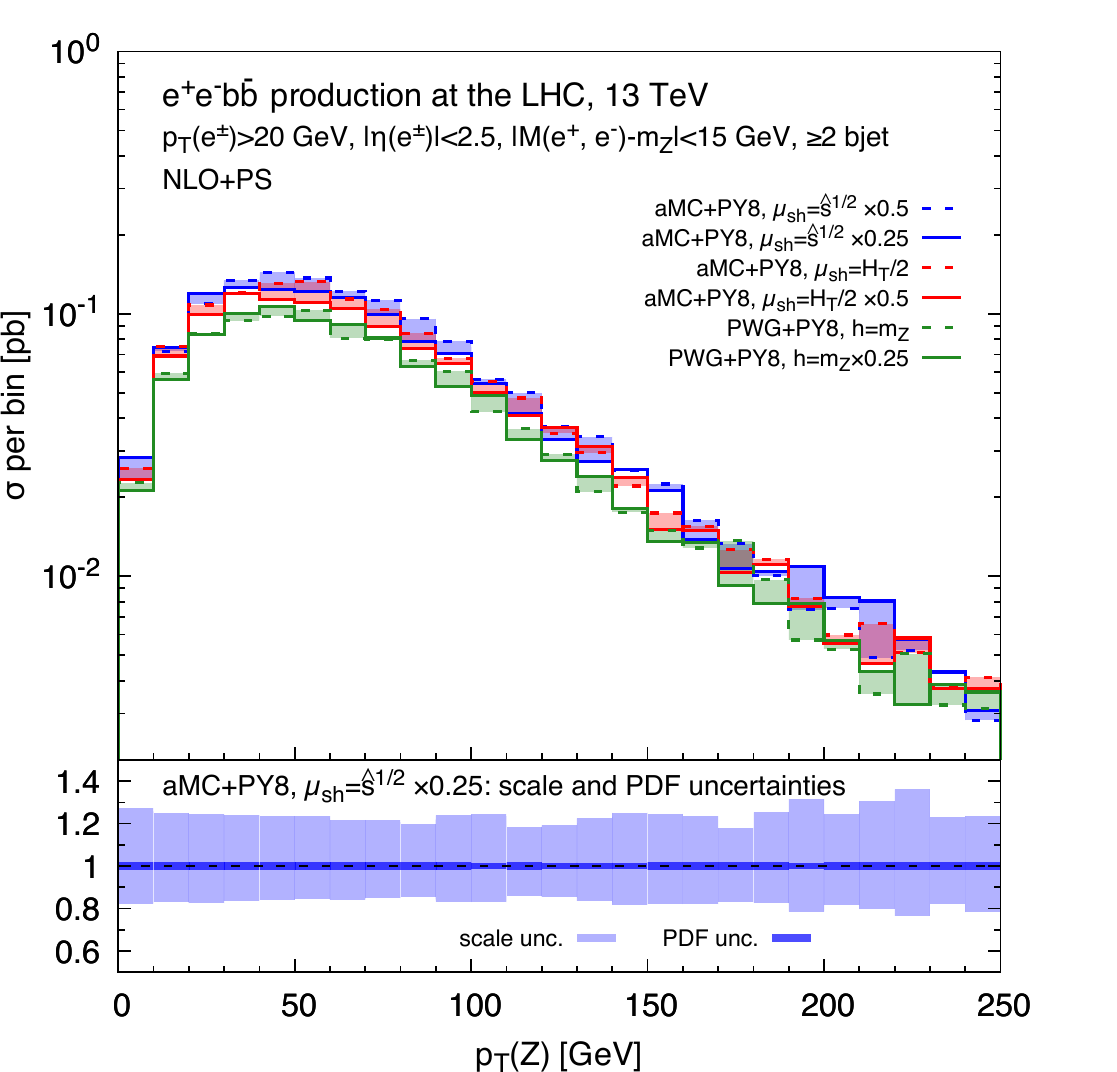}
\includegraphics[width=0.48\textwidth,angle=0,clip=true,trim={0.65cm 6.7cm 0.cm 0.4cm}]{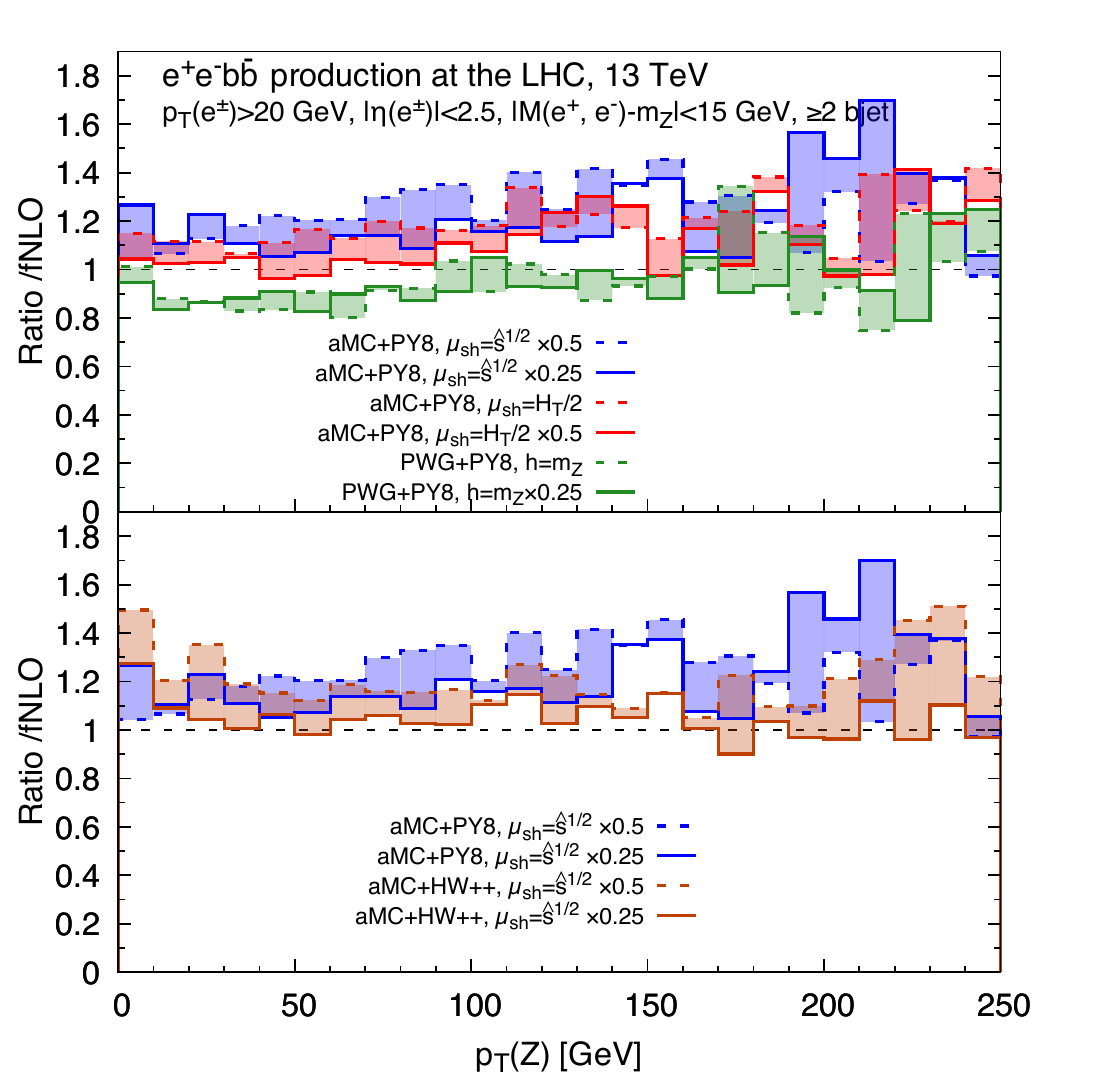}
\caption{\label{fig:ptZ2jets}
$\ptz$ distribution in association with at least 2 $b$ jets.
}
\end{figure}
In Figure~\ref{fig:ptZ1jets}
we show the results obtained for the $\ptz$ distribution, where the
lepton pair is detected in association with at least 1 $b$ jet.
The size of the higher-order corrections,
with respect to fixed NLO,
is positive and of ${\cal O}(+10\%)$ for $\ptz>50$ GeV.
In the limit $\ptz\to 0$, the choice in \amcnlo of the variable used
to select $\mush$ is very important, yielding positive (${\cal O}(+20\%)$)
or negative (${\cal O}(-20\%)$) effects with $\hat s$ or $H_T/2$ respectively.
At moderate or large $\ptz$ (above 100 GeV),
there is a good shape agreement among the different matched predictions,
while differences in normalisation reflect
those the 1-jet bin in Figure~\ref{fig:nbjetbjet}.
In Figure~\ref{fig:ptZ2jets}
we show the results obtained for the $\ptz$ distribution,
where the lepton pair is detected in association with at least 2 $b$ jets.
Corrections with respect to fixed NLO span from +30\% for \amcnlonospace+\pythia
down to -20\% for \powhegboxnospace+\pythia
but they are rather flat in shape.
This behaviour is associated to the positive impact of QCD-PS corrections
in the value of the 2 $b$-jets multiplicity.\\
In both Figures \ref{fig:ptZ1jets} and \ref{fig:ptZ2jets}
we observed a fair compatibility of the predictions computed with the different
options of matching scheme and PS,
once the differences in normalisation are accounted for.

\subsection{Invariant mass of the two hardest $b$ jets}
\begin{figure}[!h]
\centering
\includegraphics[width=0.48\textwidth,angle=0,clip=true,trim={0.65cm 6.7cm 0.cm 0.4cm}]{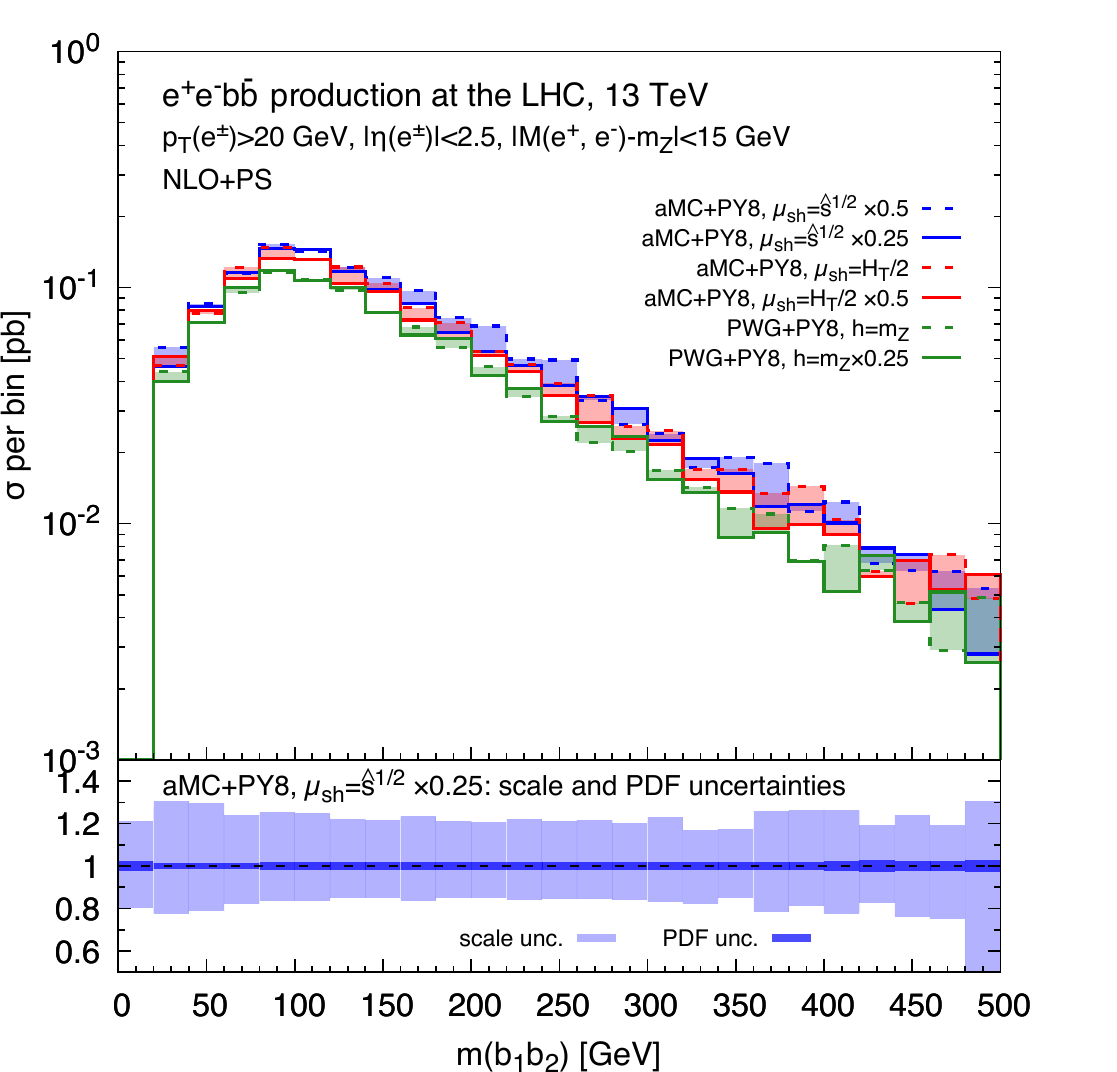}
\includegraphics[width=0.48\textwidth,angle=0,clip=true,trim={0.65cm 6.7cm 0.cm 0.4cm}]{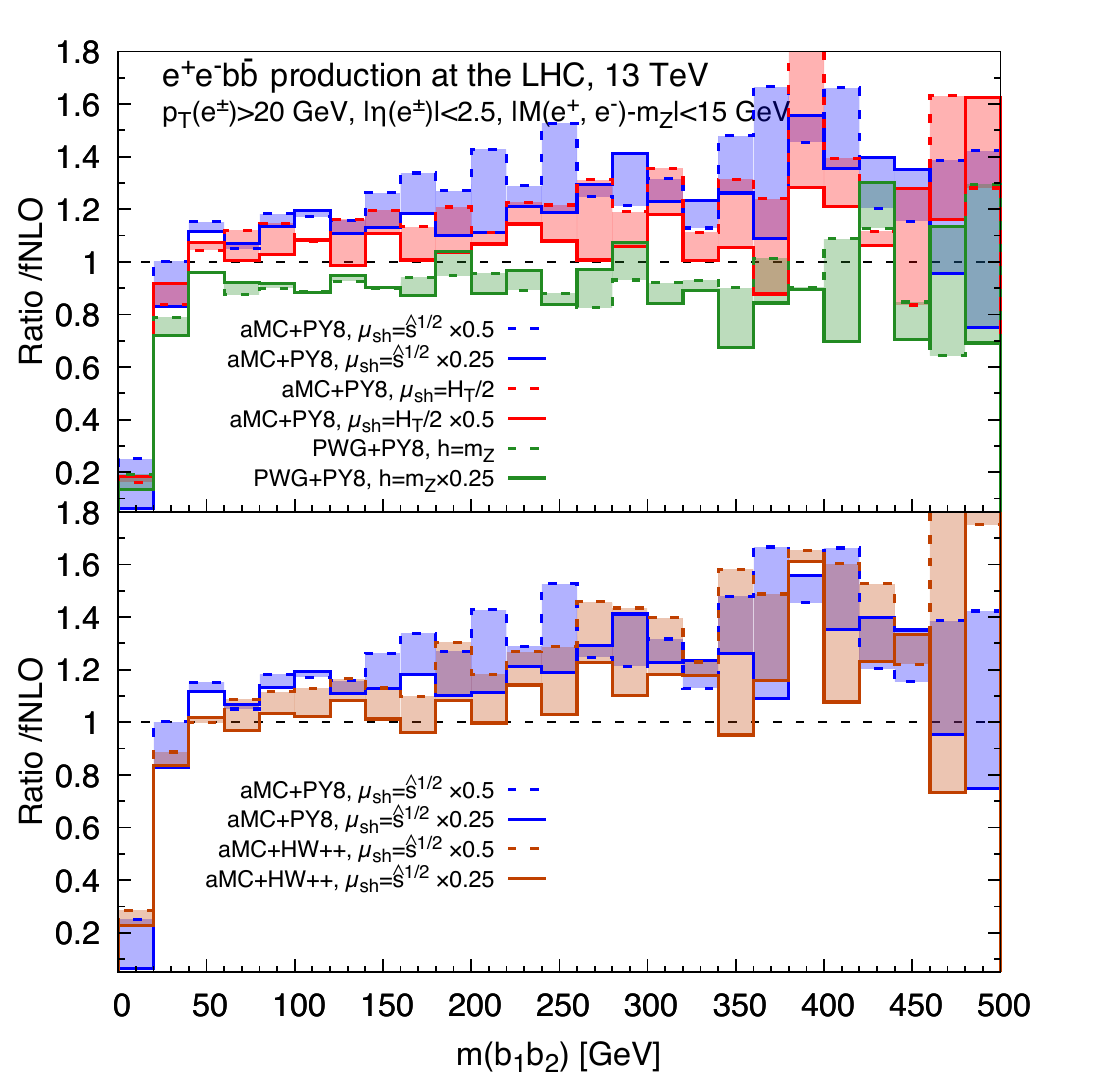}
\caption{\label{fig:minvbjetpair}
Invariant mass distribution of the hardest $b$-jet pair.
}
\end{figure}
In Figure~\ref{fig:minvbjetpair} we consider a final state
with at least 2 $b$ jets and study the invariant-mass distribution
of the hardest $b$-jet pair.
This observable shows a great sensitivity to the details of the matching,
in particular on secondary $g\to b\bar b$ splittings generated by the PS.
At very low invariant masses we observe a large negative correction
in all matching schemes,
due to the definition of $b$ jet and to the action of the QCD-PS:
at fixed NLO the $b$ jets contain, beside the $b$ quarks,
at most one additional parton and
the jet mass is therefore rather close to the $b$-quark mass;
the inclusion of additional partons via QCD-PS rapidly increases
the total jet mass,
with a consequent migration of events to the larger dijet-mass bins
and a corresponding depletion of the first ones.
At larger invariant masses,  for $m(b_1b_2)>50$ GeV,
we observe that the PS corrections obtained with \amcnlo are positive,
with the predictions matched to \pythia reaching the +$40\%$ level
when $m(b_1b_2)\sim 500$ GeV.
The effect of matching to \herwig is milder and flatter
with respect to fixed NLO, while the \powhegbox predictions lie below it.
In Figure~\ref{fig:minvbjetpair}
the differences between the various matching options are
a consequence of the jet definition,
because the largest fraction of the radiative effects due to collinear emissions is integrated in the jet cone.
Looking at the uncertainty bands due to shower-scale (in \amcnlonospace)
and $h$ variations (in the \powhegboxnospace),
we notice that the latter are visibly smaller than the former.
A similar behaviour has been observed for the same observable
in the context of the $t\bar t b \bar b$ implementation
in the \powhegboxnospace~\cite{Jezo:2018yaf}.

\subsection{Invariant mass of the two hardest $B$ hadrons with or without tagged $b$ jets}
\begin{figure}[!h]
\centering
\includegraphics[width=0.45\textwidth,angle=0,clip=true,trim={0.65cm 6.7cm 0.cm 0.4cm}]{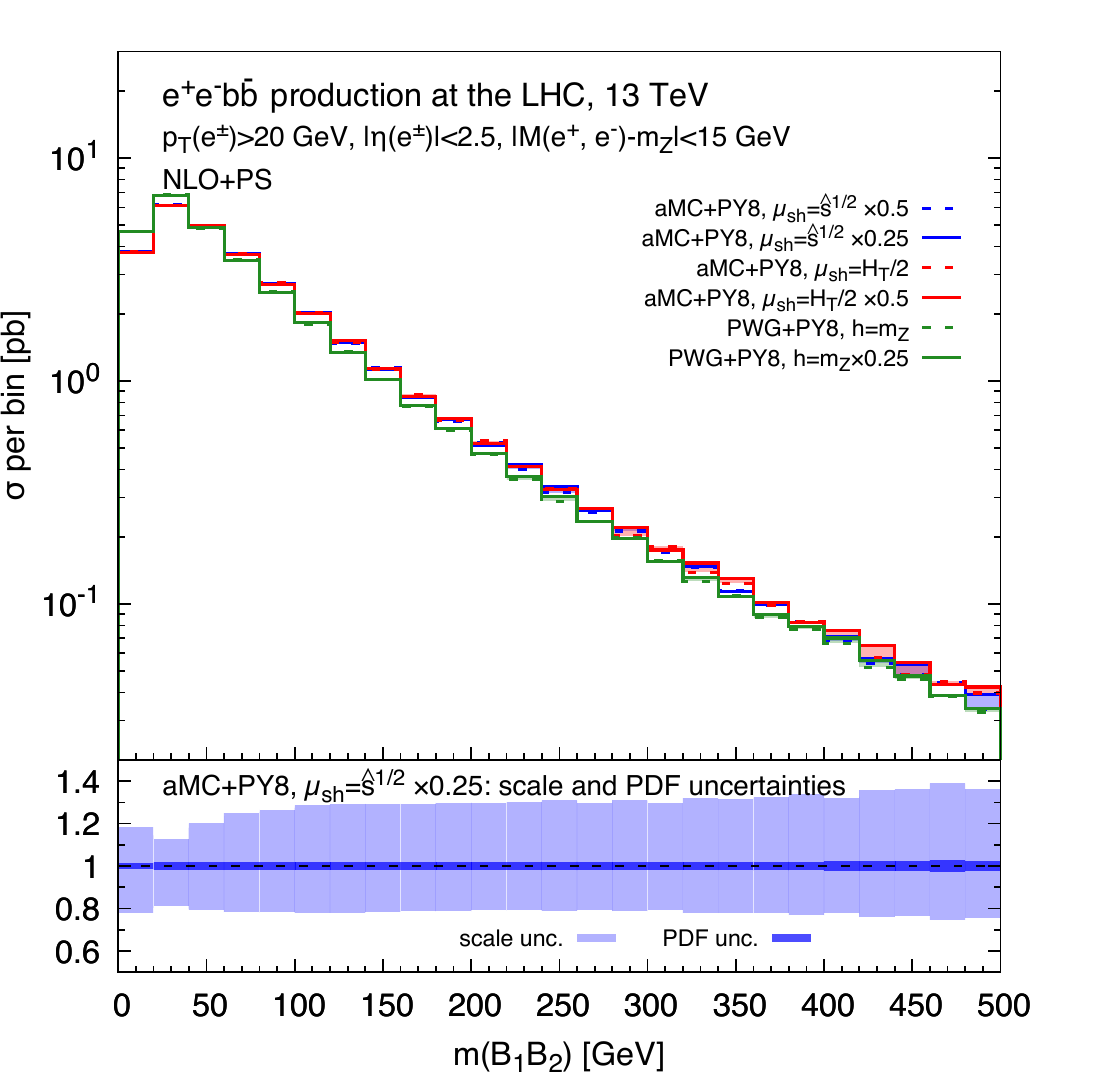}
\includegraphics[width=0.45\textwidth,angle=0,clip=true,trim={0.65cm 6.7cm 0.cm 0.4cm}]{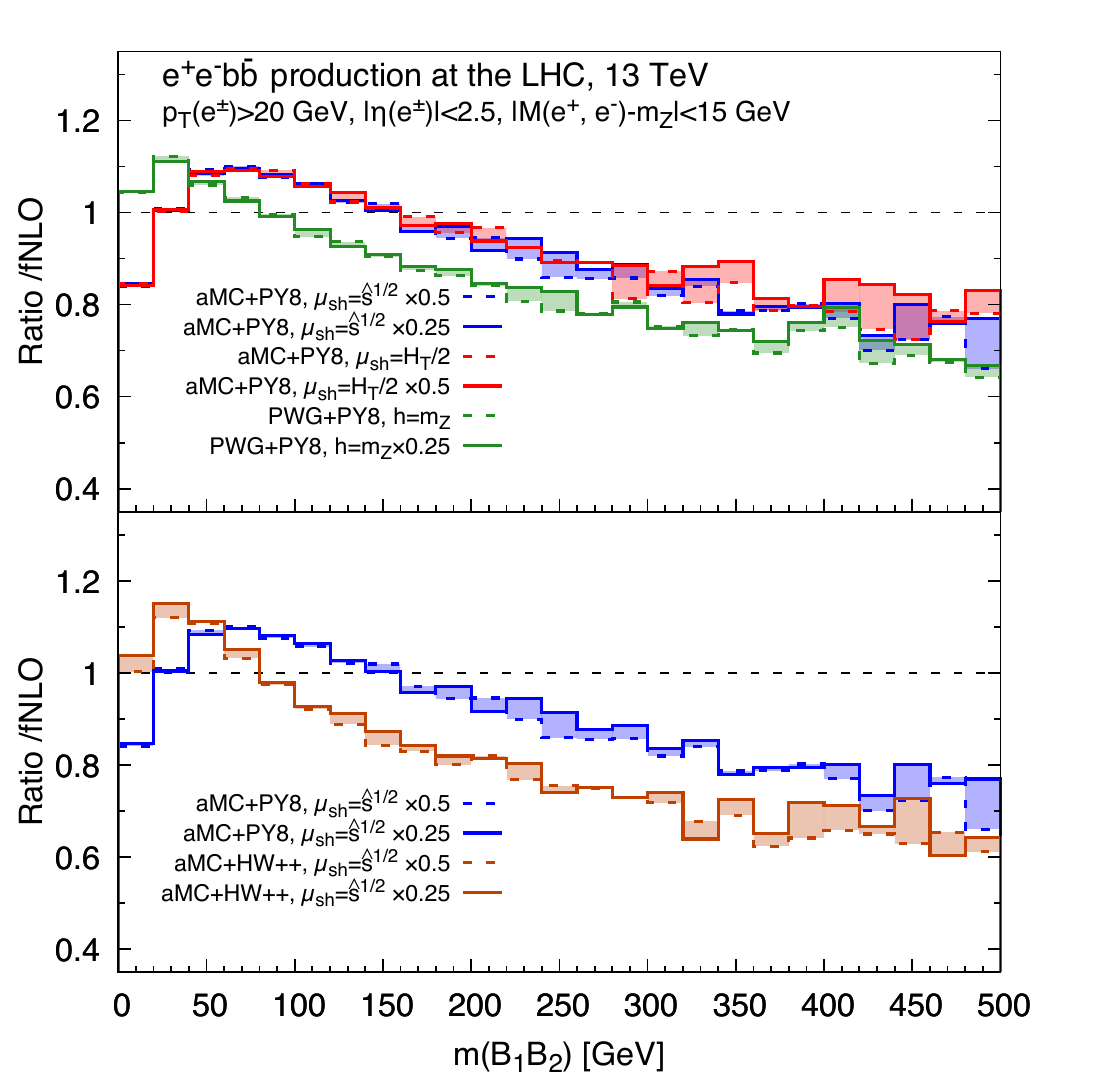}
\caption{\label{fig:minvBBpair0jets}
Invariant mass distribution of the $B$ hadrons pair in association with at least 0 $b$ jets.
}
\end{figure}
\begin{figure}[!h]
\centering
\includegraphics[width=0.45\textwidth,angle=0,clip=true,trim={0.65cm 6.7cm 0.cm 0.4cm}]{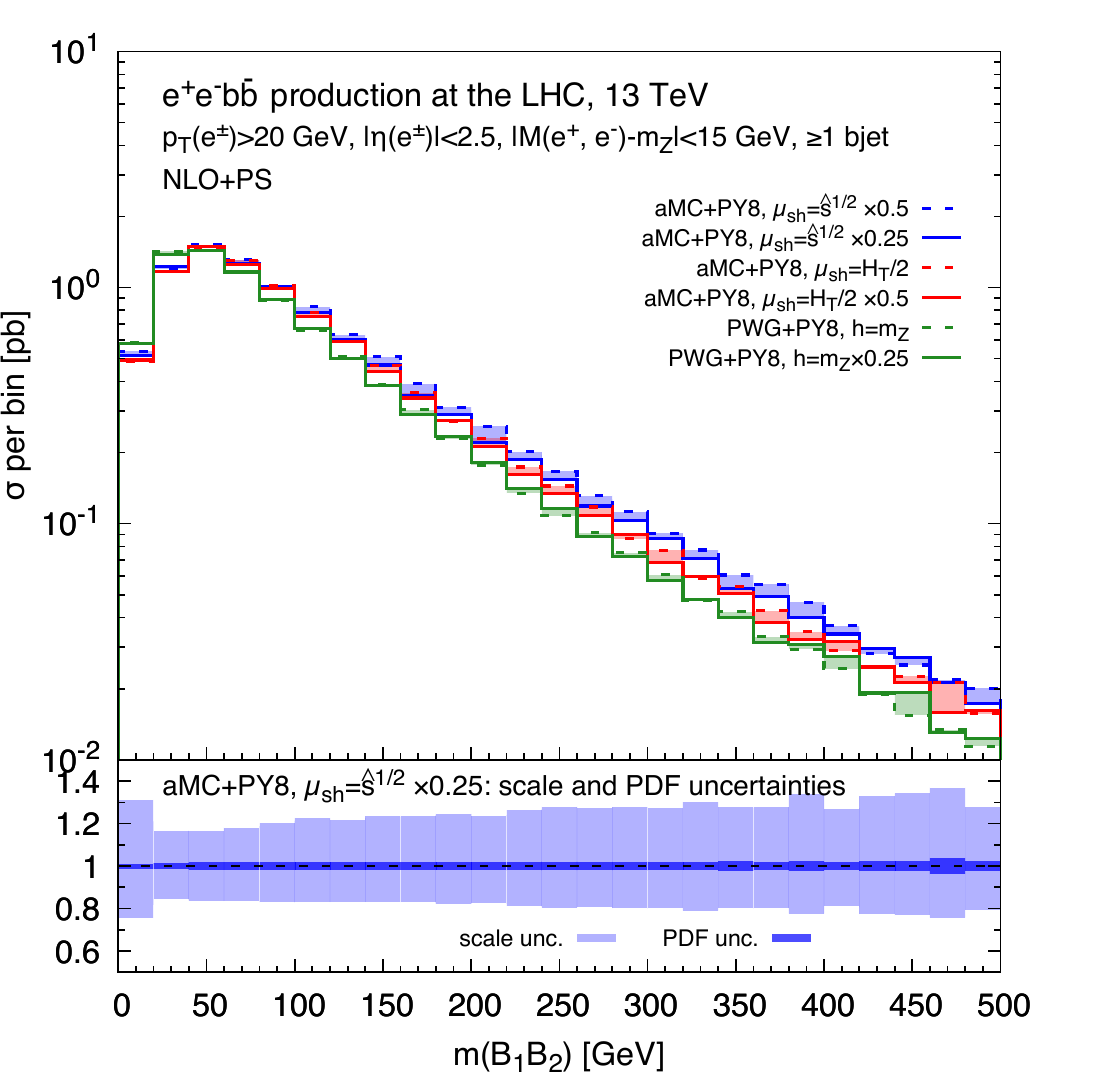}
\includegraphics[width=0.45\textwidth,angle=0,clip=true,trim={0.65cm 6.7cm 0.cm 0.4cm}]{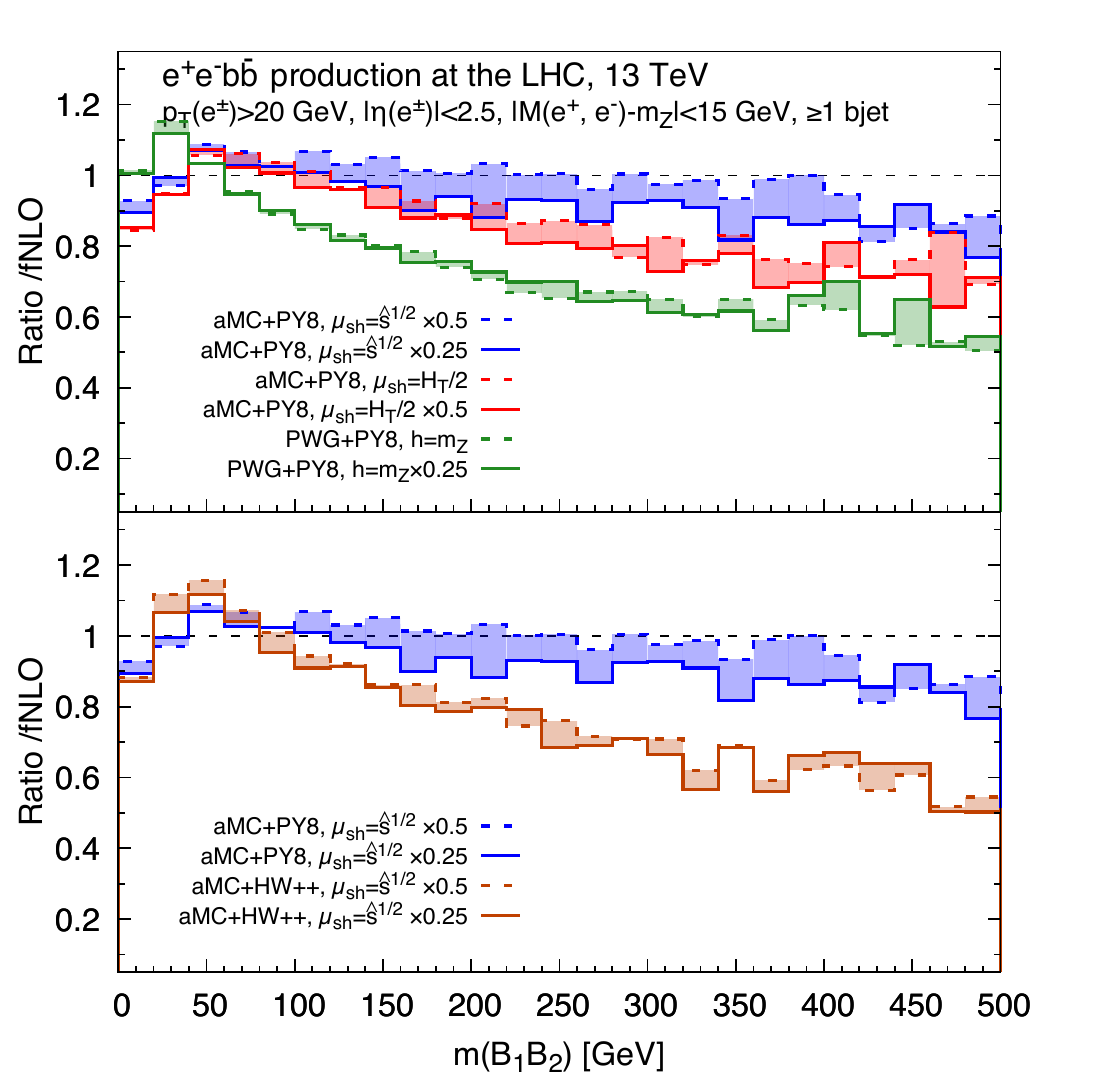}
\caption{\label{fig:minvBBpair1jets}
Invariant mass distribution of the $B$ hadrons pair in association with at least 1 $b$ jet.
}
\end{figure}
\begin{figure}[!h]
\centering
\includegraphics[width=0.45\textwidth,angle=0,clip=true,trim={0.65cm 6.7cm 0.cm 0.4cm}]{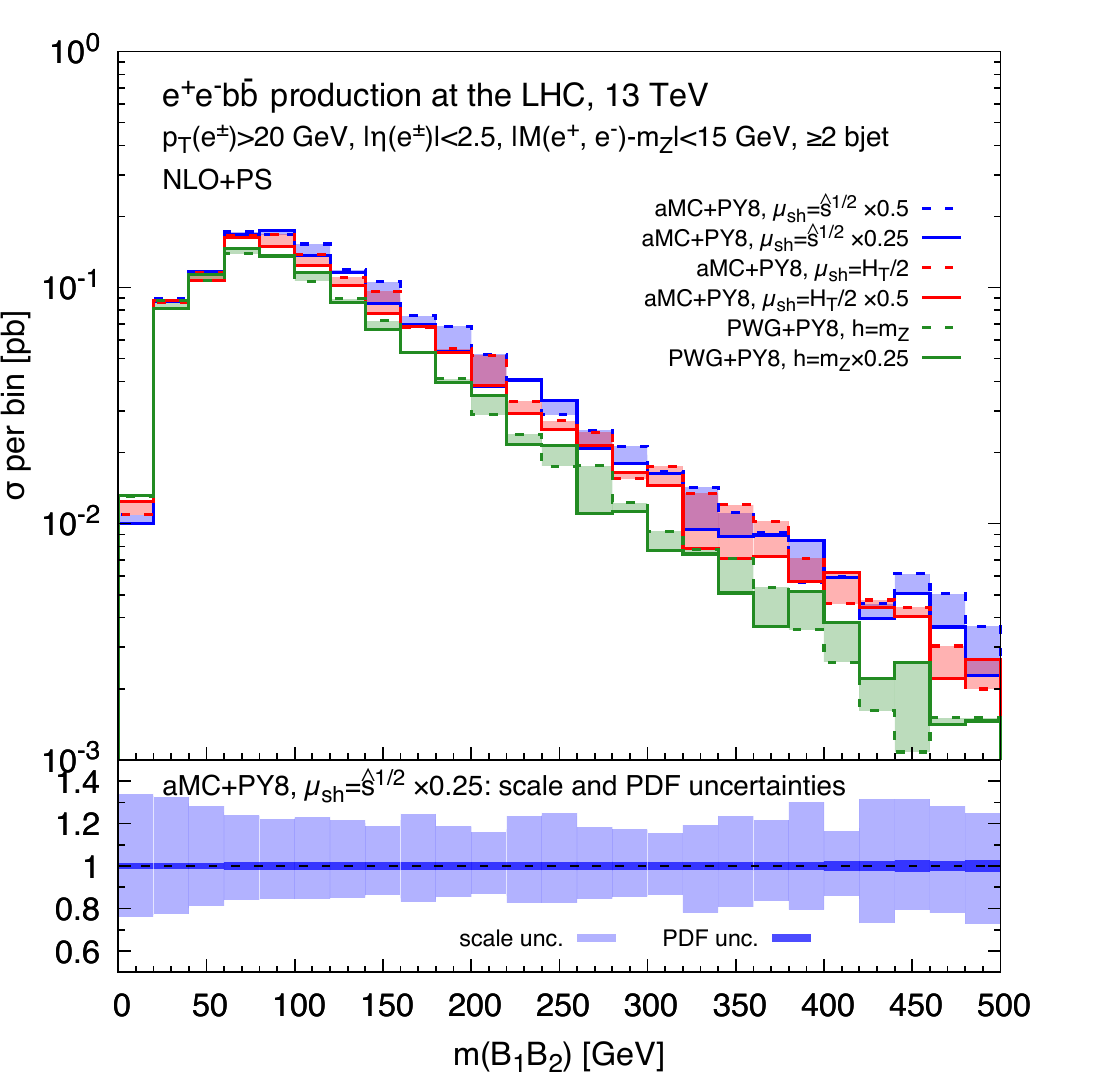}
\includegraphics[width=0.45\textwidth,angle=0,clip=true,trim={0.65cm 6.7cm 0.cm 0.4cm}]{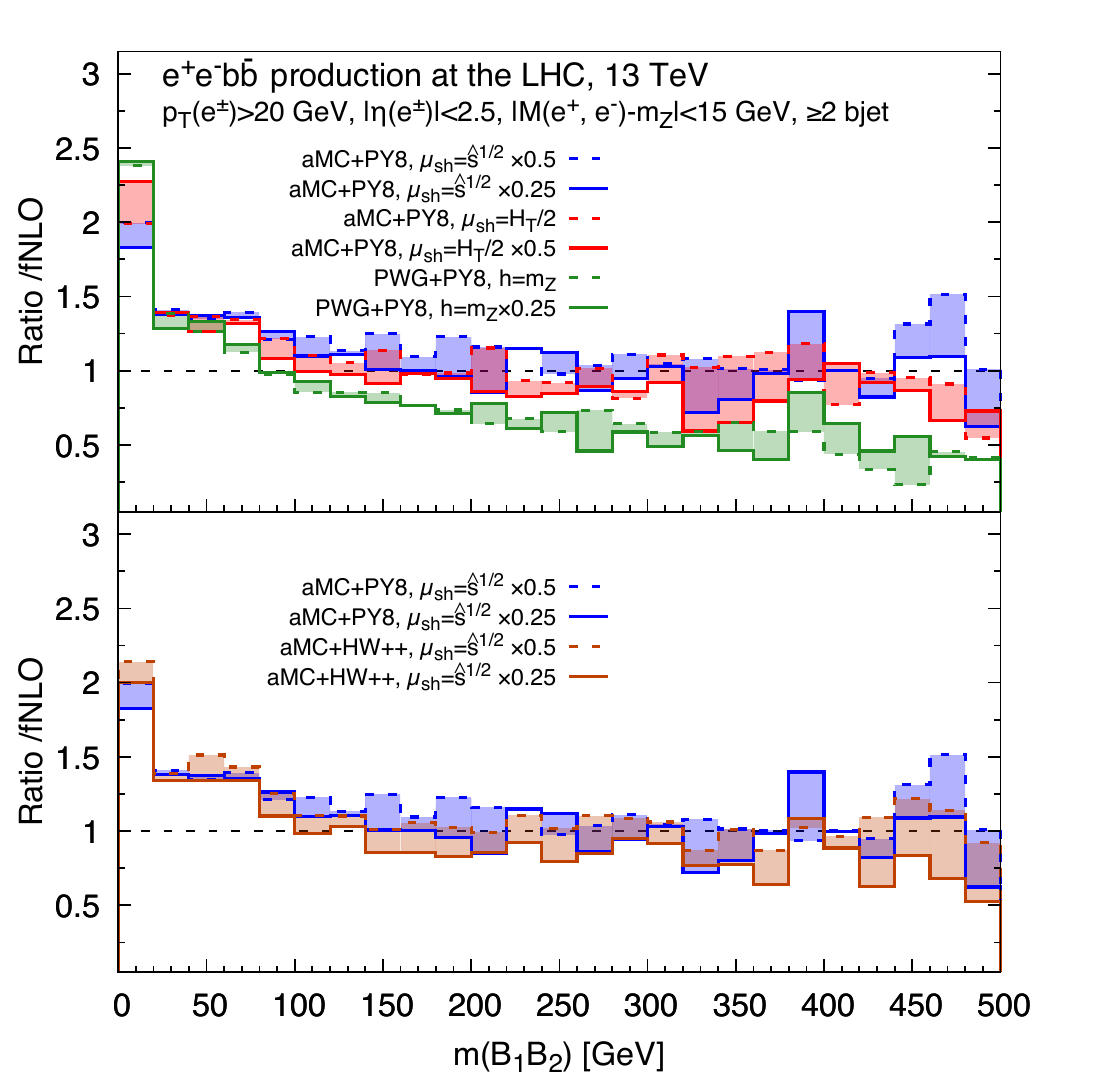}
\caption{\label{fig:minvBBpair2jets}
Invariant mass distribution of the $B$ hadrons pair in association with at least 2 $b$ jets.
}
\end{figure}
In Figures \ref{fig:minvBBpair0jets}, \ref{fig:minvBBpair1jets} and
\ref{fig:minvBBpair2jets}
we study a more exclusive observable
with respect to the one of Figure~\ref{fig:minvbjetpair},
i.e. we consider the production of a pair of $B$ hadrons
and plot the invariant-mass distribution of the pair made by the
two hardest $B$ hadrons in the event,
in events characterised by the presence of
at least 0, 1 or 2 $b$ jets respectively. We do not require that the
two hardest $B$ hadrons belong to any of the tagged jets,
nor we ask that they satisfy any cut in order to be detected.

In the case where no $b$ jet is explicitly requested,
shown in Figure~\ref{fig:minvBBpair0jets},
we observe that the \amcnlonospace+\pythia results are largely independent
of the choice of the variable and of the interval used to extract $\qsh$,
but that there is a strong sensitivity to the QCD-PS model,
with differences between \pythia and \herwig at the $20\%$ level.
Curiously enough, \amcnlonospace+\herwig is rather similar to \powhegboxnospace+\pythianospace.
All matched predictions are considerably softer than those at fixed NLO,
in which the $B$ hadrons are replaced by the $b$ quarks, because
of the loss of energy due to the fragmentation of the latter into the former.

In the case with at least 1 $b$ jet, shown in Figure~\ref{fig:minvBBpair1jets},
we observe in the \amcnlo results that
the choice of the variable used to extract $\qsh$,
namely $\hat s$ vs $H_T/2$,
yields differences at the $10-20\%$ level at large invariant masses.
From the \amcnlonospace+\pythia histograms one can appreciate
the fact that, once the matching scheme and the PS are fixed,
the pattern of the predictions
closely follows those of the shower-scale distribution shown
in Figure~\ref{fig:showerscale},
with the hardest prediction corresponding to the largest shower-scale.
The differences between \pythia and \herwig are sizeable through the whole
invariant mass spectrum, both in shape and in size of the corrections.
We stress that these effects are due to terms beyond NLO-QCD
in the perturbative expansion.
As in the case without explicitly asking extra $b$ jets, predictions
with \amcnlonospace+\herwig are close to those with \powhegboxnospace+\pythianospace.

Finally, in the case with at least 2 $b$ jets,
shown in Figure~\ref{fig:minvBBpair2jets},
we observe in the \amcnlo results that there is a good agreement
between the different options of matching fixed- and all-orders results
and between \pythia and \herwignospace.
We also observe the large size of the radiative effects
in the first two invariant mass bins,
where the higher orders enhance the cross section up to a factor $+120\%$,
while at large invariant masses the corrections range from
being  negative (-20\%) to being compatible with zero.
This large correction is explained as due to the appearance via QCD-PS
of events where both $B$ hadrons belong to the same jet
(because of secondary $g\to b \bar b$ splittings) and
turn out to be the hardest pair,
but, at the same time, have a small invariant mass.
For what concerns the \powhegbox predictions, they fall below and are
manifestly softer than fixed NLO for values of the invariant mass
starting at $100\,\gev$,
with a depletion of rate that can reach $-50\%$ at $m(B_1,B_2) \sim 500\, \gev$.

\subsection{$y-\phi$ distance of the two hardest $b$ jets}
\begin{figure}[!h]
\centering
\includegraphics[width=0.48\textwidth,angle=0,clip=true,trim=0.65cm 6.7cm 0.cm 0.4cm]{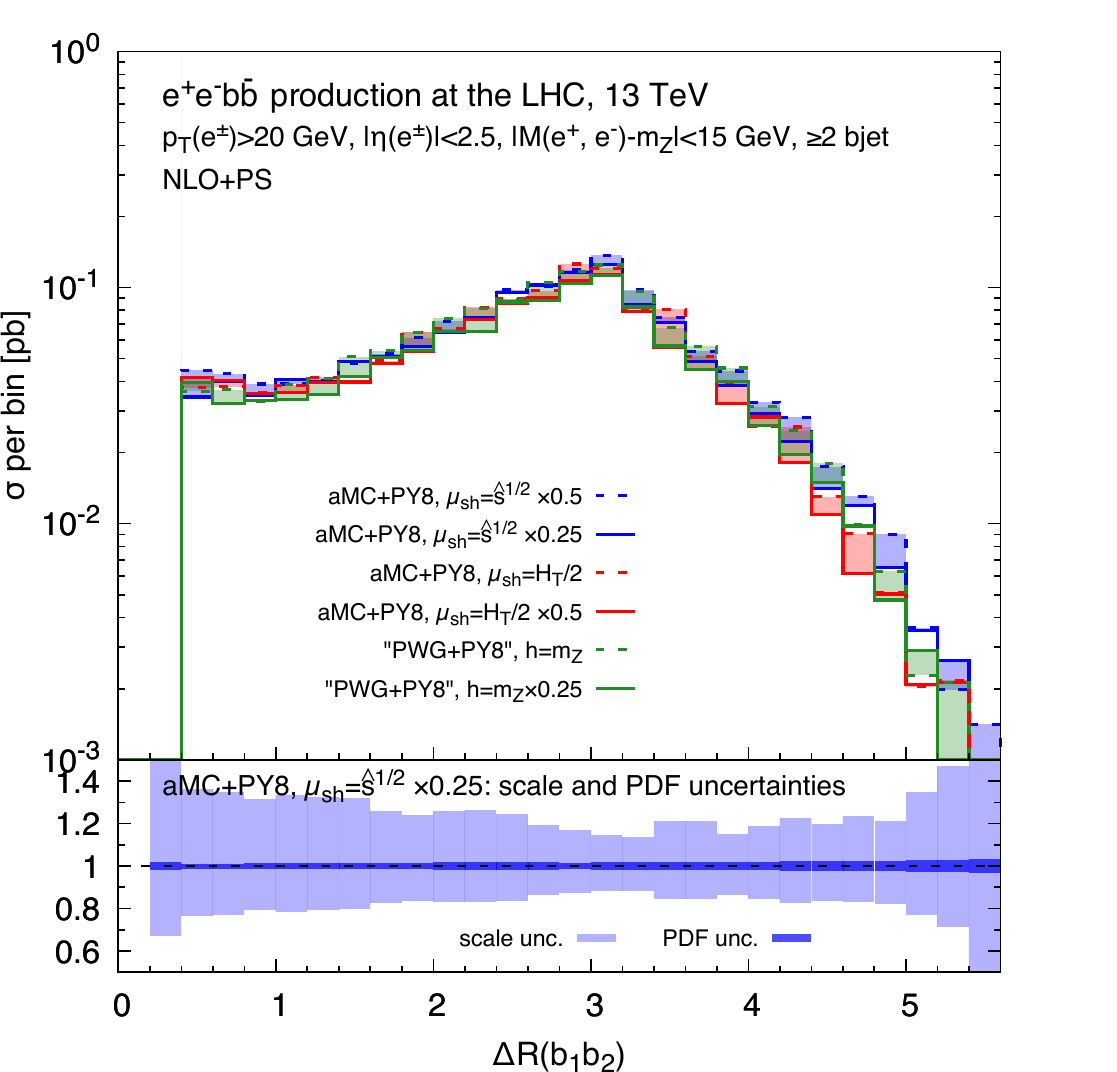}
\includegraphics[width=0.48\textwidth,angle=0,clip=true,trim=0.65cm 6.7cm 0.cm 0.4cm]{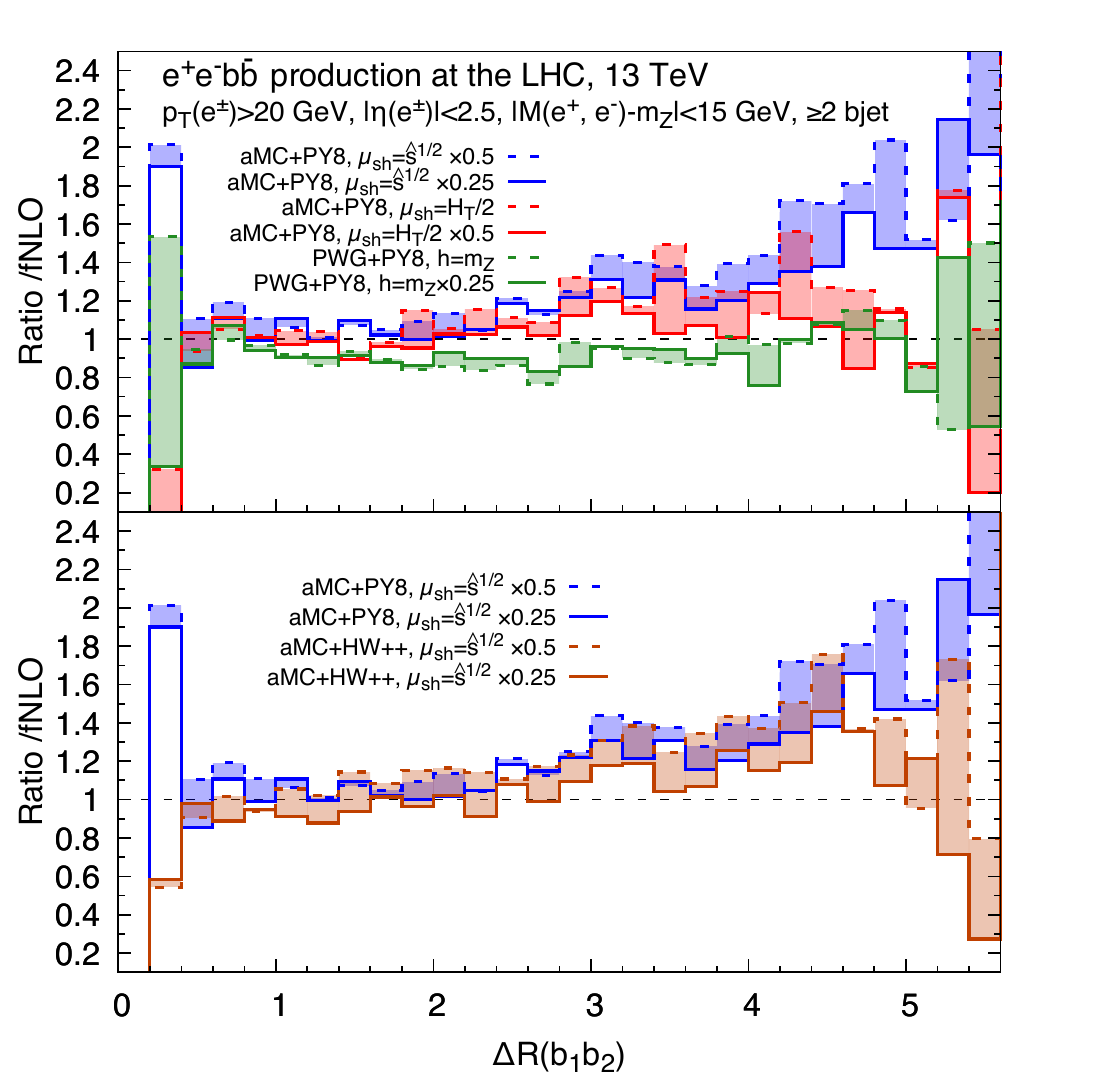}
\caption{\label{fig:deltarbjetpair}
$\Delta R$ distribution of the hardest $b$-jet pair.
}
\end{figure}
We introduce the distance
$\Delta R(ij)\equiv \sqrt{ (y_i-y_j)^2 + (\phi_i-\phi_j)^2   }$
between particles $i$ and $j$,
whose rapidity and azimuthal angle are denoted with $y$ and $\phi$, and show, in Figure~\ref{fig:deltarbjetpair},
the distribution for the distance between the two hardest $b$ jets.
For this observable, matched predictions can display large differences. More in detail,
while the \powhegboxnospace+\pythia\ prediction is rather close to the fixed-order one, with almost no visible shape distortion, the \amcnlo ones show sizeable
deviations, particularly for $\Delta R(b_1,b_2) > \pi$. In this region, the \amcnlonospace+\pythia prediction with the largest shower scale can lead to rates
which are larger than the fixed-order predictions by a factor 1.5-2. This is partially mitigated by the
choice of smaller values for the shower scale $\sim H_T$ or
by matching with \herwignospace. In fact, in these two cases, predictions show a rather similar behaviour: up to $\Delta R(b_1,b_2) = \pi$ they lie quite close
to the fixed-order one; starting from $\Delta R(b_1,b_2) = \pi$ the rate is enhanced with respect to the fixed-order one, up to a factor 1.4 at
$\Delta R(b_1,b_2)=4.5$. Finally, for very large $\Delta R(b_1,b_2)$, these matched predictions seem suppressed with respect to the fixed-order one, although
the statistics for this kinematic region is quite poor.

\subsection{$y-\phi$ distance of the two hardest $B$ hadrons with or without tagged $b$ jets}
\begin{figure}[!h]
\centering
\includegraphics[width=0.48\textwidth,angle=0,clip=true,trim=0.65cm 6.7cm 0.cm 0.4cm]{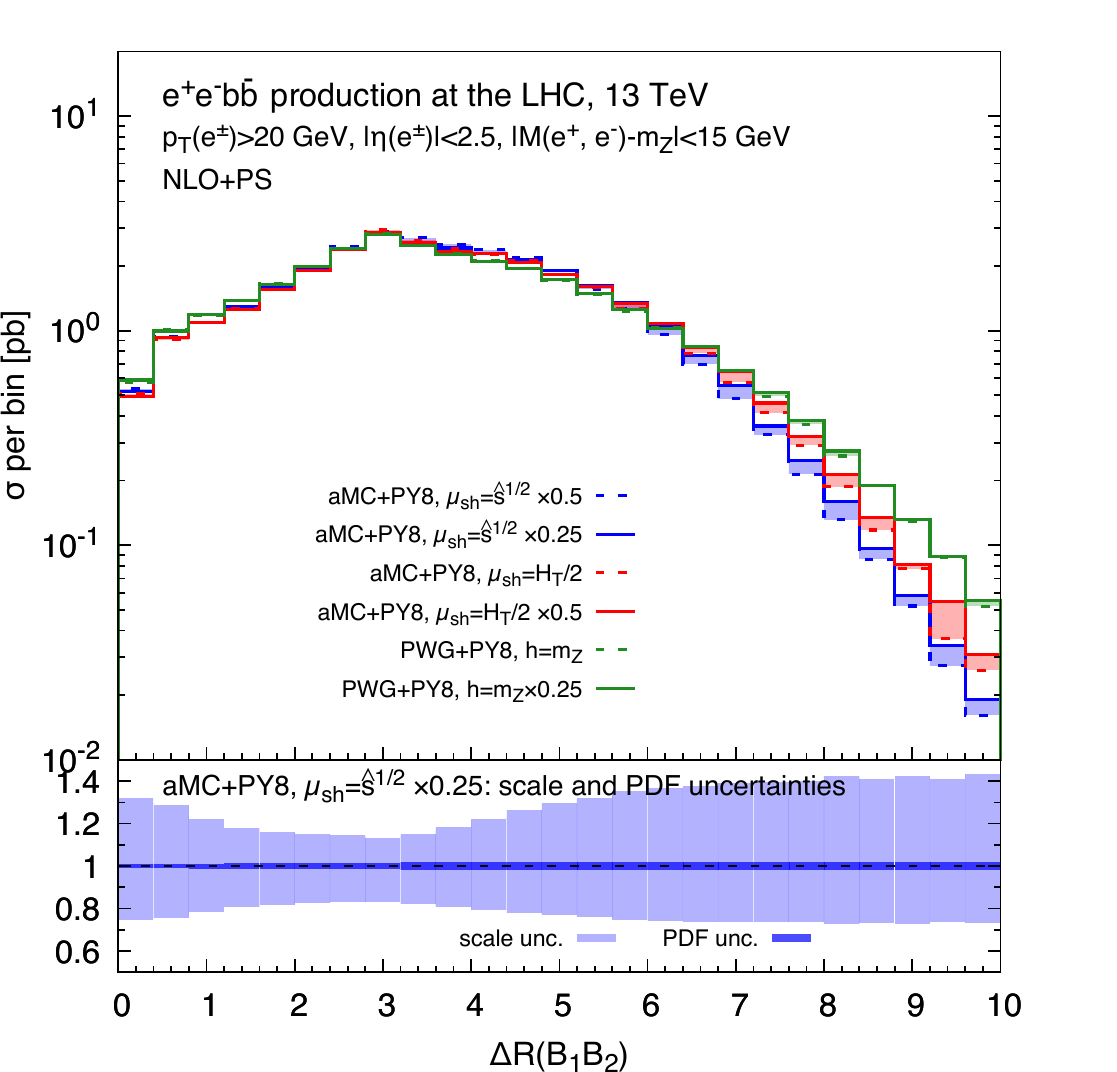}
\includegraphics[width=0.48\textwidth,angle=0,clip=true,trim=0.65cm 6.7cm 0.cm 0.4cm]{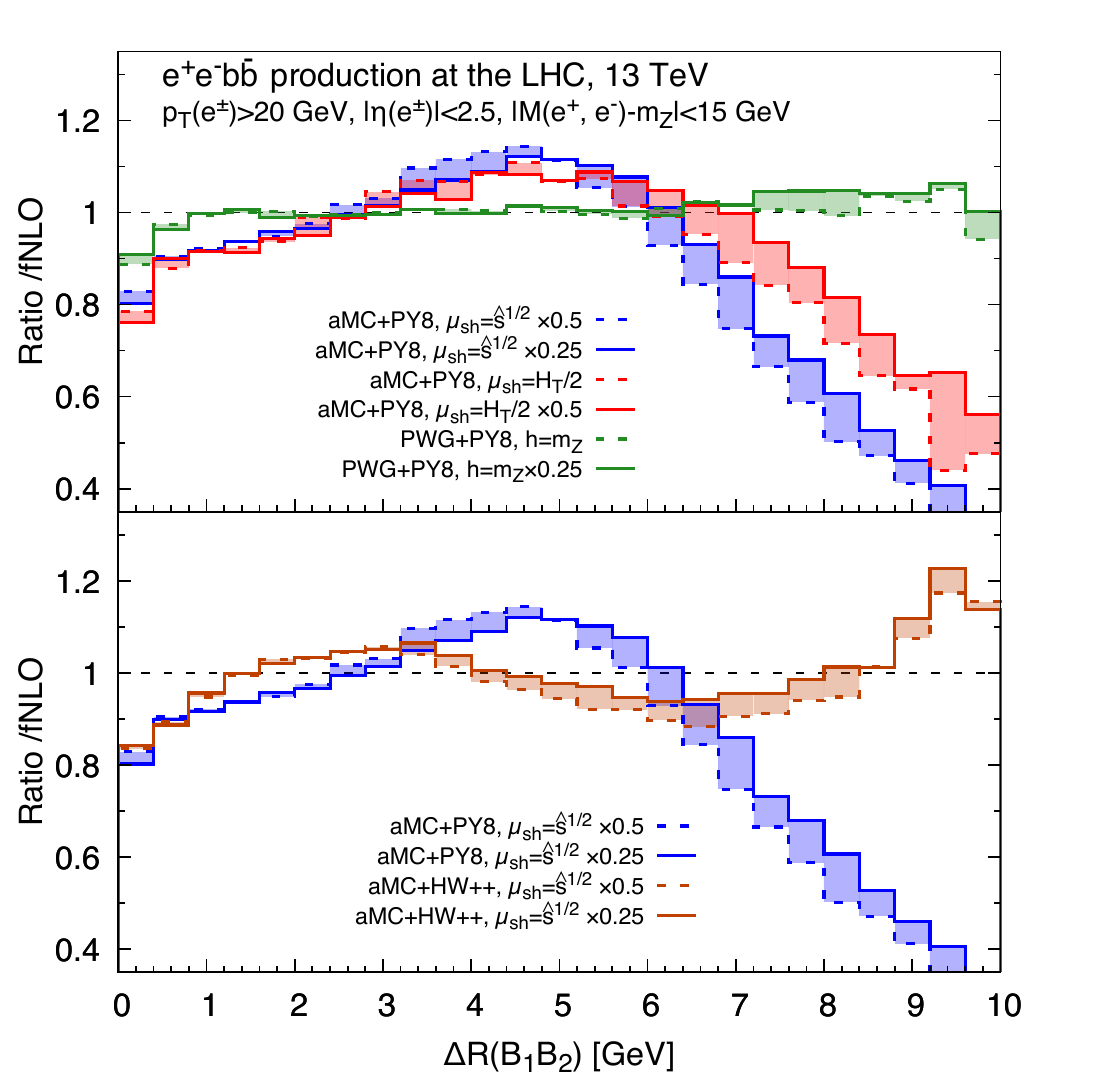}
\caption{\label{fig:deltarBBpair0jets}
$\Delta R$ distribution of the $B$ hadrons pair in association with at least 0 $b$ jets.
}
\end{figure}
\begin{figure}[!h]
\centering
\includegraphics[width=0.48\textwidth,angle=0,clip=true,trim=0.65cm 6.7cm 0.cm 0.4cm]{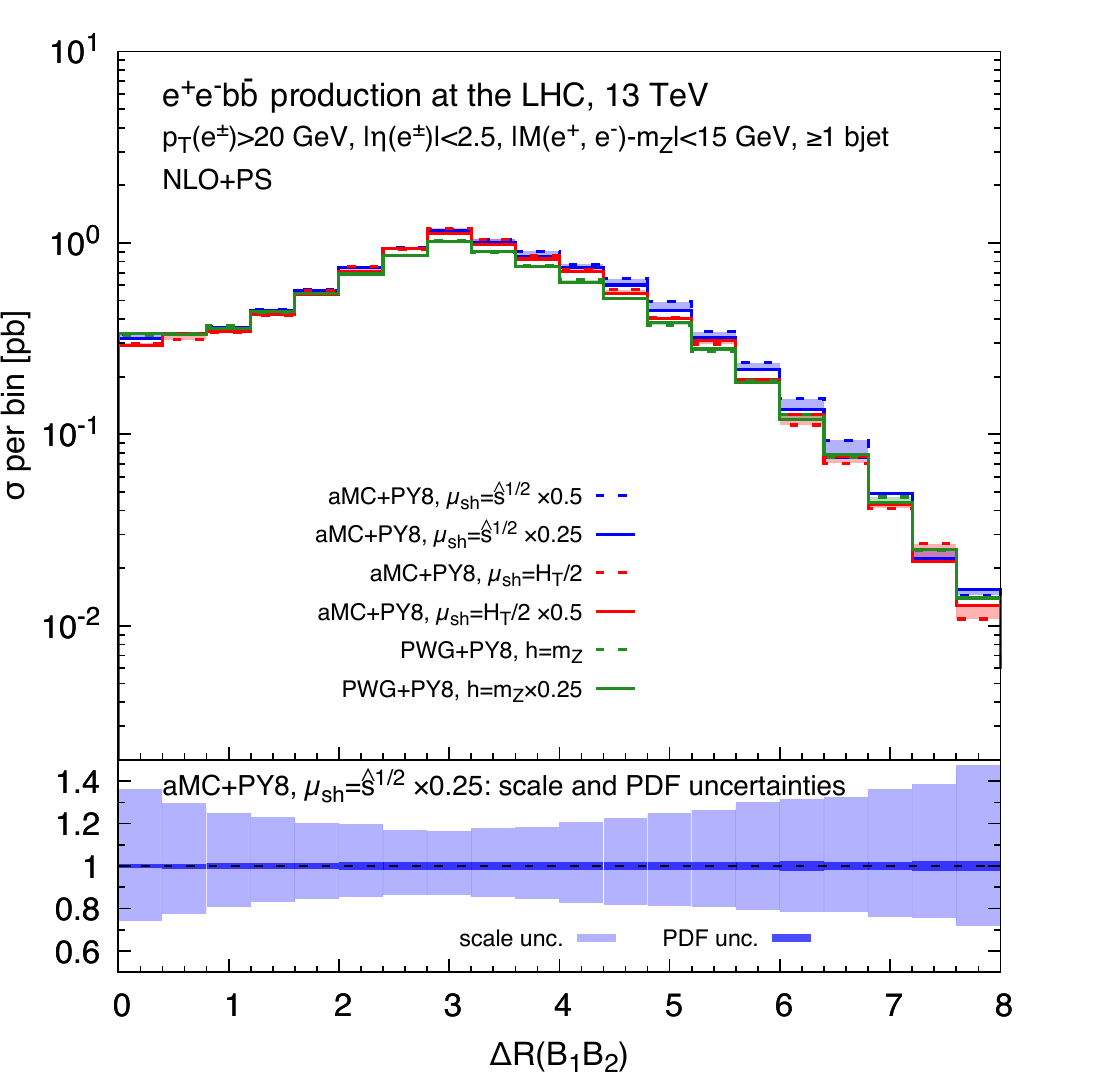}
\includegraphics[width=0.48\textwidth,angle=0,clip=true,trim=0.65cm 6.7cm 0.cm 0.4cm]{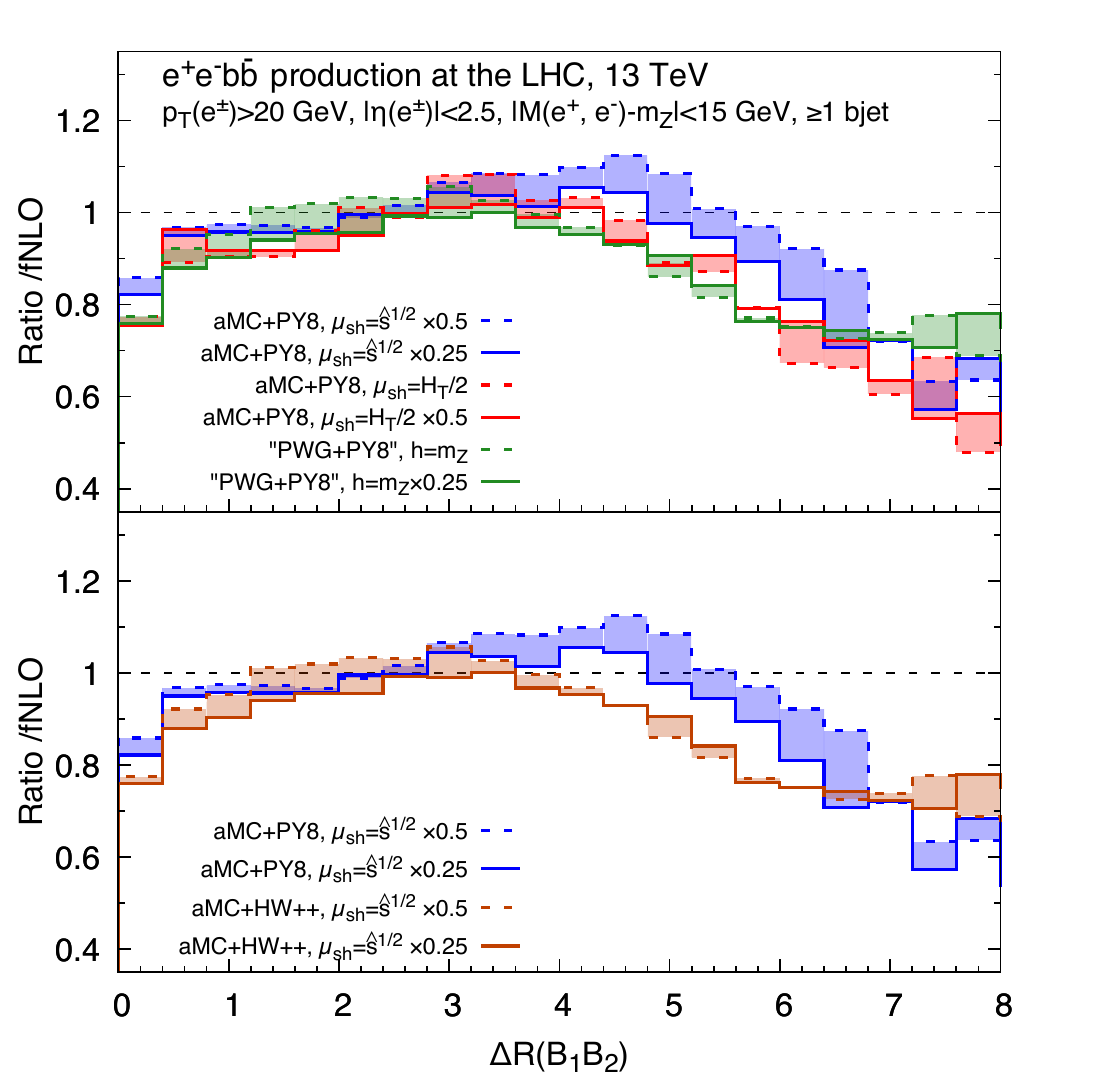}
\caption{\label{fig:deltarBBpair1jets}
$\Delta R$ distribution of the $B$ hadrons pair in association with at least 1 $b$ jet.
}
\end{figure}
\begin{figure}[!h]
\centering
\includegraphics[width=0.48\textwidth,angle=0,clip=true,trim=0.65cm 6.7cm 0.cm 0.4cm]{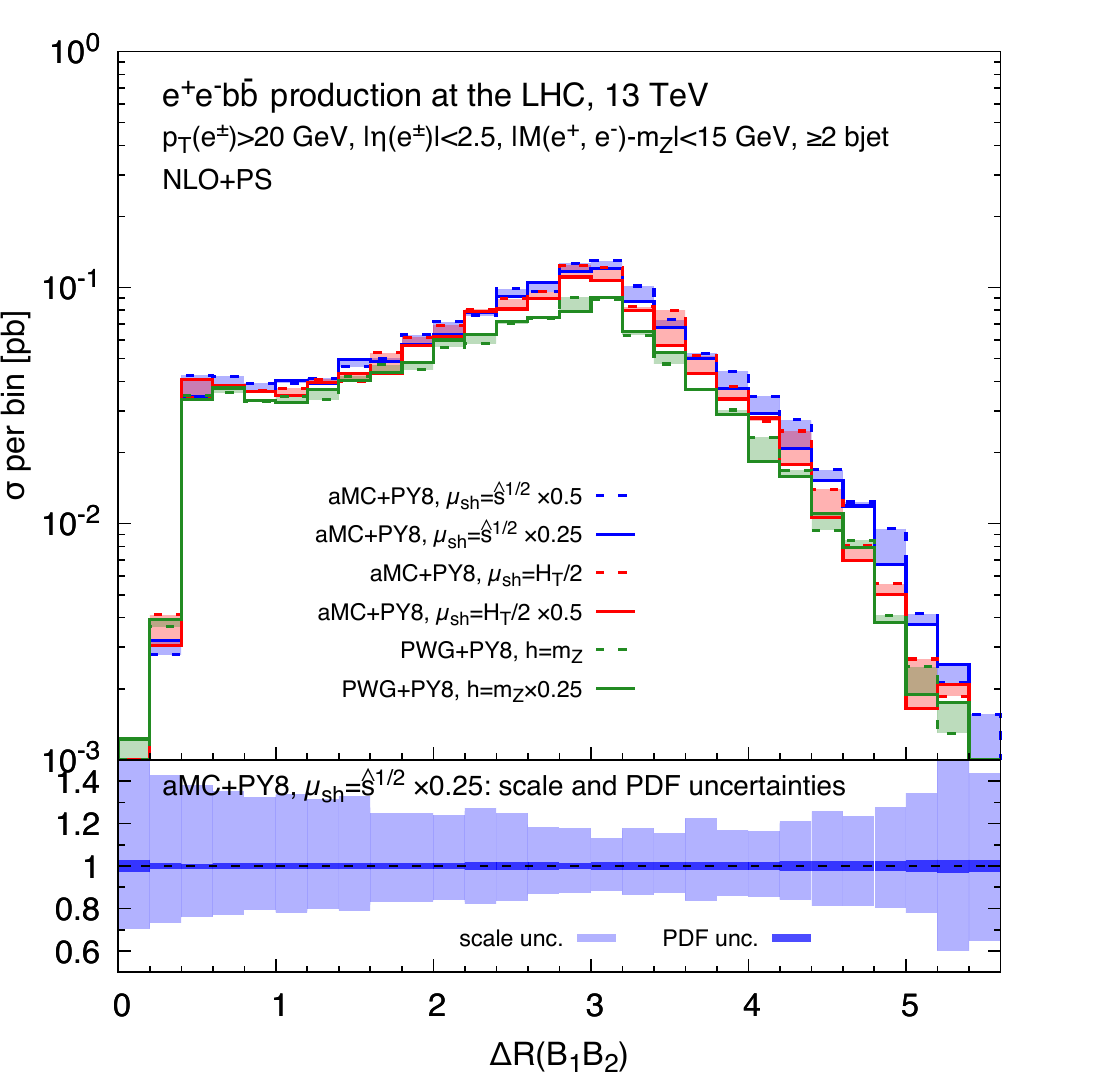}
\includegraphics[width=0.48\textwidth,angle=0,clip=true,trim=0.65cm 6.7cm 0.cm 0.4cm]{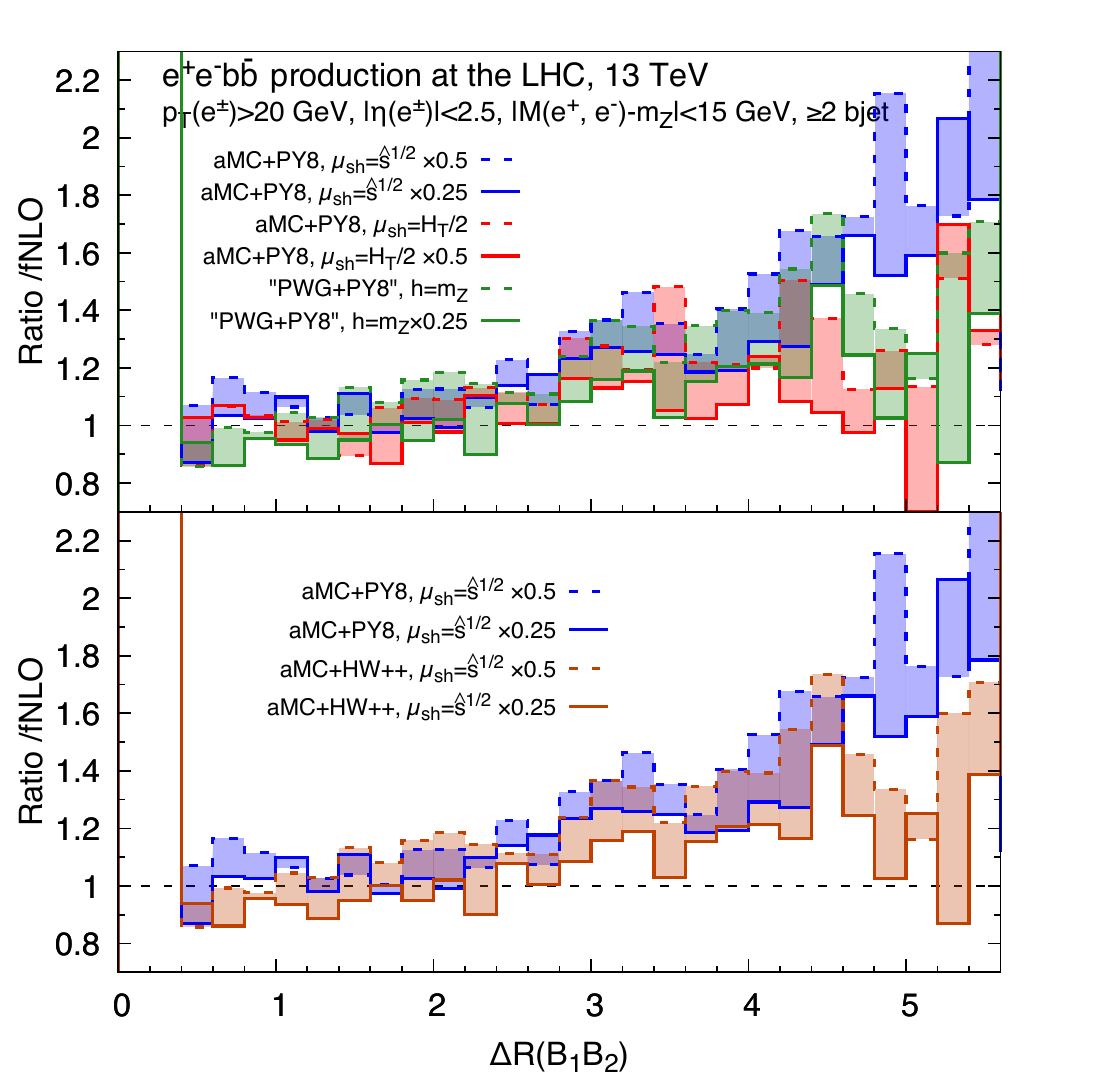}
\caption{\label{fig:deltarBBpair2jets}
$\Delta R$ distribution of the $B$ hadrons pair in association with at least 2 $b$ jets.
}
\end{figure}
We show in Figures \ref{fig:deltarBBpair0jets},\,\ref{fig:deltarBBpair1jets} and
\ref{fig:deltarBBpair2jets},
the distributions with respect to the distance $\Delta R(B_1,B_2)$ between
the two hardest $B$ hadrons,
in presence of an increasing number of $b$ jets (at least 0, 1 and 2).
As it has been the case for the corresponding invariant-mass distributions
(Figures~\ref{fig:minvBBpair0jets}-\ref{fig:minvBBpair2jets}) we do not require that the two hardest $B$ hadrons
belong to any of the tagged jets, nor any condition for tagging the $B$ hadrons is required.\\
When no $b$ jet is explicitly required (Figure~\ref{fig:deltarBBpair0jets}), the two $B$ hadrons
can reach quite large distances ($\Delta R\sim 10$) keeping sizeable rates. However,
for such large distances, the two $B$ hadrons are typically in an extreme forward-backward configuration, in kinematic regions where no or poor detector
coverage exists. Nevertheless, it remain interesting to study how different predictions behave. Looking at the distributions, we observe
important discrepancies among the various predictions: if we compare to the fixed-order distribution, the prediction matched with the \powhegbox shows
the smallest deviations, which remain well below 10\% over all the range that we display. Conversely, the \amcnlonospace-matched predictions show quite large
discrepancies: when \pythia\ is employed, the prediction is suppressed at small and large distances ($\Delta R < 3$ and $\Delta R > 6-7$), while it
is mildly enhanced (up to +10\%) at intermediate distances. While at small and moderate distances the behaviour of the \pythianospace-matched
predictions is only marginally
dependent on the choice of shower scale, and the departure from the fixed-order prediction reaches at most 20\% at very small distances, at large
distances such a dependence is apparent, with larger shower scales leading to bigger suppressions, with the predictions suppressed by a factor two
or more with respect to the fixed-order one.
If instead \amcnlonospace+\herwig is employed,
the behaviour is even more complicated, but overall the deviations with respect to the fixed-order predictions are smaller
than with \pythianospace: at very small distances the \herwignospace-matched prediction lies below the fixed-order one, with a suppression of 20\%. For distances in the
range $1< \Delta R (B_1,B_2) < 4$, the matched prediction lies 5\% above the fixed-order one, while in the range $4< \Delta R (B_1,B_2) < 9$ it is again
below, with a suppression between 10\% and 15\%. Finally, at very large distances $\Delta R (B_1,B_2) > 9$, the matched prediction returns
$\sim20\%$ above the fixed-order one.
This enhancement of the \amcnlonospace+\herwig prediction has been also observed for $b\bar b H$ associated
production~\cite{Wiesemann:2014ioa} and for charged-Higgs production in association with
a top quark~\cite{Degrande:2015vpa}.\\
Requiring at least one $b$ jet partially mitigates these discrepancies: the \powhegbox prediction is very similar to the \amcnlonospace+\herwig one, and both are
also similar to the one obtained with \amcnlonospace+\pythia and with a shower scale $\sim H_T$, up to $\Delta R (B_1, B_2)=6$ (for larger distances, the latter
prediction predicts a suppressed rate with respect to the former ones). When these matched predictions are compared to the fixed-order one, the behaviour
is not much different from the case without extra jets: predictions are suppressed (up to -20\%) for small distances ($\Delta R (B_1, B_2)<1$), for
intermediate distances ($1< \Delta R (B_1, B_2)<4$) they behave similarly to the fixed-order one, while at larger distances they are again suppressed
(-30\% for the \powhegbox and \amcnlonospace+\herwig and up to -50\% for \amcnlonospace+\pythia\ with $\qsh \sim H_T$). Finally, the \amcnlonospace+\pythia with
$\qsh \sim \sqrt {\hat s}$ follows the other matched predictions up to $\Delta R (B_1, B_2)=3.5$. It then keeps growing with respect to the
fixed-order one for about one unit of distances, where the enhancement with respect to the fixed-order prediction reaches +10\%, finally for large
distances it predicts
suppressed rates with respect to the fixed order one, with the suppression reaching -40\%. For this last prediction, a variation of $\qsh$ by a factor two
can have an effect as large as 10\% for $\Delta R (B_1, B_2)>4$, while for the other predictions the shower-scale dependence is much smaller.\\
Finally, when two $b$ jets are required, Figure~\ref{fig:deltarBBpair2jets}, the $\Delta R (B_1B_2)$
distributions closely follow the corresponding counterparts
for the distance between the two $b$ jets, shown in Figure~\ref{fig:deltarbjetpair}.

\subsection{Transverse momentum distributions of the two hardest $b$ jets}
\begin{figure}[!h]
\centering
\includegraphics[width=0.48\textwidth,angle=0,clip=true,trim=0.65cm 6.7cm 0.cm 0.4cm]{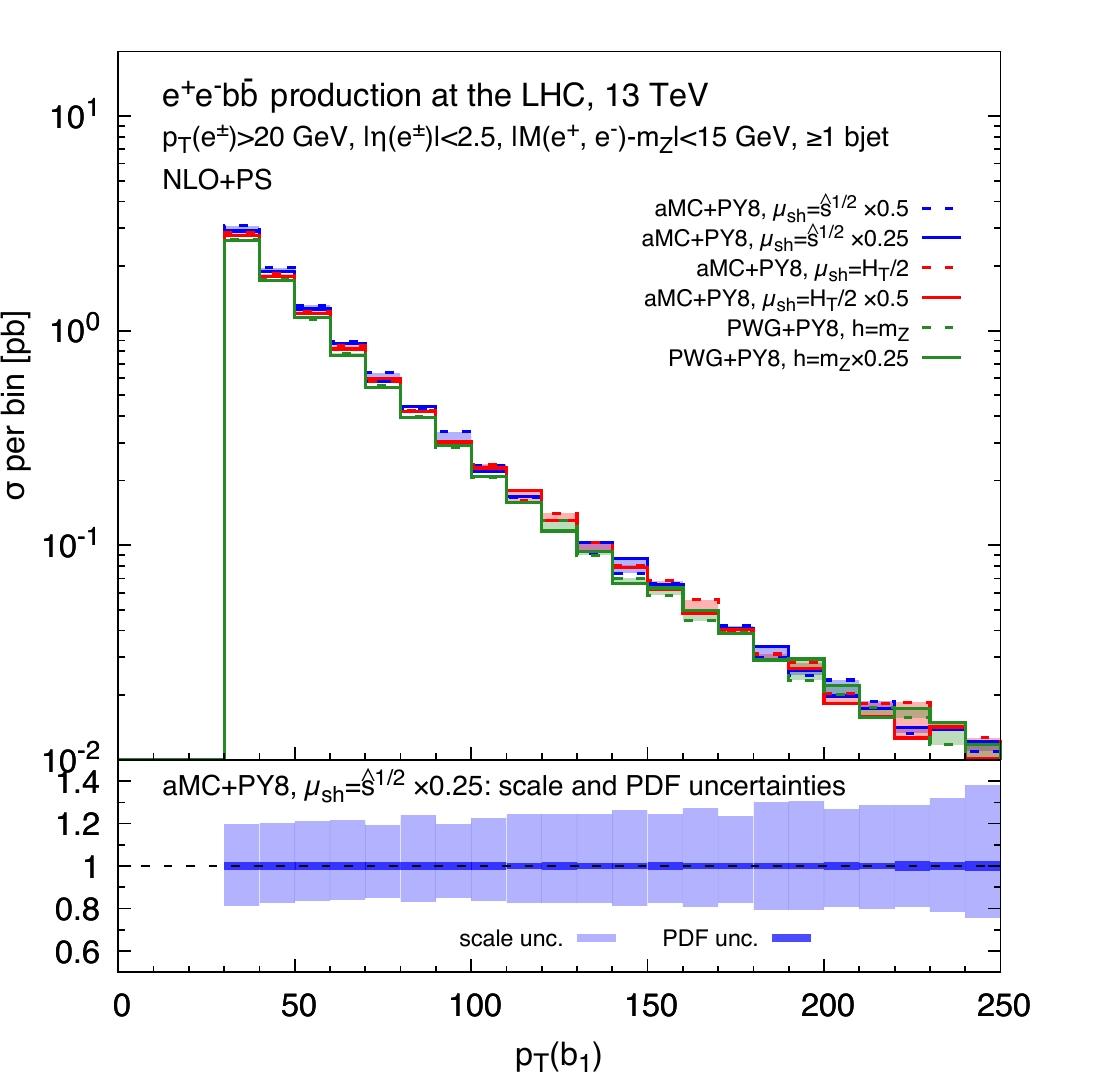}
\includegraphics[width=0.48\textwidth,angle=0,clip=true,trim=0.65cm 6.7cm 0.cm 0.4cm]{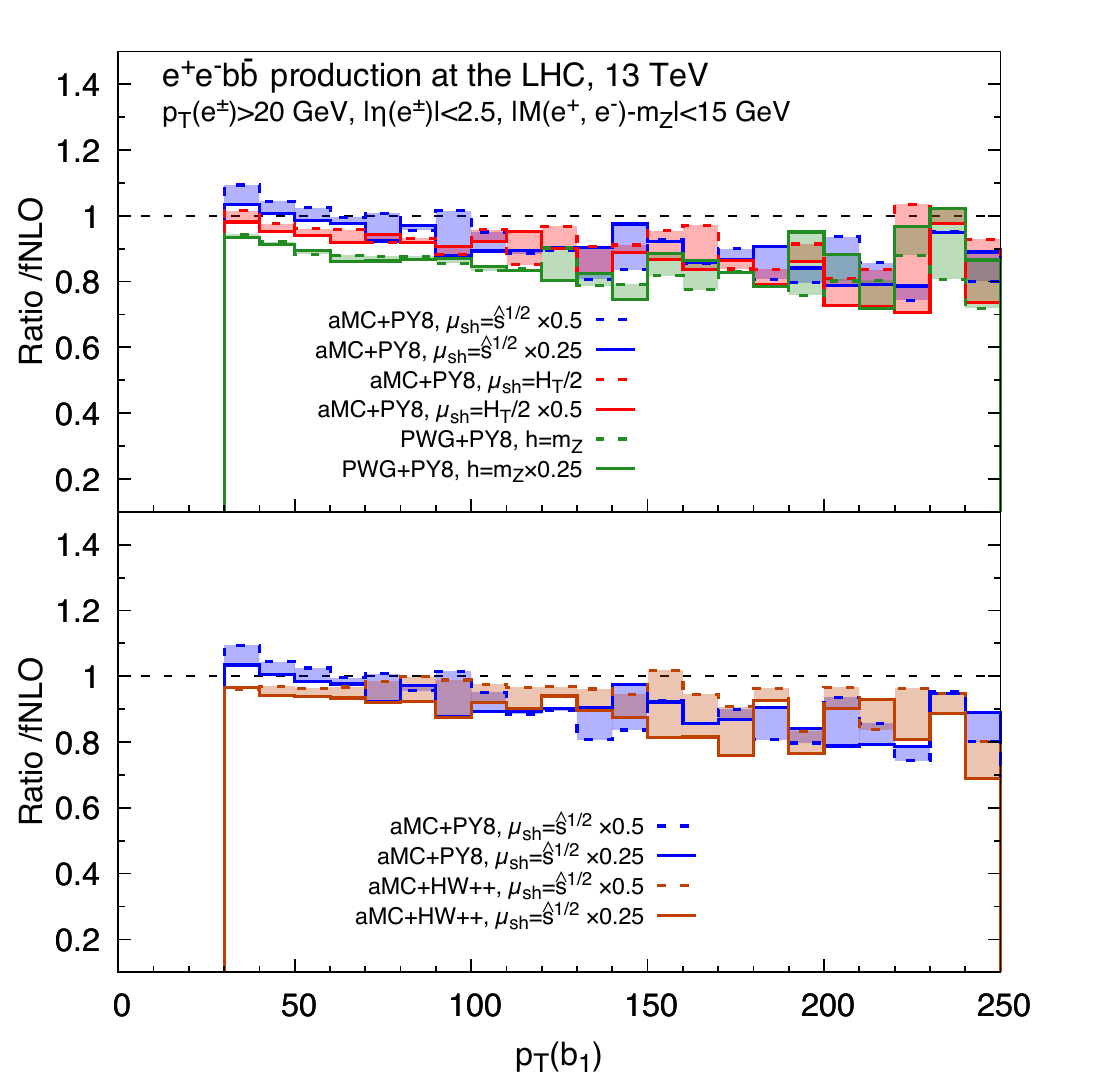}
\caption{\label{fig:ptbjet1}
Transverse momentum distribution of the hardest $b$ jet.
}
\end{figure}
\begin{figure}[!h]
\centering
\includegraphics[width=0.48\textwidth,angle=0,clip=true,trim=0.65cm 6.7cm 0.cm 0.4cm]{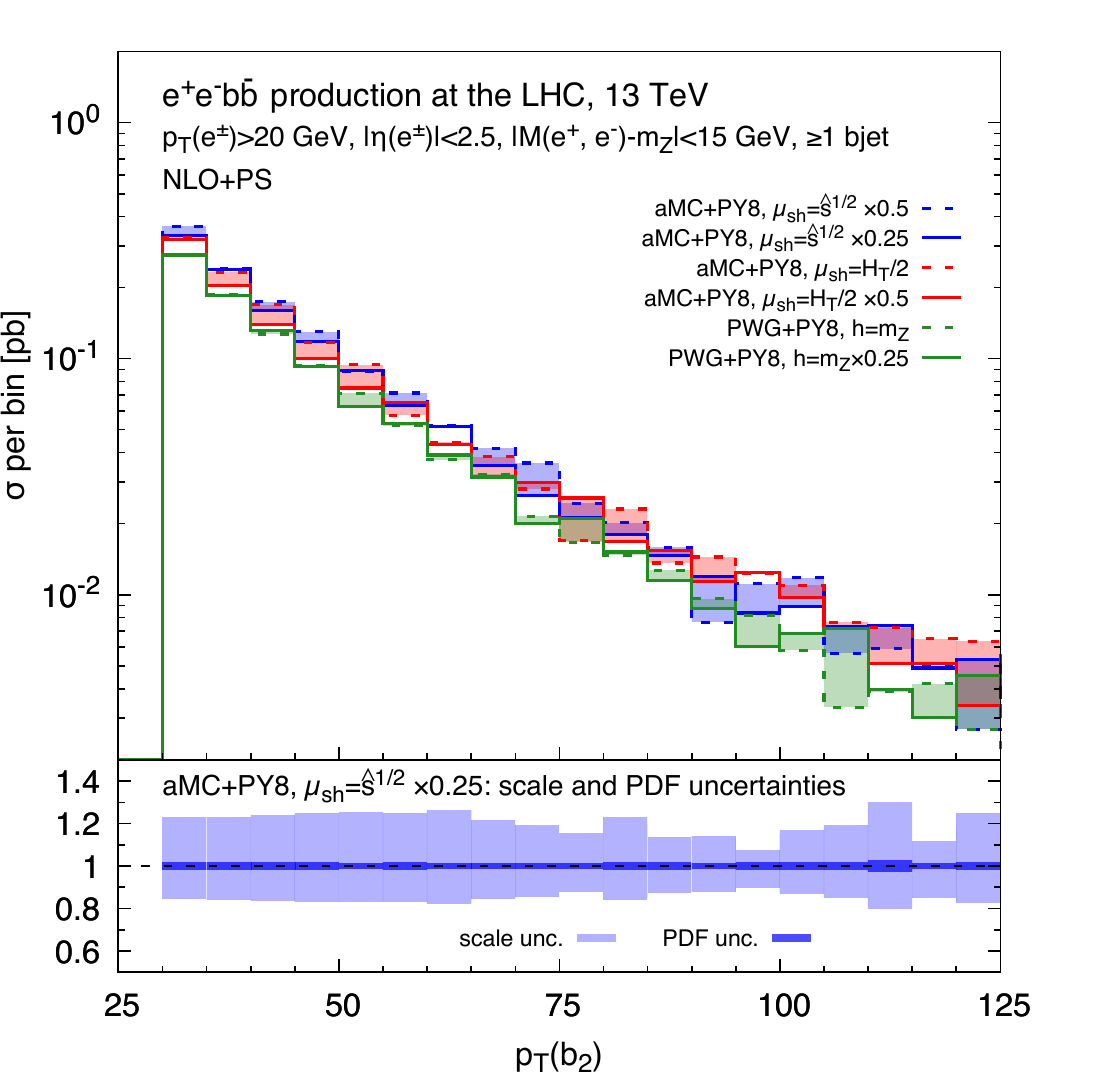}
\includegraphics[width=0.48\textwidth,angle=0,clip=true,trim=0.65cm 6.7cm 0.cm 0.4cm]{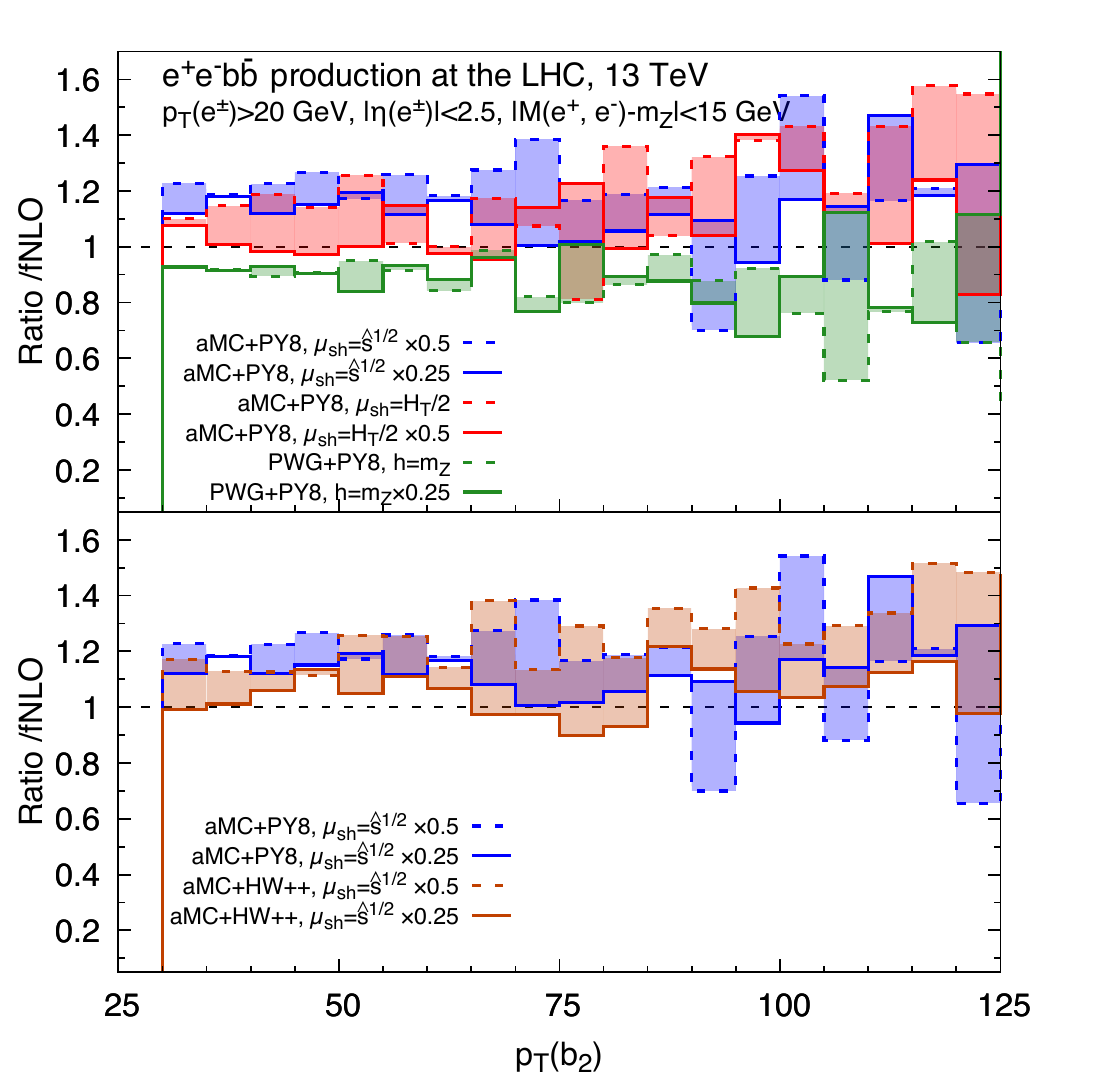}
\caption{\label{fig:ptbjet2}
Transverse momentum distribution of the second-hardest $b$ jet.
}
\end{figure}
In figures \ref{fig:ptbjet1},\,\ref{fig:ptbjet2}
we show the transverse momentum
of the hardest and the second hardest $b$ jets. For the transverse momentum of the hardest jet, the general behaviour of
 matched computations is to be softer than the fixed-order prediction, and this effect is more pronounced for predictions matched with
 \pythia\ than for the ones matched with \herwignospace. For small values of the transverse momentum, differences among the matched simulations
are moderate (at the level of 10\%) and reflect the pattern observed for the one-jet multiplicity displayed in Figure~\ref{fig:nbjetbjet}, while for
larger values such differences are mitigated.\\
For the second-hardest $b$ jet, no visible distortions of the matched spectra with respect to the fixed-order one can be appreciated, and differences
in rate reflect those of the two-jet bin in Figure~\ref{fig:nbjetbjet}.

\subsection{Pseudo-rapidity distributions of the two hardest $b$ jets}
\begin{figure}[!h]
\centering
\includegraphics[width=0.48\textwidth,angle=0,clip=true,trim=0.65cm 6.7cm 0.cm 0.4cm]{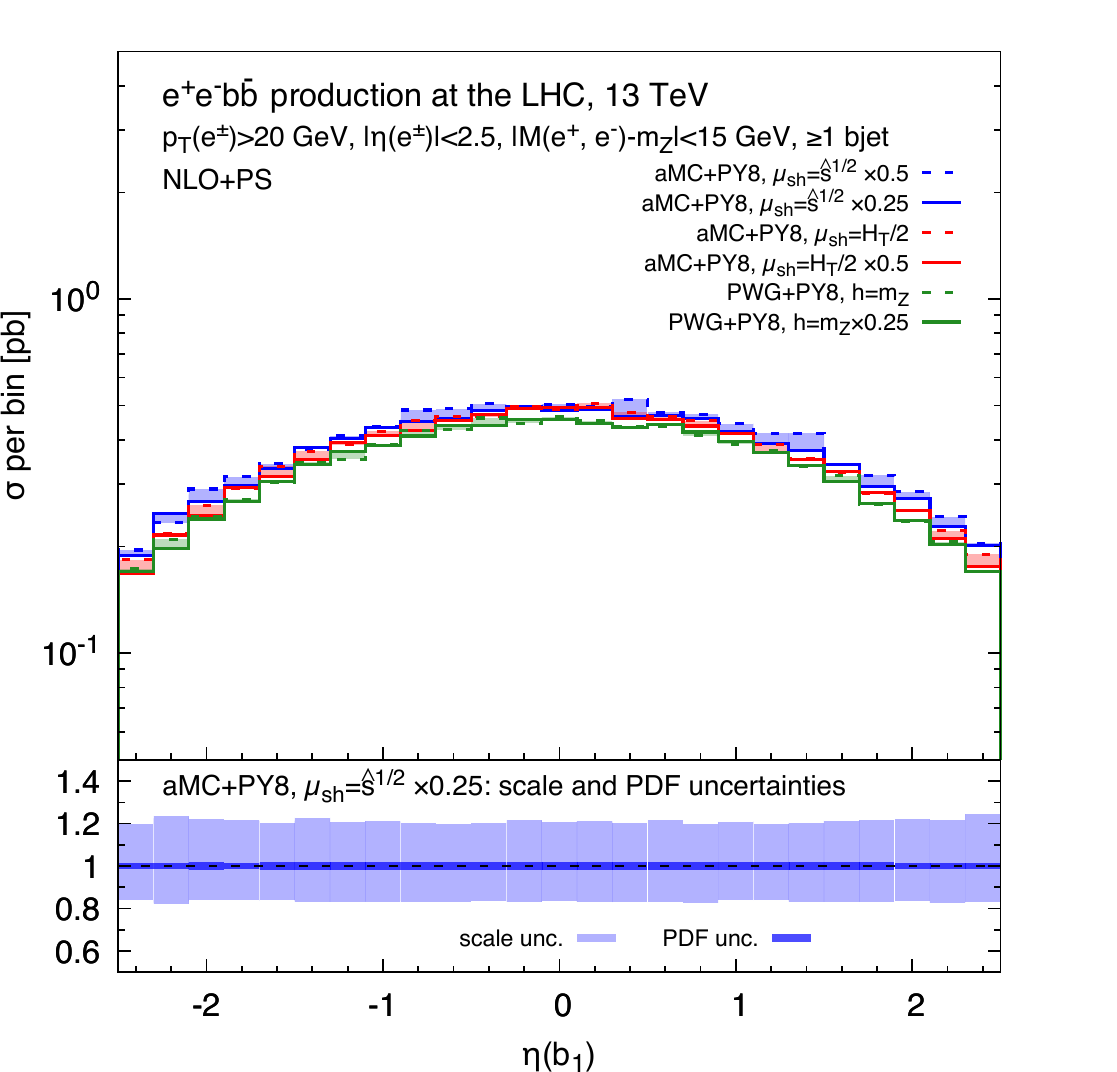}
\includegraphics[width=0.48\textwidth,angle=0,clip=true,trim=0.65cm 6.7cm 0.cm 0.4cm]{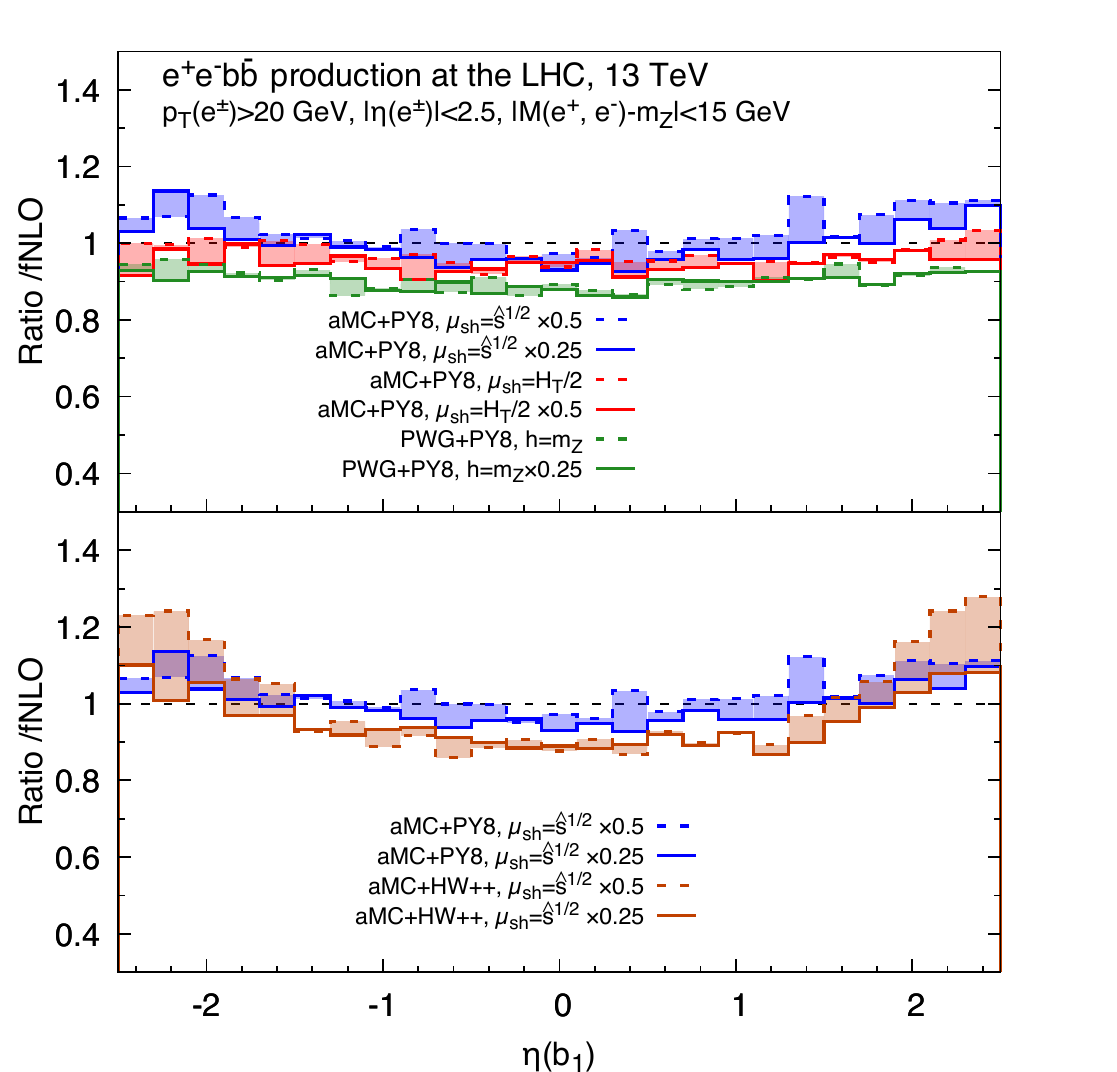}
\caption{\label{fig:etabjet1}
Pseudo-rapidity distribution of the hardest $b$ jet.
}
\end{figure}
\begin{figure}[!h]
\centering
\includegraphics[width=0.48\textwidth,angle=0,clip=true,trim=0.65cm 6.7cm 0.cm 0.4cm]{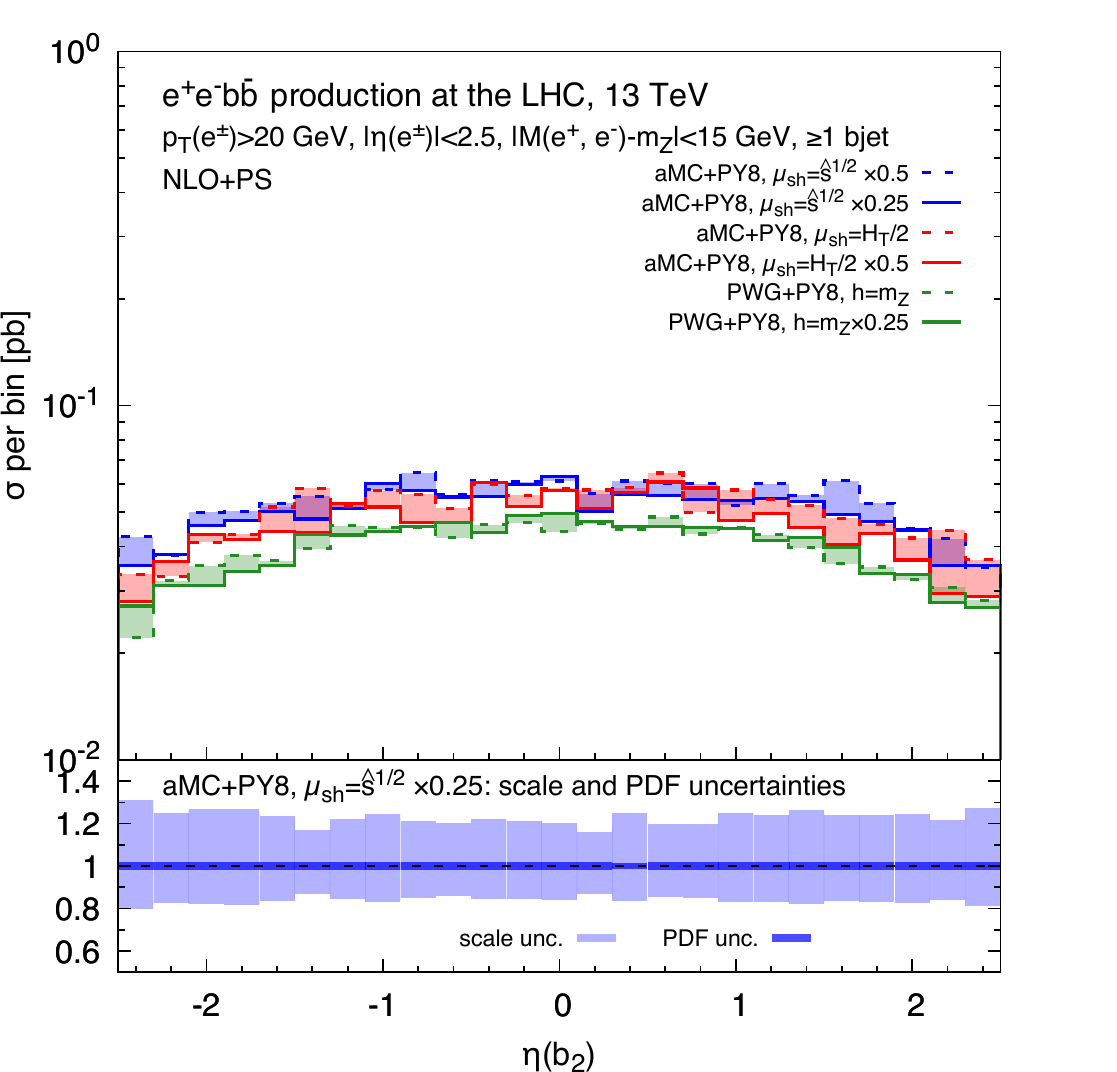}
\includegraphics[width=0.48\textwidth,angle=0,clip=true,trim=0.65cm 6.7cm 0.cm 0.4cm]{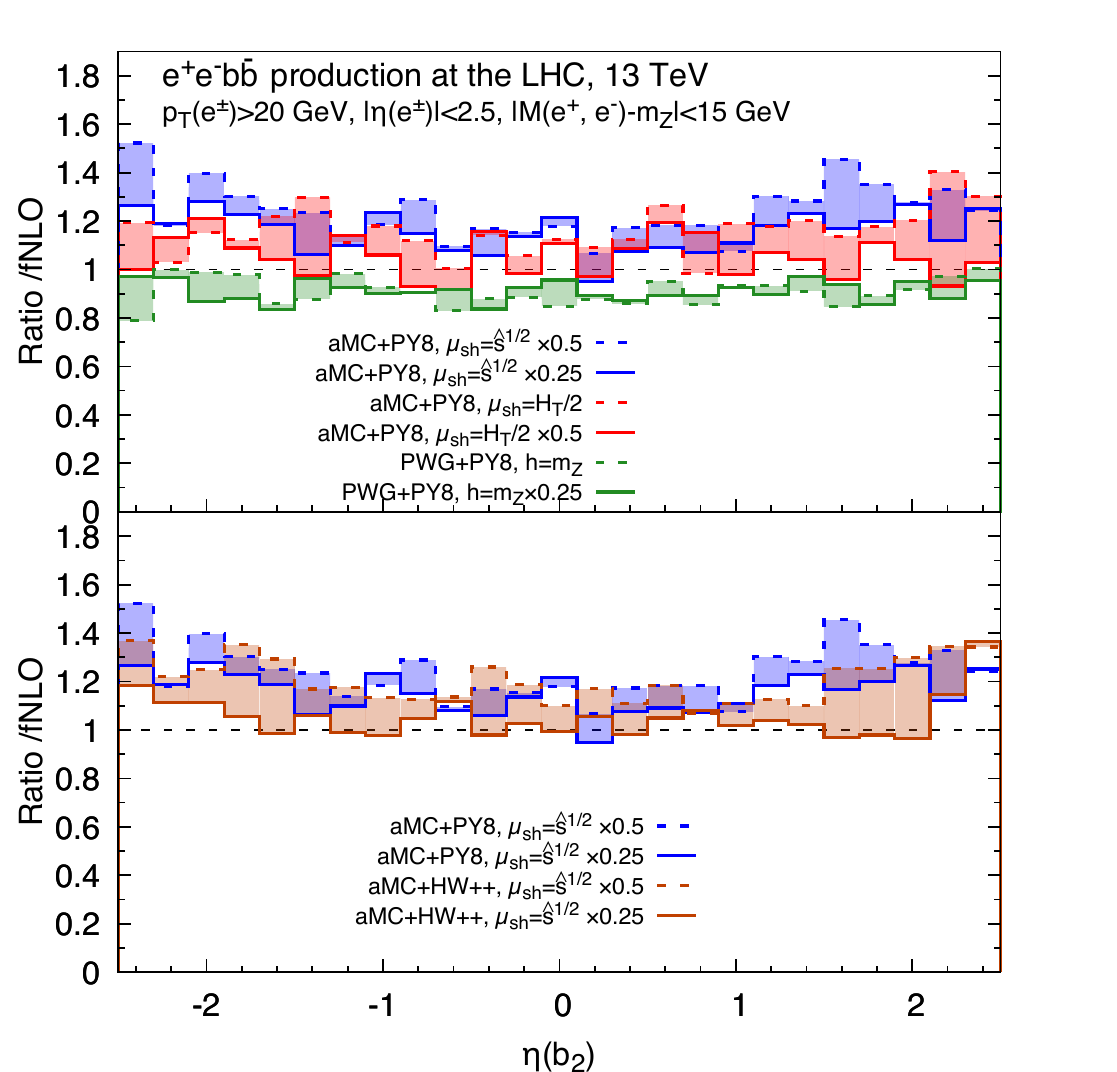}
\caption{\label{fig:etabjet2}
Pseudo-rapidity distribution of the second-hardest $b$ jet.
}
\end{figure}
In Figures \ref{fig:etabjet1},\,\ref{fig:etabjet2}
we show the pseudo-rapidity distributions
of the hardest and the second hardest $b$ jets. As it has been the case for their transverse-momentum counterpart, differences
in rate reflect the one-jet and two-jet bins of Figure~\ref{fig:nbjetbjet}. Besides these differences, it is worth to note that matched predictions
have the general tendency to populate more the forward and backward regions with respect to the fixed-order ones. Such a tendency is more pronounced
for the first jet than for the second, and when larger values of the shower scale ($\sim \hat s$) are employed, in particular for \herwignospace.

\clearpage{}

\bibliographystyle{JHEP}
\bibliography{bmvz}

\end{document}